\documentclass[number,preprint,3p]{elsarticle}

\usepackage{graphicx}
\usepackage{float}
\usepackage{setspace}
\usepackage{subcaption}
\usepackage{color}
\usepackage{xurl}
\usepackage{booktabs}
\usepackage{multirow}
\usepackage{tabularx}
\usepackage{array}
\newcolumntype{Y}{>{\raggedright\arraybackslash}X}
\usepackage[table]{xcolor}
\usepackage{caption}
\usepackage{nicefrac}
\usepackage{siunitx}
\usepackage{tcolorbox}
\usepackage{tablefootnote}
\usepackage[perpage]{footmisc} % restart footnote numbering on each page
\usepackage{epstopdf}
\usepackage{soul}

\usepackage{amsmath, amssymb, amsthm, mathtools, bm, dsfont, bbm}

\DeclareMathOperator{\erf}{erf}

\newcommand{\Prob}[1]{\mathbb{P}\left(#1\right)}
\newcommand{\E}[1]{\mathbb{E}\left[#1\right]}

\newcommand{\vect}[1]{\boldsymbol{#1}}

\usepackage{algorithm}


\usepackage[unicode,colorlinks]{hyperref}

\makeatletter
\gdef\emailauthor#1#2{\stepcounter{ead}%
  \g@addto@macro\@elseads{\raggedright%
    \let\corref\@gobble\def\@@tmp{#1}%
    \eadsep{\ttfamily\href{mailto:\expandafter\strip@prefix\meaning\@@tmp}%
      {\expandafter\strip@prefix\meaning\@@tmp}}%
    (#2)\def\eadsep{\unskip,\space}}%
}
\makeatother


\makeatletter
\newcommand{\MaketitleBoxNoRules}{%
  \resetTitleCounters
  \def\baselinestretch{1}%
  \begin{\elsarticletitlealign}%
    \def\baselinestretch{1}%
    \Large\@title\par\vskip18pt
    \ifdoubleblind
      \vspace*{2pc}
    \else
      \normalsize\elsauthors\par\vskip10pt
      \footnotesize\itshape\elsaddress\par\vskip36pt
    \fi
  \end{\elsarticletitlealign}%
}
\makeatother

\makeatletter
\newcommand{\ResetELSFrontmatter}{%
  \gdef\@title{}%
  \gdef\@subtitle{}%
  \gdef\elsauthors{}%
  \gdef\elsaddress{}%
  \gdef\@elsauthors{}%
  \gdef\@elsaddress{}%
  \gdef\@elseads{}%
  \gdef\@elsuads{}%
  \@ifundefined{@cornotes}{}{\gdef\@cornotes{}}%
  \@ifundefined{@cortext}{}{\gdef\@cortext{}}%
  \@ifundefined{@corref}{}{\let\@corref\@empty}%
  \setcounter{affn}{0}%
  \setcounter{ead}{0}%
  \setcounter{cnote}{0}%
  \setcounter{fnote}{0}%
  \setcounter{tnote}{0}%
  \resetTitleCounters
}
\makeatother

\makeatletter
\newcommand{\tableofcontentsSI}{%
  \section*{\LARGE \contentsname}%
  \@starttoc{sitoc}%
}
\makeatother

\renewcommand{\figurename}{Fig.}
\renewcommand{\thefootnote}{\fnsymbol{footnote}}
\usepackage{bibunits}
\defaultbibliographystyle{plainnat}
\biboptions{numbers,sort&compress,square}
\journal{arXiv}

\newcommand{\SupplNotesHeading}{%
  \section*{\LARGE Supplementary Notes}%
  \addcontentsline{sitoc}{section}{Supplementary Notes}%
  \setcounter{section}{0}%
  \renewcommand{\thesection}{Supplementary Note~\arabic{section}}%
}

\newcommand{\SupplMethodsHeading}{%
  \section*{\LARGE Supplementary Methods}%
  \addcontentsline{sitoc}{section}{Supplementary Methods}%
  \setcounter{section}{0}%
  \renewcommand{\thesection}{Supplementary Method~\arabic{section}}%
}

\newcommand{\SupplFiguresHeading}{%
  \section*{\LARGE Supplementary Figures}%
  \addcontentsline{sitoc}{section}{Supplementary Figures}%
}

\makeatletter
\newcommand{\suppnote}[1]{%
  \refstepcounter{section}%
  \protected@edef\@currentlabel{\arabic{section}}% what \ref prints
  \section*{Supplementary Note \arabic{section}. #1}% what is shown in the title
  \addcontentsline{sitoc}{subsection}{Supplementary Note \arabic{section}. #1}%
}

\newcommand{\suppmethod}[1]{%
  \refstepcounter{section}%
  \protected@edef\@currentlabel{\arabic{section}}% what \ref prints
  \section*{Supplementary Method \arabic{section}. #1}%
  \addcontentsline{sitoc}{subsection}{Supplementary Method \arabic{section}. #1}%
}
\makeatother

\begin{document}

% =========================
% Main paper front page
% =========================
\begin{frontmatter}

\title{Phase Transitions in Collective Damage of Civil Structures \\under Natural Hazards}

\author[1]{Sebin Oh}
\author[2]{Jinyan Zhao}
\author[3]{Raul Rincon}
\author[3]{Jamie E. Padgett}
\author[1]{Ziqi Wang\corref{cor1}}
\ead{ziqiwang@berkeley.edu}
\cortext[cor1]{Corresponding author}

\address[1]{Department of Civil and Environmental Engineering, University of California, Berkeley, CA, United States of America}
\address[2]{Department of Mechanical and Civil Engineering, California Institute of Technology, Pasadena, CA, United States of America}
\address[3]{Department of Civil and Environmental Engineering, Rice University, Houston, TX, United States of America}

\begin{abstract}
The fate of cities under natural hazards depends not only on hazard intensity but also on the coupling of structural damage, a collective process that remains poorly understood. Here we show that urban structural damage exhibits phase-transition phenomena. As hazard intensity increases, the system can shift abruptly from a largely safe to a largely damaged state, analogous to a discontinuous phase transition in statistical physics. Higher diversity in the building portfolio smooths this transition, but multiscale damage clustering traps the system in a highly volatile regime that suppresses the emergence of predictable Gaussian statistics. The observed phenomenological patterns are interpreted through an effective random-field Ising model, with its key parameters translated into tangible engineering variables such as hazard demand, structural diversity, and modeling uncertainty. These insights establish a phase-aware description of collective damage, revealing that common engineering simplifications can distort the inherent phase of real urban portfolios. Critically, such phase distortions can substantially bias hazard loss estimates, producing relative errors of up to 50\% in exceedance probabilities under moderate earthquakes---equivalent to a several-fold gap in anticipated repair costs. The phase-aware description turns the collective behavior of civil infrastructure damage into actionable diagnostics for urban risk assessment and planning.
\end{abstract}
    

\end{frontmatter}

\begin{bibunit}
% ---------------------------------------------------
% Main text
% ---------------------------------------------------
\section{Introduction}
\noindent
Cities are increasingly exposed to natural hazards whose impacts extend far beyond individual buildings. The 2023 Kahramanmaraş earthquakes ($M_w$ 7.8 and 7.5) severely damaged over $280{,}000$ buildings and displaced millions, causing losses exceeding US\$34 billion~\cite{world_bank_global_2023,ocha_humanitarian_2023, dilsiz_steer_2023}. Similarly, the 2024 Category 4 Hurricane Helene unleashed catastrophic flooding across the southeastern United States, resulting in losses estimated at US\$78.7 billion~\cite{noaa_ncei_billion_dollar_disasters_2025, brennan_tropical_2024}. Such extensive impacts necessitate a move beyond parcel-level analysis toward regional assessments that capture the emergence of collective vulnerabilities~\cite{sadhasivam_building_2025, robinson_use_2018, buldyrev_catastrophic_2010, vespignani_fragility_2010, meena_emergent_2023}. 

Facilitating this shift, rapid advances in computational modeling and data availability have enabled the prediction of regional disaster impacts at unprecedented granularity~\cite{mckenna_nheri-simcenterr2dtool_2025,zsarnoczay_open-source_2025, van_de_lindt_interdependent_2023}. Fine-grained building inventories, physics-based hazard models, and field reconnaissance data can now be integrated seamlessly into probabilistic analysis frameworks that evaluate hazard impacts at building-level resolution~\cite{zhong_regional_2024, dahal_high-fidelity_2025, zsarnoczay_open-source_2025, angeles_advancing_2023}. This has spurred a widespread focus on fidelity: improving model accuracy, refining component-level physics, and accelerating simulations~\cite{dahal_high-fidelity_2025, gupta_machine_2025, xu_regional-scale_2023, sheibani_accelerated_2021, jenkins_physics-based_2023, yabe_toward_2022, zhao_integrated_2026}, implicitly assuming that increasing local precision automatically yields a better understanding of global resilience.

Yet a fundamental question persists: even if every structure in a city could be modeled with absolute fidelity, would the governing principles of urban resilience emerge? We contend they would not. Beyond the barriers of computational cost, the sheer granularity of such an approach risks obscuring damage patterns behind a fog of detail. This mirrors the philosophy of statistical physics: the laws of thermodynamics emerge not from tracking individual molecular trajectories, but from identifying how micro-interactions organize into macro-behavior~\cite{ma_modern_2019, kadanoff_statistical_2000}. Similarly, natural hazards engineering must transcend the pursuit of ever-finer resolution to uncover the universal organizing principles of collective structural response---insights that brute-force, high-resolution simulation alone can never yield.

Here we uncover the underlying physics that governs collective responses in city-scale building portfolios. Using large-scale nonlinear time-history analyses coupled with stochastic and physics-based earthquake models, we identify two \emph{phase transitions} in structural damage. In this context, a phase corresponds to a distinct pattern of collective damage, and a phase transition marks a qualitative shift between such patterns. First, in regions where buildings have similar structural capacities (low structural diversity), the system exhibits a \emph{synchronized state}, where regional damage can shift abruptly from largely safe to widespread failure, analogous to a first-order (discontinuous) phase transition. Second, increasing diversity does not yield a predictable Gaussian aggregate response; instead, the inherent spatial clustering of structures traps cities in a highly volatile regime, analogous to a Griffiths phase with significant realization-to-realization variability~\cite{vojta_rare_2006,moretti_griffiths_2013}. In this \emph{volatile state}, engineering predictions become highly sensitive to modeling details, undermining their reliability. Conventional portfolio-level simplifications can inadvertently toggle a city model between synchronized and volatile regimes, substantially biasing the repair-cost exceedance probabilities critical for urban hazard preparedness and recovery planning. Consequently, establishing both the validity and fundamental limits of engineering predictions requires diagnosing these underlying regimes, calling for \emph{phase-awareness} in city-scale hazard risk assessment. While the framework is posed for natural hazards in general, the quantitative evidence developed in this study focuses on seismic hazards, where the required capacity and demand models are most mature; extensions to other hazards are outlined in the Discussion.

\newpage
\section{Collective behavior in city-scale structural damage}
\begin{figure}[H]
    \centering
    \includegraphics[width=1.0\linewidth]{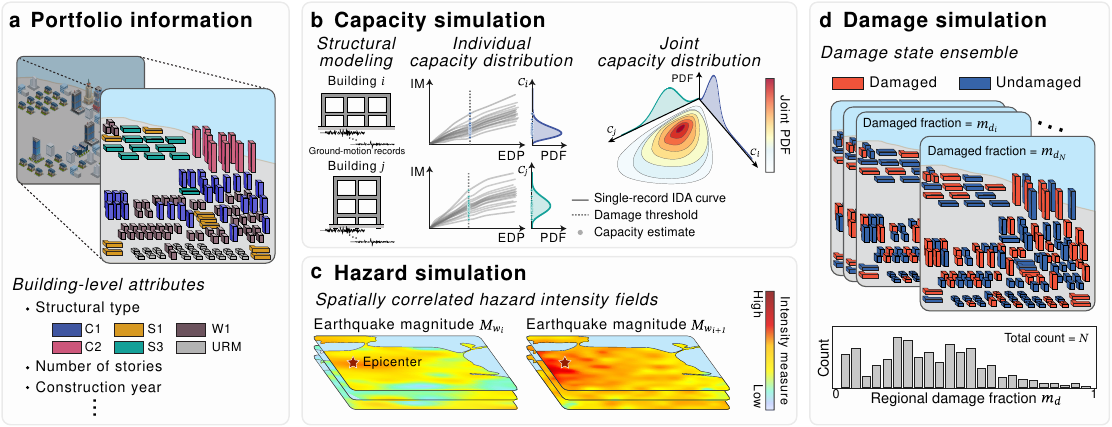}
    \caption{
    \textbf{Regional-scale probabilistic simulation of structural damage under natural hazards.}
    \textbf{a}, Building portfolio information is assembled by collecting building-level attributes, such as structural type, number of stories, and construction year. Structural-type labels follow the Hazus~6.1 taxonomy~\cite{federal_emergency_management_agency_fema_hazus_2024}.
    \textbf{b}, Each building is represented by a multi-degree-of-freedom structural model based on its inventory attributes. Incremental dynamic analysis (IDA)~\cite{vamvatsikos_incremental_2002} with empirical ground-motion records provides capacity samples $C_i$ represented by the ground-motion intensity measure (IM) at which building $i$ reaches its code-specified threshold for engineering demand parameter (EDP). The resulting marginal probability density functions (PDFs) for all buildings are combined with inter-building dependence inferred from the IDA-derived samples to construct the joint capacity distribution. The cumulative distribution function of $C_i$ defines the building fragility curve.
    \textbf{c}, For each hazard scenario with magnitude $M_{w_i}$, spatially correlated hazard intensity fields are generated using a ground-motion prediction equation (GMPE) to specify the intensity measure distribution at each site and an intra-event spatial correlation model to account for spatial dependence.
    \textbf{d}, For each realization, buildings are classified as damaged if the sampled demand exceeds their capacities, and as undamaged otherwise. Repeated realizations form a damage-state ensemble for the region under the given hazard scenario. The regional damage fraction $m_d$ is the fraction of damaged buildings in each realization, and its histogram summarizes the ensemble distribution of regional impact. Variants of this capacity--demand workflow include alternative ways to construct capacity--demand distributions or fully physics-based structural response simulations with sampled hazard excitations.
    }
    \label{fig:1}
\end{figure}

\noindent
Milpitas, California, a residential city in the San Francisco Bay Area, serves as the study region, with earthquakes considered as the representative hazard for the city. The analysis focuses on multistory buildings (two or more stories), which dominate regional seismic risk due to their greater vulnerability and damage consequences. The regional simulations combine fragility-based building capacities with spatially correlated seismic demands, assign building-level damage, and aggregate the results into a regional damage fraction (Fig.~\ref{fig:1}). Repeated realizations yield the ensemble damage distributions analyzed below (see Methods for implementation details; Supplementary Note~1 discusses the rationale for using a probabilistic framework).

\begin{figure}[H]
    \centering
    \includegraphics[width=1.0\linewidth]{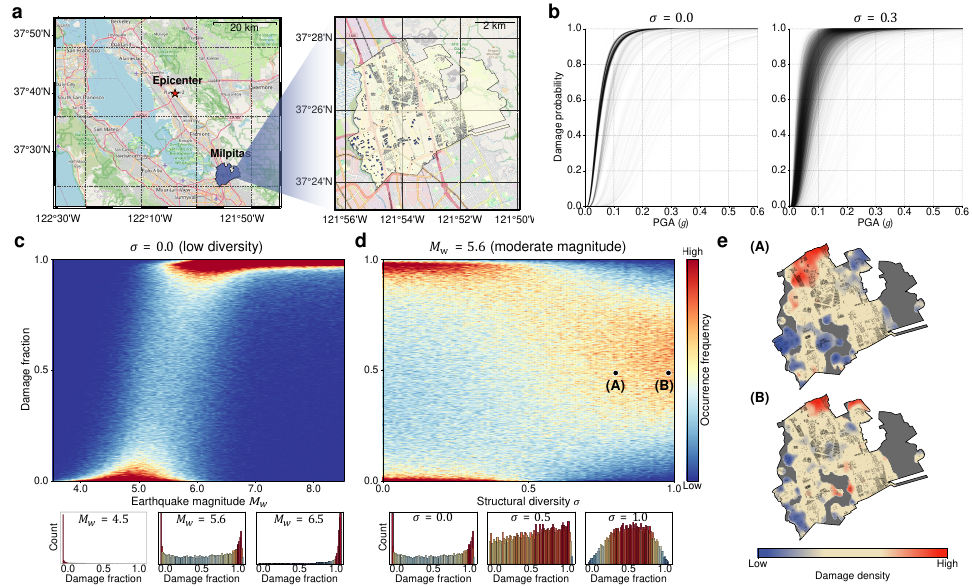}
    \caption{
    \textbf{Collective structural responses and transition behavior in a city-scale building portfolio.}
    \textbf{a}, Study region and building distribution. Milpitas (San Francisco Bay Area), near the modeled earthquake epicenter (left). Locations of the $5{,}943$ multistory buildings analyzed (right).
    \textbf{b}, Fragility curves for the Milpitas portfolio (left) and a counterpart with greater dispersion in structural capacity (right), representing a city with higher structural diversity. Here, $\sigma$ quantifies the additional heterogeneity imposed on structural capacity relative to the real-inventory baseline, with $\sigma=0$ recovering the unperturbed portfolio. A fragility curve probabilistically defines structural capacity in terms of peak ground acceleration (PGA), expressed in units of gravitational acceleration, $g$.
    \textbf{c}, Evolution of damage-fraction distributions with increasing earthquake magnitude at relatively low structural diversity, showing an abrupt collective transition from a collectively safe to a collectively damaged state. The color scale denotes occurrence frequency. Snapshots at $M_w = 4.5$, $5.6$, and $6.5$ are shown below. At moderate magnitudes ($M_w \approx 5$--$6$), the distributions become bimodal, with two peaks corresponding to largely undamaged and largely damaged regional outcomes.
    \textbf{d}, Evolution of the distributions with increasing structural diversity at a moderate magnitude ($M_w = 5.6$). Greater diversity smooths the transition and suppresses bistability, yielding a continuous transition to an extended critical-like state characterized by non-Gaussian statistics. The histograms below show snapshots at $\sigma=0$, $0.5$, and $1.0$. Markers (A) and (B) indicate the $\sigma=0.8$ and $1.0$ cases, respectively, mapped in panel~\textbf{e}.
    \textbf{e}, Damage clustering at $(M_w,\sigma)=(5.6,0.8)$ and $(5.6,1.0)$. Colors indicate damage density (low to high); areas without buildings are masked (gray). Damage states form clusters across multiple spatial scales, rather than the approximately uniform yellow field without pronounced red or blue clusters expected in a fully disordered regime.
    }
    \label{fig:2}
\end{figure}

As earthquake intensity for Milpitas (Fig.~\ref{fig:2}a) increases, the response shifts from a collectively safe to a collectively damaged state (Fig.~\ref{fig:2}c). At low magnitudes, most buildings remain undamaged; at high magnitudes, nearly all are damaged. While this trend is expected, the abruptness of the transition is striking: around $M_w = 5\text{--}6$, the collective state switches suddenly from safe to damaged. Within this range, the damage-fraction distributions become bimodal, peaking at the two extremes, indicating that the region tends to be either largely safe or largely damaged rather than partially so. This synchronization emerges under moderate-intensity hazards, where engineering decisions are most consequential yet most uncertain. Unlike weak or extreme events, where appropriate actions are relatively straightforward (ignore or fully mobilize), moderate events require more finely calibrated decisions. Yet their polarized outcomes call for substantially different levels of response, complicating preparedness planning. Extended Data Figs.~\ref{ext_fig:heatmap_1st_milpitas_all} and~\ref{ext_fig:phase_diagrams_fullmilpitas_sf} show the results for the full building portfolio, including single-story buildings.

Different cities may exhibit distinct collective behavior under similar hazard conditions. To examine how the transition depends on portfolio heterogeneity (structural diversity), we systematically perturb building capacities using a phenomenological control parameter $\sigma$ (Fig.~\ref{fig:2}b; see Methods for implementation details). $\sigma$ represents the \emph{extra} dispersion imposed on the inventory-derived building capacities and is not a city-specific calibrated parameter. The perturbations are applied independently across buildings, increasing unstructured capacity heterogeneity without introducing additional spatial coherence. The resulting portfolios retain the real inventory's spatial layout, and $\sigma=0$ recovers the unperturbed portfolio.

At high structural diversity, the bimodality observed at moderate magnitudes ($M_w = 5.6$ in Fig.~\ref{fig:2}d) is suppressed, smoothing the transition from collectively safe to collectively damaged states (Extended Data Fig.~\ref{ext_fig:heatmaps_milpitas_sigma_0}). Synchronization vanishes around $\sigma\approx0.5$, where the damage-fraction distribution flattens and the collective damage outcomes become highly volatile. Notably, increasing diversity does not drive the system immediately into a fully disordered regime where strong randomness yields an approximately Gaussian aggregate; instead, it remains in a volatile regime with persistently high variability and non-Gaussian statistics.

To summarize how the collective response varies with earthquake magnitude and structural diversity, we construct a heatmap of the most probable regional damage fraction (the mode), serving as an empirical phase diagram (Fig.~\ref{fig:3}a). Two synchronized states emerge: a collectively safe state (near-zero damage) and a collectively damaged state (near-complete damage). At low $\sigma$, the transition between them is sharp, marked by a narrow phase-coexistence band where the damage-fraction distribution is bimodal. Beyond a critical diversity $\sigma_c\approx0.5$, the transition smooths into a broad volatile regime characterized by high-variance, non-Gaussian statistics. The real northeastern San Francisco portfolio, which spans a diverse mix of residential and commercial districts and is therefore inherently more heterogeneous, exhibits a similarly broadened transition (Extended Data Figs.~\ref{ext_fig:phase_diagrams_fullmilpitas_sf} and~\ref{ext_fig:heatmap_1st_sanfrancisco}), demonstrating that greater heterogeneity weakens synchronization not only in the controlled perturbation analysis but also in an actual building portfolio.

Specific quantitative details, such as the exact shape of the transition boundary or the extent of the volatile regime, vary with modeling assumptions, but the core phenomenological features remain robust. Supplementary sensitivity analyses confirm that this collective behavior consistently emerge across a spectrum of modeling choices, including non-parametric KDE-based fragility curves, alternative empirical ground motion models for seismic demand, and physics-based seismic and structural simulations using three-dimensional seismic wave propagation and multi-degree-of-freedom structural models (Supplementary Figs.~1--5). This consistency indicates that the observed regimes are not artifacts of specific modeling choices, but rather inherent regularities governing the interplay between hazards and structural systems.

These observed shifts in collective behavior mirror phase transitions in ferromagnetic systems, where heterogeneity modulates the abruptness of the transition. The resulting heatmap that summarizes the dominant collective-response regimes of the urban portfolio thus serves as an empirical phase diagram, motivating the statistical physics-based interpretation formalized in the next section.

\newpage
\section{Ising model interpretation for phase transitions}\label{section:ising_interpretation}
\noindent
The Ising model family provides a parsimonious statistical description for the collective behavior of interacting components with discrete states~\cite{goldenfeld_lectures_2018,theodorakis_random-field_2014,oh_longrange_2024}. We map our empirical results of city-scale structural damage to an effective random-field Ising model (RFIM). The corresponding Hamiltonian, analogous to the unnormalized negative log-likelihood function, is expressed as
\begin{equation}\label{eq:RFIM_Hamiltonian}
    \mathcal{H}(\{s_i\}) = -\sum_{i} (H + h_i)s_i - \sum_{i<j} J_{ij}s_i s_j\,.
\end{equation}
Here, $s_i\in\{-1(\text{undamaged}),1(\text{damaged})\}$ denotes the binary damage state of the $i$th building. $H$ represents the regional mean effective hazard demand, while $h_i$ captures local disorder arising from site-to-site variations across the portfolio, including differences in structural capacity and hazard demand. The interaction term $J_{ij}$ characterizes pairwise dependence between building states, reflecting shared hazard sources and correlated structural vulnerabilities that give rise to collective damage. 

This effective model captures the fundamental competition between cooperative coupling and local disorder, encoded by $J_{ij}$ and $h_i$, respectively. A global hazard demand ($H>0$) pushes all structures toward the damaged state, and positive coupling ($J_{ij}>0$) favors buildings sharing the same state. Conversely, spatial heterogeneity in $h_i$ helps the system resist the synchronization tendency. Table~\ref{tab:rfim_mapping} summarizes the engineering correspondences of the model.

The system magnetization $m=\frac{1}{N}\sum_{i=1}^N s_i$, or equivalently the regional damage fraction $m_d=(m+1)/2$, is the key quantity used to match regional damage simulations to an effective RFIM. Supplementary principal component analyses confirm that the regional damage fraction captures the primary mode of portfolio damage predicted by engineering models (Supplementary Method~1 and Supplementary Figs.~6--11). 

Temperature is a key control parameter in the RFIM and has an engineering analog in regional hazard-impact simulations. Regional simulations deterministically map a given realization of structural capacity and hazard demand to a binary outcome: damaged if demand exceeds capacity, and undamaged otherwise. Each realization therefore corresponds to a zero-temperature limit. Introducing a nonzero temperature allows us to recast the deterministic damage-assignment rule as a stochastic limit-state model that probabilistically classifies damage, with the damage probability expressed as $1 /(1+\exp({(\mathrm{Capacity} - \mathrm{Demand})}/T))$. In this formulation, the temperature-like hyperparameter $T$ serves as a phenomenological proxy for residual, unstructured uncertainty not already encoded in the engineering risk models and is not calibrated to a specific city or dataset. The results from nonzero-temperature simulations are summarized in Extended Data Figs.~\ref{ext_fig:heatmap_temp} and~\ref{ext_fig:phase_diagram_temperature} as a complementary sensitivity analysis to examine how the parameter smooths the undamaged-to-damaged transition; see Supplementary Note~2 for further discussion and Supplementary Method~2 for implementation details.

\begin{figure}[H]
    \centering
    \includegraphics[width=1.0\linewidth]{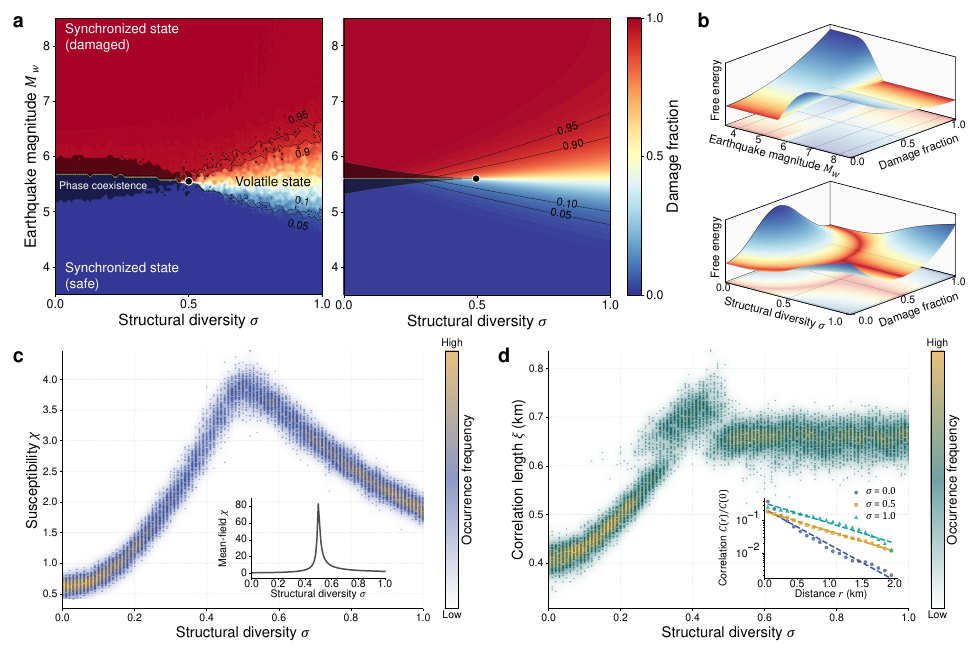}
    \caption{
    \textbf{Random-field Ising-model analysis of collective structural damage transitions.}
    \textbf{a}, Phase diagrams showing how the dominant collective damage pattern changes with earthquake magnitude and structural diversity, comparing regional simulations for Milpitas (left) with the RFIM reconstruction (right). Color denotes the most-probable damage fraction at each point, and contours indicate isodamage levels. The reconstruction successfully captures the distinct collective states, the phase-coexistence band, and the critical point (black circle).
    \textbf{b}, Mean-field free-energy landscapes of the fitted RFIM at $\sigma = 0$ (top) and $M_w = 5.6$ (bottom), which act as effective landscapes whose minima correspond to the most probable regional states. Increasing $M_w$ drives an abrupt switch between minima, whereas increasing $\sigma$ flattens and merges them, smoothing the abrupt collective response into a continuous crossover. Projections provide an RFIM-based reconstruction of Figs.~\ref{fig:2}c,d.
    \textbf{c}, Susceptibility peak and persistent volatility beyond criticality. Near the critical magnitude $M_{w_c}\approx 5.6$, we compute susceptibility $\chi$ from 50 independent portfolio realizations at each $\sigma$ (see Methods) and summarize its distribution as a background heatmap. Here, $\chi$ quantifies the macroscopic sensitivity of the regional damage fraction to variations in effective hazard demand. It exhibits a pronounced but finite peak at $\sigma_c \approx 0.5$. The depleted density (absence of yellow) over $\sigma\in[0.3,0.6]$ indicates strong realization-to-realization variability, consistent with a broadened critical regime (Griffiths-like behavior). The inset shows the mean-field susceptibility of the fitted RFIM, which drops sharply beyond $\sigma_c$; by contrast, the empirical susceptibility decays slowly, highlighting the persistent volatility of the collective damage state. The mean-field curve is evaluated near criticality to avoid divergence.
    \textbf{d}, Correlation-length peak and persistence through the volatile regime. Using the same portfolio realizations as in panel \textbf{c}, we estimate connected two-point correlations $C(r)$ (see Methods) and infer the correlation length $\xi$ by fitting the decay $C(r)/C(0)\propto e^{-r/\xi}$ versus distance $r$ (inset). Physically, $\xi$ measures the spatial reach of damage dependency. The correlation length remains elevated for $\sigma>\sigma_c$, indicating that $C(r)$ decays slowly with distance and that damage correlations are not confined to nearby buildings but persist over a broad range of spatial separations.
    }
    \label{fig:3}
\end{figure}

In the zero-temperature limit, the RFIM admits an analytical mean-field description of the regional damage fraction. Using the mean-field formulation, we map the engineering control parameters of our regional simulation—moment magnitude \(M_w\) and imposed structural heterogeneity \(\sigma\)—to the mean effective hazard demand \(H\) and disorder strength \(\Delta\) of the RFIM, respectively, with the coupling parameters \(J_{ij}\) absorbed during fitting. The disorder strength \(\Delta\) summarizes the local disorder \(h_i\) under the Gaussian model \(h_i\sim\mathcal N(0,\Delta^2)\). The resulting mappings enable the mean-field RFIM to reproduce the main features of the empirical phase diagram (Fig.~\ref{fig:3}a), including the sharp transition and phase-coexistence band at low diversity and the smoother transition at higher diversity. Full details are provided in Methods, Supplementary Note~3, Supplementary Method~3, and Extended Data Fig.~\ref{ext_fig:rfim_param_est}.

Free energy provides a compact representation of equilibrium behavior in statistical physics~\cite{goldenfeld_lectures_2018, kadanoff_statistical_2000, ma_modern_2019}, manifesting in regional simulations as an effective landscape whose minima correspond to the most probable regional damage fractions (Methods). The resulting free-energy landscapes clearly illustrate the abrupt, first-order-like response at low structural diversity and the smooth, continuous transitions at high structural diversity (Fig.~\ref{fig:3}b). Supplementary Figs.~12--14 further demonstrate the transferability of this effective RFIM description to held-out city portfolios; see Supplementary Method~4 for details.

Within the RFIM framework, our empirical findings in Fig.~\ref{fig:2} admit clear physical interpretations. When structural diversity is relatively low, effective couplings dominate local disorder, placing the city in an ordered state where buildings respond in synchronization to hazards; increasing hazard intensity then drives a sharp transition. At high structural diversity, one might expect local disorder to overwhelm coupling, yielding a fully disordered phase with Gaussian statistics. Yet, even at $\sigma$ values where the dispersion of fragility curves becomes unrealistically broad (Extended Data Fig.~\ref{ext_fig:frag_evolution}), the aggregate damage distribution remains non-Gaussian and highly volatile. This persistent volatility is characterized by two key metrics: a slowly decaying susceptibility and an elevated correlation length (Figs.~\ref{fig:3}c,d; see also Supplementary Method~5 and Supplementary Figs.~15--16). Susceptibility here quantifies the system's macroscopic sensitivity to external excitation; a high, slowly decaying susceptibility indicates that a small change in effective hazard demand can trigger disproportionately large fluctuations in the regional damage fraction. Meanwhile, the correlation length measures the spatial reach of damage dependence.

We attribute the persistent volatility to localized building domains within the urban portfolio. Nearby buildings experience similar hazard demands and often share similar capacities due to zoning regulations and construction cohorts; domains with a similar capacity--demand balance therefore produce local clusters of damage (Fig.~\ref{fig:2}e). Because planning and construction practices vary across neighborhoods, the resulting localized domains span a range of spatial scales. Field data from the 2024 Noto Peninsula earthquake further confirm such local clustering (Supplementary Fig.~17 and Supplementary Method~6). This extended non-Gaussian behavior serves as an urban analogue to a Griffiths-like phase in statistical physics. While prior studies have noted that spatial coherence amplifies portfolio tail risks~\cite{weatherill_exploring_2015}, our findings reveal a new consequence: in the volatile regime, the persistence of multiscale correlations indicates that aggregated risk metrics remain intrinsically sensitive to modeling details, limiting reliable predictions.

Overall, the RFIM provides a statistical physics-based characterization of the regional damage patterns. Its mean-field solution yields simplified, quantitative mappings from the engineering control variables to the effective RFIM parameters, thereby providing an analytical description of the collective-damage phases. However, the mean-field approximation does not capture the subtle Griffiths-like behavior observed for $\sigma>\sigma_c$ in the regional simulations; instead, it predicts a fully disordered phase with Gaussian statistics. While the precise topology of the diagram varies across urban configurations---for instance, cities with skewed or multimodal disorder distributions can produce asymmetric boundaries or sequential transitions (Extended Data Figs.~\ref{ext_fig:heatmap_1st_milpitas_all} and~\ref{ext_fig:phase_diagrams_fullmilpitas_sf})---the macroscopic patterns remain robust. 

\newpage
\begin{table}[H]
\centering
\footnotesize
\begingroup
\captionsetup{labelfont={color=black,bf}, textfont={color=black}}
\arrayrulecolor{black}
\setlength{\tabcolsep}{3.5pt}

\caption{
\textbf{Correspondence between the regional damage simulations and the effective RFIM.}}
\label{tab:rfim_mapping}

\begin{tabularx}{\linewidth}{@{}p{0.24\linewidth}YY@{}}
\toprule
Role & Regional simulation & Effective RFIM \\
\midrule

% First row
Microscopic state & Structural damage $s_i'\in\{0,1\}$ & Spin state $s_i\in\{-1,+1\}$ \\

% Second row
Order parameter & Regional damage fraction $m_d=\tfrac{1}{N}\sum_i s_i'$ & Magnetization $m=\tfrac{1}{N}\sum_i s_i$ \\

% Third row
External excitation & Hazard intensity (e.g., $M_w$) & External field $H$ \\

% Fourth row
System disorder & Imposed heterogeneity $\sigma$ & Disorder strength $\Delta$ \\

% Fifth row
Spatial coherence & Spatial correlation in hazard demand and structural capacity & Spin--spin coupling $J_{ij}$ (or mean-field $J$) \\

% Sixth row
Residual stochasticity & Stochastic damage assignment & Temperature $T$ \\

\bottomrule
\end{tabularx}

\endgroup
\end{table}

\newpage
\section{Phase-aware urban hazard risk assessment}\label{section:eng_implications}
\noindent
Despite recent advances in regional-scale risk and resilience analysis for natural hazards, current engineering practices and design codes remain centered on individual structures, with limited consideration of their collective behavior at the city scale. For example, risk and loss assessment frameworks such as the \textit{Hazus Earthquake Model Technical Manual}, published by the Federal Emergency Management Agency (FEMA), classify buildings by coarse categories such as structural type, material, and construction year, and then assign capacities by class~\cite{federal_emergency_management_agency_fema_hazus_2024}. While efficient, such coarse categorization can impose artificial homogeneity within each class and obscure the diversity that governs the collective response ~\cite{rincon_fragility_2024}. Likewise, although spatial correlations of hazard intensities have long been recognized and incorporated into risk assessment guidelines~\cite{Jayaram2009CorrelationIntensities, goda_spatial_2008}, correlations among structural capacities, arising from shared materials, typologies, or construction practices~\cite{ghosh_seismic_2014, baker_model_2024, Kang2021EvaluationAnalysis, xiang_structure--structure_2024, you_spatial_2024}, have only recently begun to receive attention and remain underrepresented~\cite{heresi_rpbee_2023}. Structural capacities are still often modeled as conditionally independent given hazard intensity~\cite{van_de_lindt_interdependent_2023}, suppressing collective behavior and biasing regional decision metrics.

\begin{figure}[H]
    \centering
    \includegraphics[width=1.0\linewidth]{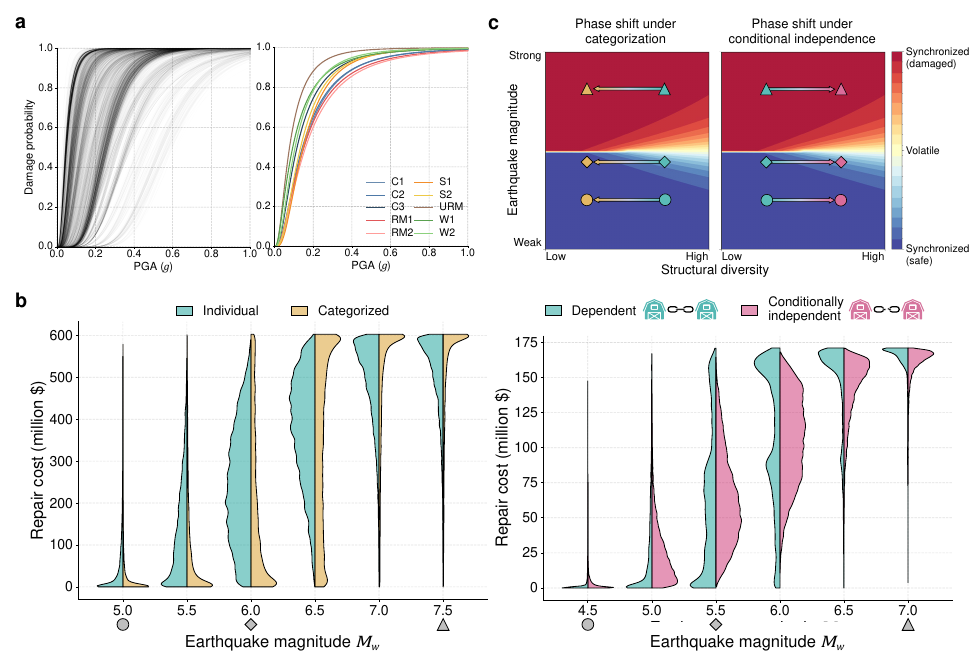}
    \caption{
    \textbf{Engineering implications of collective behavior and phase awareness.}
    \textbf{a}, Fragility curves for the northeastern San Francisco building portfolio analyzed with (colored) and without (gray) coarse structural-type categorization. Categories for the coarse structural type classification are defined according to Hazus~6.1~\cite{federal_emergency_management_agency_fema_hazus_2024}.
    \textbf{b}, Regional repair-cost distributions illustrating two common engineering simplifications: coarse structural-type categorization for the San Francisco portfolio (left) and conditional-independence assumption for the Milpitas portfolio (right). In San Francisco, replacing individual fragility curves (teal) with category-based curves (gold) imposes artificial homogeneity and reshapes the total repair-cost distributions, promoting synchronized collective behavior. In Milpitas, assuming conditionally independent structural capacities (pink) instead of explicitly dependent capacities (teal) suppresses collective alignment in damage and broadens the repair-cost distributions. In both cases, these distributional changes mainly alter quantile-based tail-risk metrics while leaving mean risk metrics largely unchanged. Marker symbols below the x-axis identify representative earthquake magnitudes used in panel~\textbf{c}.
    \textbf{c}, Phase-diagram interpretation of the engineering simplifications. Background colors denote collective-risk phases; the two synchronized regimes correspond to collectively damaged (red) and collectively safe states (blue), respectively. Each simplification changes the inferred phase-diagram location of the same underlying portfolio by altering its effective heterogeneity: coarse categorization (left) reduces apparent diversity (moving the portfolio leftward), whereas the conditional-independence assumption (right) increases apparent diversity (moving it rightward). Under weak or strong hazards, these shifts do not change the qualitative phase (circle and triangle markers), but at moderate intensities they can push the portfolio across the phase boundary (diamond markers). Such phase misrepresentation systematically distorts loss predictions, particularly in the tail of the distribution, where risk-informed decisions are most sensitive. Marker symbols correspond to the representative magnitudes in panel~\textbf{b}.
    }
    \label{fig:4}
\end{figure}

Common engineering assumptions can substantially bias regional repair-cost exceedance probabilities and upper-tail loss estimates even when mean repair costs remain similar. Fig.~\ref{fig:4} quantifies the practical stakes of these assumptions by comparing regional repair-cost distributions in the northeastern part of San Francisco, subjected to the same earthquake scenarios as the Milpitas case, with and without coarse structural categorization. Compared with the Milpitas building portfolio, the selected San Francisco region, which includes commercial districts such as the Financial District and Mission Bay and residential zones such as Pacific Heights and Hayes Valley, exhibits a more diverse building portfolio (Extended Data Fig.~\ref{ext_fig:region_maps_donuts}). Enforcing coarse categories narrows the portfolio’s effective capacity spread and exacerbates collective clustering, changing the shape of the distribution and rendering spurious first-order-like transition behavior. Fig.~\ref{fig:4} further contrasts analyses with and without the conditional-independence assumption for the Milpitas portfolio; neglecting inter-building correlation obscures collective alignment in damage and leads to a biased distribution shape.

In both cases, mean repair costs remain similar, yet the tails alter markedly; exceedance probabilities for repair costs are biased by up to 50\% at moderate earthquake magnitudes ($M_w \approx 5.5$--$6.0$), and the 1\% quantile repair cost is misestimated by up to one to two orders of magnitude, ranging from a 33-fold underestimation under coarse categorization in San Francisco to a 30-fold overestimation under conditional independence in Milpitas (Supplementary Figs.~18 and 19). Since natural hazard preparedness, resource prepositioning, and recovery planning are informed by tail-risk measures (e.g., exceedance probabilities, value-at-risk, or conditional value-at-risk), these assumption-driven shifts are decision-critical. The distributional change is most pronounced at moderate hazard intensities, where design and policy choices are both consequential and challenging, in contrast to weak or extreme hazards for which decisions are relatively straightforward.

Viewed through the lens of collective phases, engineering simplifications do not merely reduce precision; they may shift a portfolio into an incorrect collective phase. Coarse structural categorization artificially synchronizes the modeled portfolio, inducing abrupt transitions, whereas the conditional-independence assumption pushes it toward the volatile regime. Phase distortion is most pronounced at moderate hazard intensities where the system is near a critical phase boundary, but becomes negligible under extreme hazards. Consequently, model validity is not absolute but contingent on the portfolio's proximity to phase boundaries. This motivates a phase-aware diagnostic framework for constructing balanced engineering models that remain faithful to the underlying collective phase while maximizing efficiency. Crucially, such a framework identifies when a system enters the volatile regime, where critical-like behavior fundamentally constrains predictability and renders aggregated risk metrics intrinsically sensitive to modeling choices.

\newpage
\section{Discussion}
\noindent
The emergence of collective damage regimes indicates that regional disaster risk follows universal laws of collective behavior. Rather than acting as a sum of isolated building failures, a city behaves as a coupled system whose macroscopic state depends jointly on hazard intensity and structural diversity. The random-field Ising model effectively captures these dynamics, demonstrating that the competition between inter-building coupling and local disorder governs whether the regional response is abrupt and synchronized, or continuous and volatile. 

These findings generalize beyond seismic risk. A similar phase-aware analysis may be relevant whenever one can define a spatially distributed demand field and a corresponding capacity or threshold field, mapping component failure to relevant outcomes such as loss of habitability, road blockage, substation outage, or other forms of functional loss. Hazards such as hurricanes, floods, and wildfires may therefore exhibit analogous collective transitions when correlated regional demand interacts with built-environment capacity. Infrastructure networks, where interactions between nodes are often causal and non-local, may exhibit even richer collective behavior~\cite{berezin_localized_2015, shao_percolation_2015, scagliarini_assessing_2025}. More broadly, in other risk-related applications where uncertainties are embedded in component-level models, the phase-transition paradigm may help reveal distinct patterns and characterize the stability of macroscopic collective responses emerging from local component interactions. Future work using scaled experiments that mimic the interplay between spatial correlation and disorder could provide experimental evidence of these predicted collective dynamics.

The statistical physics perspective also has important implications for urban risk assessment and planning. Simplifications that distort hazard correlations or structural diversity do not merely reduce precision; they can inadvertently shift a region across a phase boundary, fundamentally altering the nature of the predicted response. This is particularly critical when the system resides in the volatile phase, where persistent multiscale correlations slow the self-averaging of fluctuations. This non-Gaussian behavior imposes an intrinsic limit on predictability, leaving tail-risk metrics used to inform preparedness and recovery planning sensitive to modest modeling changes. Embedding phase awareness through diagnostics such as critical points or free-energy landscapes offers a path toward balanced engineering models that remain computationally practical while respecting the underlying physics of collective behavior.

More broadly, situating urban risk within the physics of critical phenomena reframes the paradigm of urban resilience modeling. Recognizing cities as complex systems operating near criticality reveals a fundamental limit to the constructionist approach: refining individual component models alone cannot capture the full risk landscape; it must be complemented by a macroscopic framework that navigates the transitions between collective stability and instability.

\newpage
\section*{\LARGE{Methods}}
\setcounter{section}{0}
\section*{Regional simulation framework}\label{methods:simframe}
\noindent
Following the overview in Fig.~\ref{fig:1}, the framework for regional-scale structural response simulations is further elaborated in Supplementary Method~7, with the specific implementation detailed in Supplementary Method~8.

\subsection*{Study regions and building inventories}\label{methods:simframe_region}
\noindent
Building-level attributes were obtained from the Natural Hazards Engineering Research Infrastructure (NHERI) database, which provides detailed inventory data for the San Francisco Bay Area, including the number of stories, construction year, and structural type~\cite{zsarnoczay_simcenter_2023}. Two representative regions were analyzed: Milpitas and northeastern San Francisco. The Milpitas dataset includes 5,943 buildings, predominantly residential, whereas the San Francisco dataset comprises 6,323 buildings spanning commercial districts such as the Financial District and Mission Bay, as well as residential neighborhoods including Pacific Heights and Hayes Valley. District boundaries were retrieved from the San Francisco Open Data Portal~\cite{sfgov_neighborhoods}. Extended Data Fig.~\ref{ext_fig:region_maps_donuts} shows maps of the two regions and the distributions of structural attributes; detailed inventory statistics and site information are provided in the Supplementary Figs.~20--23.

\subsection*{Structural modeling}\label{methods:simframe_strmodel}
\noindent
For each building, a multi-degree-of-freedom (MDOF) shear model was constructed using the retrieved building attributes, following established simulation frameworks~\cite{lu_open-source_2020,mckenna_nheri-simcenterr2dtool_2025}. Each model reflected the building’s material, typology, construction year, number of stories, height, and plan area, while each story was modeled as a zero-length element with a hysteretic uniaxial material capable of reproducing nonlinear stiffness degradation and pinching behavior. Story-wise stiffness and strength parameters were derived from the Hazus~6.1 Earthquake Model database~\cite{federal_emergency_management_agency_fema_hazus_2024} according to structural type and design era, and implemented in the open-source finite-element platform OpenSeesPy~\cite{mckenna_opensees_2011,zhu_openseespy_2018}. The models were subjected to ground motions through a uniform base-excitation pattern applied at the structures' base supports, and the peak inter-story drift ratio was extracted as the primary engineering demand parameter (EDP). 

Incremental dynamic analysis (IDA)~\cite{vamvatsikos_incremental_2002} was adopted to estimate structural capacities using 100 empirical ground-motion records from the NGA-West2 database~\cite{ancheta_nga-west2_2014}. Each record was incrementally scaled until the structure reached the threshold EDP specified in Hazus~6.1~\cite{federal_emergency_management_agency_fema_hazus_2024} for the corresponding structural type, construction era, and selected damage state, yielding 100 capacity estimates per building. We adopted the \emph{slight} damage state because it allows the onset of collective damage to be observed at plausible hazard intensities. For more severe damage states, the corresponding fragility curves would generally be more heterogeneous, potentially narrowing or eliminating the phase-coexistence band and placing the portfolio in the volatile regime. The phase diagram reported here is therefore conditional on the slight-damage threshold. A conceptual extension to the simultaneous representation of multiple damage states is discussed in Supplementary Note~4.

Following standard practice in earthquake fragility modeling~\cite{baker_efficient_2015}, a lognormal distribution was fitted to the resulting set of 100 capacity estimates for each building, and its cumulative distribution function was used as the corresponding building-level fragility curve. Regional correlations among structural capacities were then computed from the simulated capacities across all buildings in the study region; its sampling stability, numerical treatment, and sensitivity to alternative dependence models are reported in Supplementary Note~5 and Supplementary Figs.~24 and~25. To assess sensitivity to the lognormal distribution, Supplementary Fig.~1 reports results obtained using non-parametric KDE-based fragility curves, which yield qualitatively similar collective behavior.

To emulate cities with more heterogeneous building portfolios, we introduced controlled additional variability into structural capacities. For each $\sigma$, we independently perturbed each building’s median capacity by a lognormal random factor whose logarithm has mean zero and standard deviation $\sigma$. This procedure effectively broadened the regional portfolio of fragility curves, as illustrated in Extended Data Fig.~\ref{ext_fig:frag_evolution}.

\subsection*{Hazard modeling}\label{methods:simframe_hazmodel}
\noindent
A strike-slip earthquake scenario on the Hayward fault, representative of the San Francisco Bay Area, was defined with an epicenter at $(37.666^{\circ}\mathrm{N},\,122.076^{\circ}\mathrm{W})$, strike $325^{\circ}$, dip $90^{\circ}$, rake $180^{\circ}$, and a 3~km depth to the top of rupture. The moment magnitude $M_w$ served as the hazard parameter. Site conditions were assigned using $V_{s30}$ values from the U.S.~Geological Survey dataset~\cite{thompson_updated_2018}. These parameters defined the input for simulating spatially correlated ground motions under a California strike-slip mechanism.

For stochastic ground-motion field generation, we followed the standard GMPE-based approach, in which site-specific intensity measures are modeled as lognormally distributed. We used the GMPE of Chiou and Youngs~\cite{chiou_update_2014}, which is applicable to active crustal regions in California, to obtain the logarithmic mean and standard deviation of PGA at each site. Fault rupture length and width were estimated from moment magnitude using the empirical scaling relations of Wells and Coppersmith~\cite{wells_new_1994}. Spatial correlation of intra-event residuals was modeled following Jayaram and Baker~\cite{Jayaram2009CorrelationIntensities}, using an exponential decay function calibrated for strike-slip mechanisms and soil conditions representative of the Bay Area. The resulting PGA fields therefore account for both inter-event residuals across earthquakes under the same scenario and spatially correlated intra-event residuals within a single earthquake.

To enhance the physical realism of the generated ground‐motion fields, the stochastically simulated intensity measures over Milpitas were cross‐validated against physics‐based simulations from the Southern California Earthquake Center (SCEC) Broadband Platform (BBP)~\cite{maechling_scec_2015}. The BBP generates broadband seismic wave synthetics over a frequency range extending beyond 20\,Hz using a combination of physics-based components to simulate fault rupture and three‐dimensional wave propagation. The same earthquake scenarios were used in both approaches. The comparison shows broad agreement in station-wise PGA levels and distributions between the two approaches (Extended Data Fig.~\ref{ext_fig:pga_comparison}; Supplementary Figs.~26--29), with the magnitude-dependent differences in spatial-correlation from the Jayaram--Baker model (Supplementary Fig.~30).

\subsection*{Damage simulation}\label{methods:simframe_damage}
\noindent
For each seismic scenario, we generated building capacities and ground-motion demands and then assigned damage deterministically. Capacities $C_i$ for building $i$ were sampled from the fitted lognormal distributions calibrated from the IDA-based fragility analysis, while preserving the IDA-derived correlation structure. Seismic demands $D_i$ were sampled from GMPE-based lognormal distributions, including both inter- and intra-event residuals. Damage was assigned according to the binary rule $s_i'=\mathds{1}\{D_i>C_i\}$, and the regional damage fraction was defined as $m_d=\frac{1}{N}\sum_{i=1}^{N} s_i'$ for $N$ buildings. We performed 10{,}000 Monte Carlo realizations at each $(M_w,\sigma)$ pair to estimate the conditional distribution of $m_d$, denoted by $p(m_d \mid M_w,\sigma)$. This ensemble size was chosen as a practical balance between sufficiently resolving the damage-fraction distribution at each grid point and maintaining computational tractability. The moment magnitude ranged from $M_w=3.5$ to $8.5$ in increments of $0.05$ (101 values). The diversity parameter ranged from $\sigma=0.00$ to $1.00$ in increments of $0.01$ (101 values).

\subsection*{Loss and cost estimation}\label{methods:simframe_cost}
\noindent
Repair costs were estimated using building-level data from the SimCenter Earthquake Testbed for the San Francisco Bay Area~\cite{zsarnoczay_simcenter_2023}, which provides building-level replacement costs and repair-cost estimates computed using the PELICUN framework~\cite{Zsarnoczay2020pelicun}. The dataset does not explicitly identify the damage state associated with the repair-cost field. Across the dataset, the repair-to-replacement cost ratio has a mean of approximately 4\% and a median of approximately 2\%, broadly consistent with the \textit{moderate}-damage repair-cost ratios tabulated in Hazus 6.1 (Table 11-2)~\cite{federal_emergency_management_agency_fema_hazus_2024}. For the occupancy classes represented in the portfolios, the same table assigns \textit{slight}-damage repair costs of approximately 0.3--0.6\% of replacement cost. We therefore divided the testbed repair-cost estimates by five and, for each realization, summed the scaled estimates over all damaged buildings, while preserving the building-to-building cost variation. Because the same factor is applied uniformly in all compared analyses, it changes the absolute monetary scale but not the ratios between corresponding repair-cost quantiles.

\section*{Construction of phase representations}\label{methods:visual}
\subsection*{Statistical heatmap construction}\label{methods:visual_heatmap}
\noindent
For the $M_w$ and $\sigma$ panels of Fig.~\ref{fig:3}a, we constructed two-dimensional histograms by stacking the empirical conditional distributions of the damage fraction $m_d$ over values of the tuning parameter. Thus, each vertical slice in the $M_w$ panel represents $p(m_d \mid M_w,\sigma=0.0)$, whereas each vertical slice in the $\sigma$ panel represents $p(m_d \mid M_w=5.6,\sigma)$. Equivalently, the resulting heatmaps can be viewed as empirical representations of the conditional joint distributions $p(m_d,M_w \mid \sigma=0.0)$ and $p(m_d,\sigma \mid M_w=5.6)$, respectively. Histogram counts were accumulated on a uniform grid with bin widths of 0.05 in $M_w$, 0.01 in $\sigma$, and $10^{-3}$ in $m_d$. To stabilize the color scale, counts were clipped at a robust upper cutoff and then normalized (95th percentile for the $M_w$ panel; 99th percentile for the $\sigma$ panel). The resulting rescaled occurrence frequencies were rendered as 100-level filled contours, with the horizontal axis denoting the tuning parameter and the vertical axis denoting $m_d$.

\subsection*{Empirical phase representation}\label{methods:visual_phase}
\noindent
For the empirical phase diagram, we estimated the most probable damage fraction $m_d^\star$ at each $(M_w,\sigma)$ grid point. To obtain a stable estimate, we constructed a 100-bin uniform histogram of $m_d$, ranked bins by count, and averaged samples within the smallest set of top bins whose cumulative count exceeded 100 (1\% of 10,000 simulations). The resulting field $m_d^\star(M_w,\sigma)$ was visualized as a 100-level filled contour map with $\sigma$ on the horizontal axis and $M_w$ on the vertical axis.

\section*{RFIM mapping and free-energy analysis}\label{methods:statphys}
\noindent The RFIM has been studied extensively for avalanche dynamics and hysteresis under quasi-static driving~\cite{sethna_crackling_2001, sethna_hysteresis_1993}; here we instead use its static equilibrium structure to characterize the damage-fraction distributions observed across simulated realizations. 
\subsection*{Mean-field approximation}\label{methods:statphys_meanfield}
\noindent
For the zero-temperature mean-field approximation, we assume independent Gaussian local fields $h_i\sim\mathcal{N}(0,\Delta^2)$ and uniform coupling $J_{ij}=J/N$, where $N$ is the number of buildings. The magnetization then satisfies the mean-field self-consistency relation (see Supplementary Method~3 for the derivation)
\begin{equation}\label{eq:RFIM_meanfield}
    m=\erf\!\left(\frac{m+H/J}{\sqrt{2}\,\Delta/J}\right)
    =\erf\!\left(\frac{m+a_1}{\sqrt{2}\,a_2}\right),
\end{equation}
where $\erf(\cdot)$ is the Gaussian error function, and $a_1 \equiv H/J$ and $a_2 \equiv \Delta/J$ are the normalized effective RFIM parameters, with the interaction scale $J$ absorbed into them.

\subsection*{Parameter mapping}\label{methods:statphys_paramest}
\noindent
We inferred the normalized effective RFIM parameters $a_1$ and $a_2$, representing $H$ and $\Delta$ in units of the coupling strength $J$, over the $(M_w,\sigma)$ grid by fitting the simulated most-probable damage field $m_d^\star(M_w,\sigma)$ to the mean-field self-consistency relation in Eq.~\eqref{eq:RFIM_meanfield}, using the corresponding magnetization $m^\star=2m_d^\star-1$. We represented $a_1(M_w)$ and $a_2(\sigma)$ by low-order polynomials (quadratic in $M_w$ and non-decreasing linear in $\sigma$) and estimated their coefficients by least squares over the grid. To stabilize the fit for $a_2$ and avoid spurious bistability, we fit it using only data from the volatile regime ($\sigma \ge 0.6$), excluding the synchronized regime, and applied light regularization to suppress unrealistically sharp transitions. The resulting fitted relations are shown in Extended Data Fig.~\ref{ext_fig:rfim_param_est}.

\subsection*{RFIM phase diagram}\label{methods:statphys_phasediagram}
\noindent
Using the fitted maps $a_1(M_w)$ and $a_2(\sigma)$, we numerically solved the mean-field self-consistency relation in Eq.~\eqref{eq:RFIM_meanfield} over the $(M_w,\sigma)$ grid. At each grid point, the stable solution defines the equilibrium magnetization $m^\star$, which we then converted to the corresponding damage fraction $m_d^\star$ for visualization. Because $a_2(\sigma)$ was fitted only over $\sigma \ge 0.6$, we extrapolated the fitted relation into the lower-diversity synchronized regime. We identified the bistable region as the set of grid points where solutions initialized from opposite states ($m_0=\pm1$) differed by more than $10^{-4}$. The critical magnitude and critical diversity, $M_{w_c}$ and $\sigma_c$, were then inferred from the fitted free-energy landscape (see Methods, \emph{Identification of critical magnitude and diversity}).

\subsection*{Free-energy landscape}\label{methods:statphys_landau}
\noindent
Using the fitted maps $a_1(M_w)$ and $a_2(\sigma)$, we evaluated the corresponding mean-field free energy (see Supplementary Method~3),
\begin{equation}\label{eq:RFIM_free_energy}
\begin{aligned}
    F(m;M_w,\sigma)
    &=\frac{1}{2}m^2-(m+a_1(M_w))\,\erf\!\left(\frac{m+a_1(M_w)}{\sqrt{2}\,a_2(\sigma)}\right) \\
    &\quad-\sqrt{\frac{2}{\pi}}\,a_2(\sigma)\,\exp\!\left(-\frac{(m+a_1(M_w))^2}{2\left(a_2(\sigma)\right)^2}\right)+C,
\end{aligned}
\end{equation}
on dense grids in $m$ for slices at fixed $M_w$ or fixed $\sigma$, with $C=0$. Because only relative differences in $F$ are physically meaningful, each slice was normalized by subtracting its minimum and dividing by the global maximum height after minimum subtraction. For visualization in Fig.~\ref{fig:3}b, we then applied a power-law normalization to emphasize the low-energy basins corresponding to equilibrium states and mapped $m$ to $m_d$.

\subsection*{Identification of critical magnitude and diversity}
\noindent
The critical magnitude $M_{w_c}$ is the point at which the equilibrium magnetization $m$ changes sign, which in the mean-field form corresponds to $a_1(M_w)=0$. From the fitted relation, this occurs at $M_{w_c}\approx 5.6$ for the benchmark Milpitas case at $\sigma=0.0$. The critical diversity $\sigma_c$, beyond which $m$ varies smoothly with increasing $M_w$ and no abrupt change occurs, is defined by the condition $\tfrac{d^2F}{dm^2}\big|_{m=0}=0$. Evaluating this condition for the mean-field free energy in Eq.~\eqref{eq:RFIM_free_energy} yields $a_{2_c}=\sqrt{2/\pi}$, corresponding to $\sigma_c\approx 0.49$.

\subsection*{Susceptibility and correlation estimation}
\noindent
For the ensemble at fixed $(M_w,\sigma)$, we first identified an equilibrium range as the minimum-width interval of the order parameter $m$ that contains 10\% of the total realizations. We then estimated the susceptibility from equilibrium fluctuations of $m$ as $\chi = N\,\mathrm{Var}(m)=N\left(\langle m^2\rangle-\langle m\rangle^2\right)$, where $\langle\cdot\rangle$ denotes an ensemble average over the equilibrium samples and $N$ is the number of buildings. This fluctuation-based definition serves as a practical proxy for the linear-response sensitivity to an external-field perturbation. Spatial dependence was quantified using the connected two-point correlation $C(i,j)=\langle s_i s_j\rangle-\langle s_i\rangle\langle s_j\rangle$. We then radially averaged $C(i,j)$ over all distinct pairs by separation $r_{ij}$, using a bin width $\Delta r$, to obtain $C(r)$, and reported the normalized correlation $C(r)/C(0)$, where $C(0)$ is the mean on-site variance. We repeated this analysis over 50 independent portfolio realizations at each $\sigma$ to obtain robust distributions of $\chi$ and $\xi$ (Supplementary Method~5).

\clearpage
\putbib[references_main]

\newpage
\section*{Acknowledgements}
\noindent The contributions of Raul Rincon and Jamie E. Padgett were supported in part by the U.S. National Science Foundation under Award No. CMMI-2227467.

\section*{Data availability}
\noindent The data that support the findings of this study are available in the paper, Supplementary Information, and from the corresponding author upon reasonable request.

\section*{Code availability}
\noindent The code used to generate the results is available at [URL that will be available once the paper is accepted].

\section*{Author contributions}
\noindent S.O. and Z.W. conceived and developed the core ideas. S.O., R.R., J.E.P., and Z.W. contributed to the study methodology. S.O. developed the software for the overall workflow, performed the simulations, curated the data, conducted the investigation and formal analyses, and produced the visualizations. R.R. provided baseline code for GMPE-based intensity-measure generation and contributed to producing Fig.~\ref{fig:1}. J.Z. developed and executed the physics-based ground-motion time-history generation. S.O. wrote the original manuscript. S.O., J.Z., R.R., J.E.P., and Z.W. reviewed and edited the manuscript. Z.W. supervised the project and provided access to high-performance computing resources.

\section*{Competing interests}
\noindent There are no competing interests to declare.

\section*{Additional information}
\noindent Supplementary information is available for this paper. Correspondence and requests for materials should be addressed to Ziqi Wang.

\newpage
\section*{\LARGE{Extended Data}}
\setcounter{figure}{0}
\renewcommand{\figurename}{Extended Data Fig.}
\setcounter{table}{0}
\renewcommand{\tablename}{Extended Data Table}

\begin{figure}[H]
    \centering
    \includegraphics[width=1.0\linewidth]{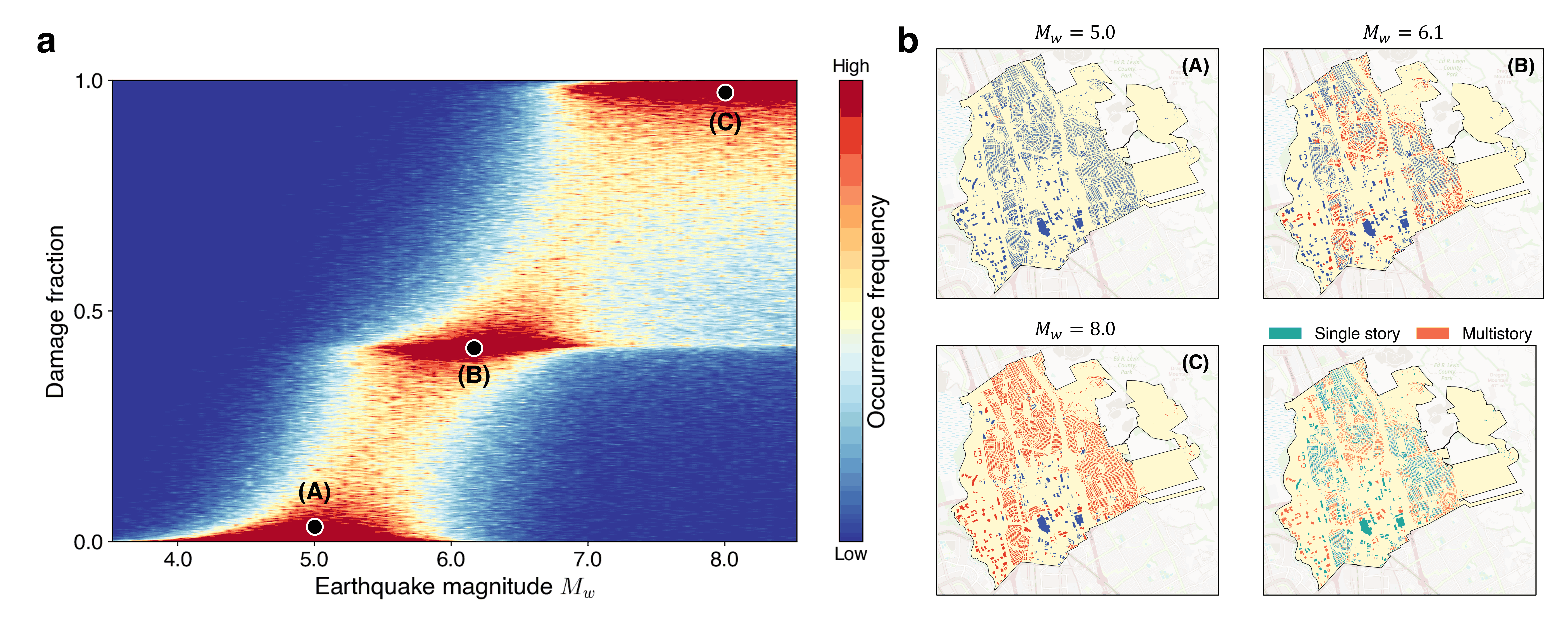}
    \caption{\textbf{Evolution of damage-fraction distributions with increasing earthquake magnitude for the full Milpitas portfolio.} 
    \textbf{a}, Heatmap of regional damage-fraction distributions versus earthquake magnitude; colors indicate relative occurrence frequency. In Milpitas, the number of stories is nearly bimodal, whereas structural type is nearly uniform across the city (Extended Data Fig.~\ref{ext_fig:region_maps_donuts}). This bimodality gives rise to three collective states (A, B, and C) separated by two sequential transitions.
    \textbf{b}, Representative spatial damage maps for the three collective states, shown with the spatial distribution of the number of stories for reference. Blue and red indicate undamaged and damaged states, respectively. Damage in the intermediate state~B aligns with multistory buildings, indicating collective damage of multistory structures. States~A and~C correspond to the collectively safe and collectively damaged states, respectively.}
    \label{ext_fig:heatmap_1st_milpitas_all}
\end{figure}

\newpage
\begin{figure}[H]
    \centering
    \includegraphics[width=1.0\linewidth]{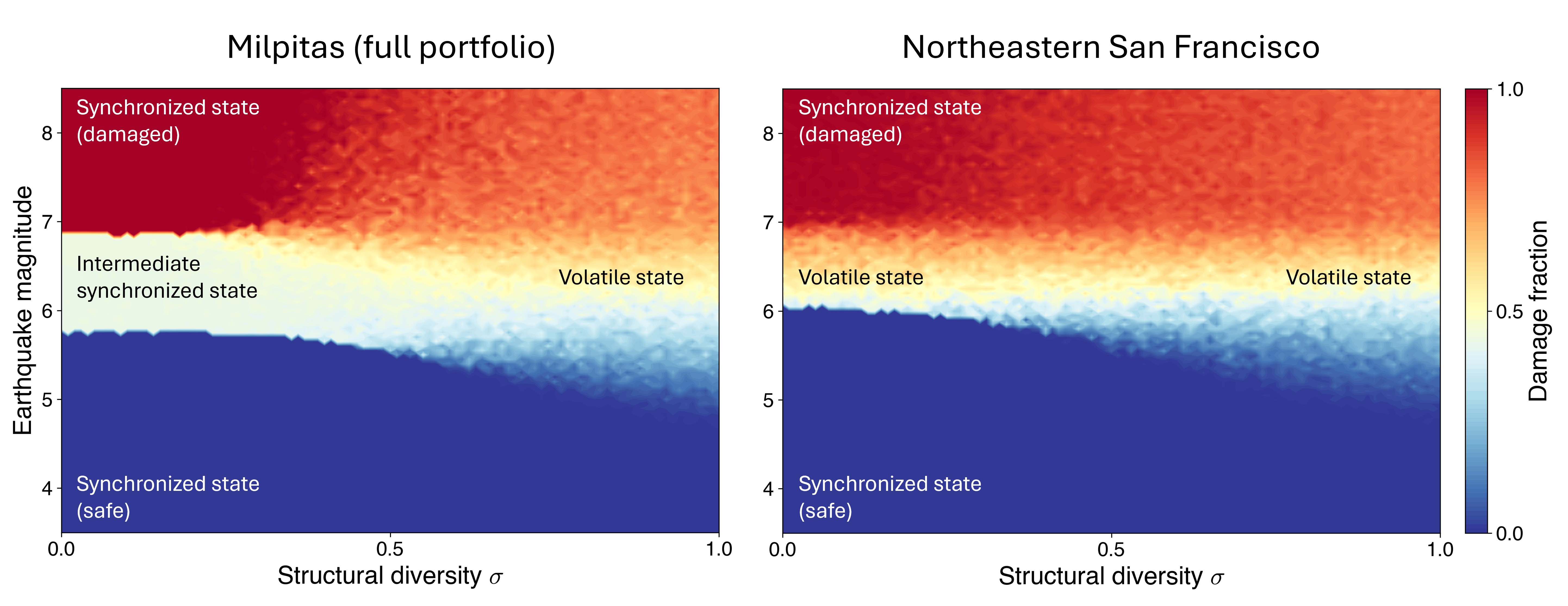}
    \caption{\textbf{Simulation-derived phase diagrams for the full Milpitas portfolio (left) and the northeastern San Francisco portfolio (right).}
    The full Milpitas portfolio exhibits three collective states at low structural diversity, consistent with the three-state behavior in Fig.~\ref{ext_fig:heatmap_1st_milpitas_all}. The San Francisco portfolio does not show a distinct transition boundary between the collectively safe and collectively damaged states; instead, the change is gradual, producing a smooth transition rather than a sharp phase boundary.}
    \label{ext_fig:phase_diagrams_fullmilpitas_sf}
\end{figure}

\newpage
\begin{figure}[H]
    \centering
    \includegraphics[width=1.0\linewidth]{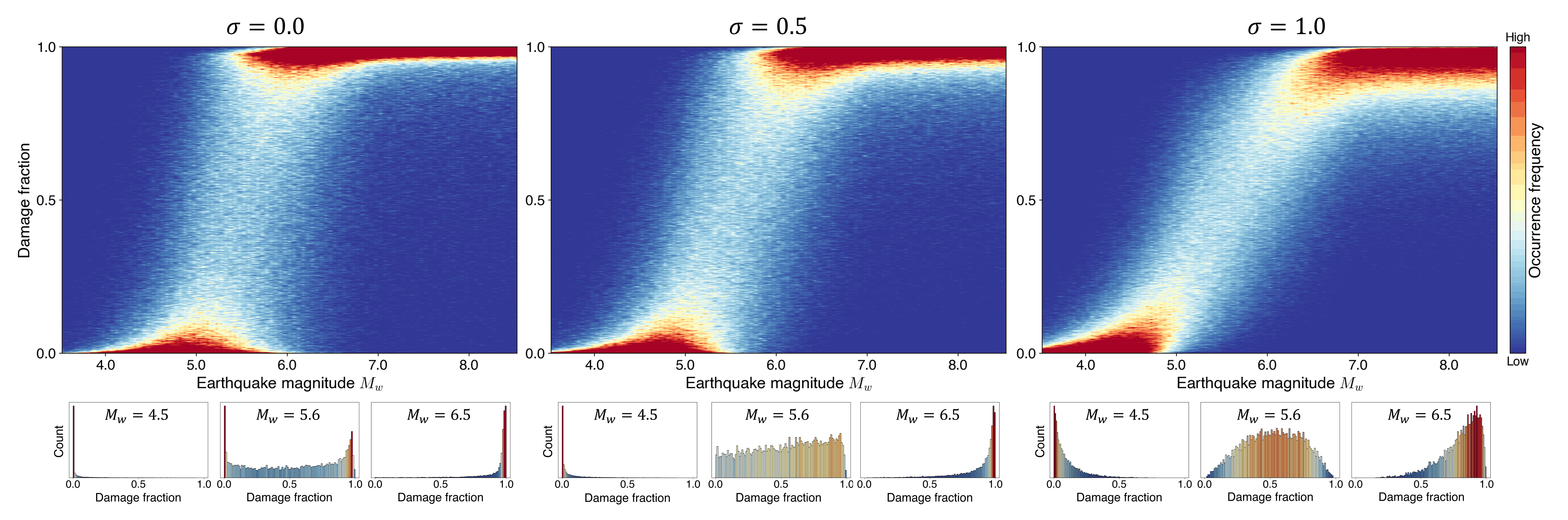}
    \caption{\textbf{Evolution of regional damage-fraction distributions at representative structural diversity levels.}
    Heatmaps show the evolution of the Milpitas damage-fraction distribution at three structural diversity levels ($\sigma = 0.0$, $0.5$, and $1.0$). Snapshots below each panel show the corresponding histograms at representative magnitudes ($M_w = 4.5$, $5.6$, and $6.5$). As structural diversity increases, the transition from the collectively safe to the collectively damaged state becomes progressively smoother, and the bimodality at intermediate magnitudes disappears.}
    \label{ext_fig:heatmaps_milpitas_sigma_0}
\end{figure}

\newpage
\begin{figure}[H]
    \centering
    \includegraphics[width=1.0\linewidth]{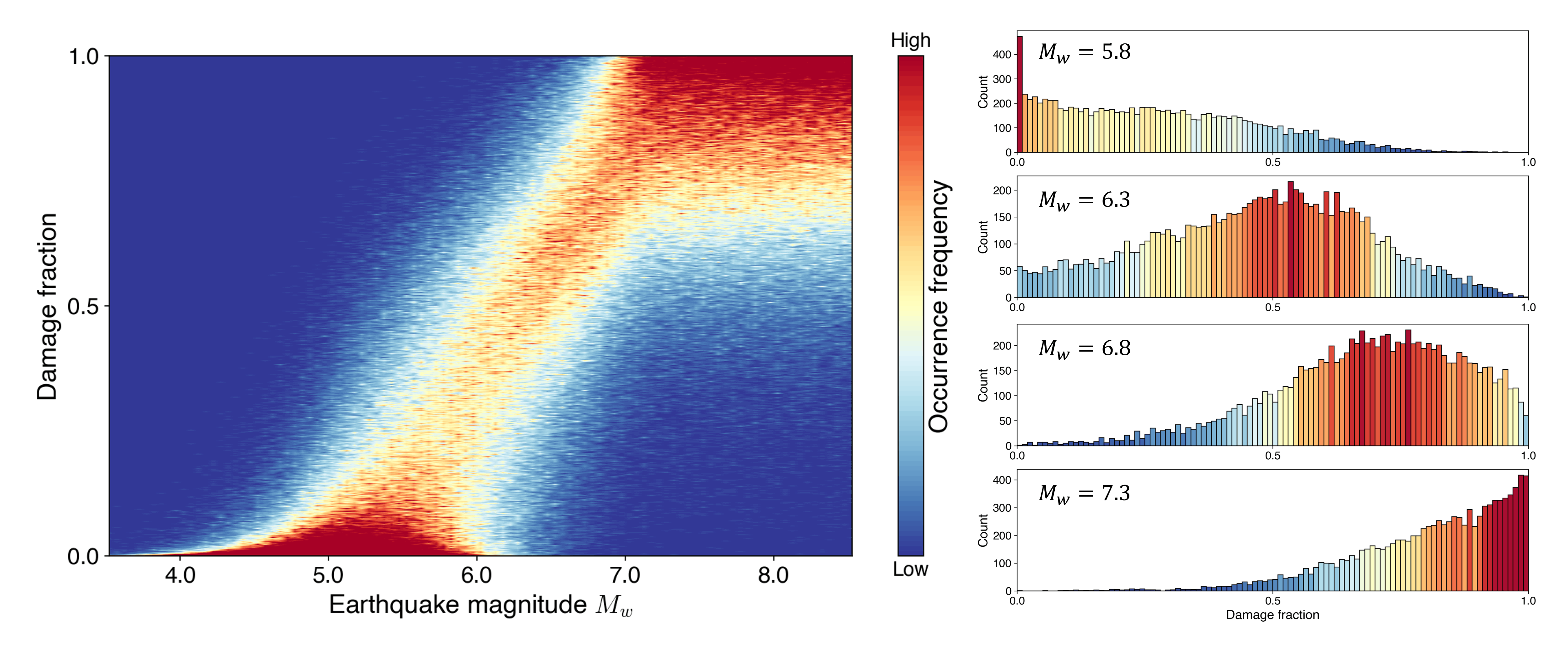}
    \caption{\textbf{Evolution of damage-fraction distributions with increasing earthquake magnitude for the northeastern San Francisco portfolio.} 
    Heatmap (left) shows regional damage-fraction distributions versus earthquake magnitude, with the color indicating relative occurrence frequency. Histograms (right) show the distributions at selected magnitudes ($M_w = 5.8$, $6.3$, $6.8$, and $7.3$), highlighting a smooth transition from the collectively safe to the collectively damaged state in this high-diversity portfolio.}
    \label{ext_fig:heatmap_1st_sanfrancisco}
\end{figure}

\newpage
\begin{figure}[H]
    \centering
    \includegraphics[width=0.8\linewidth]{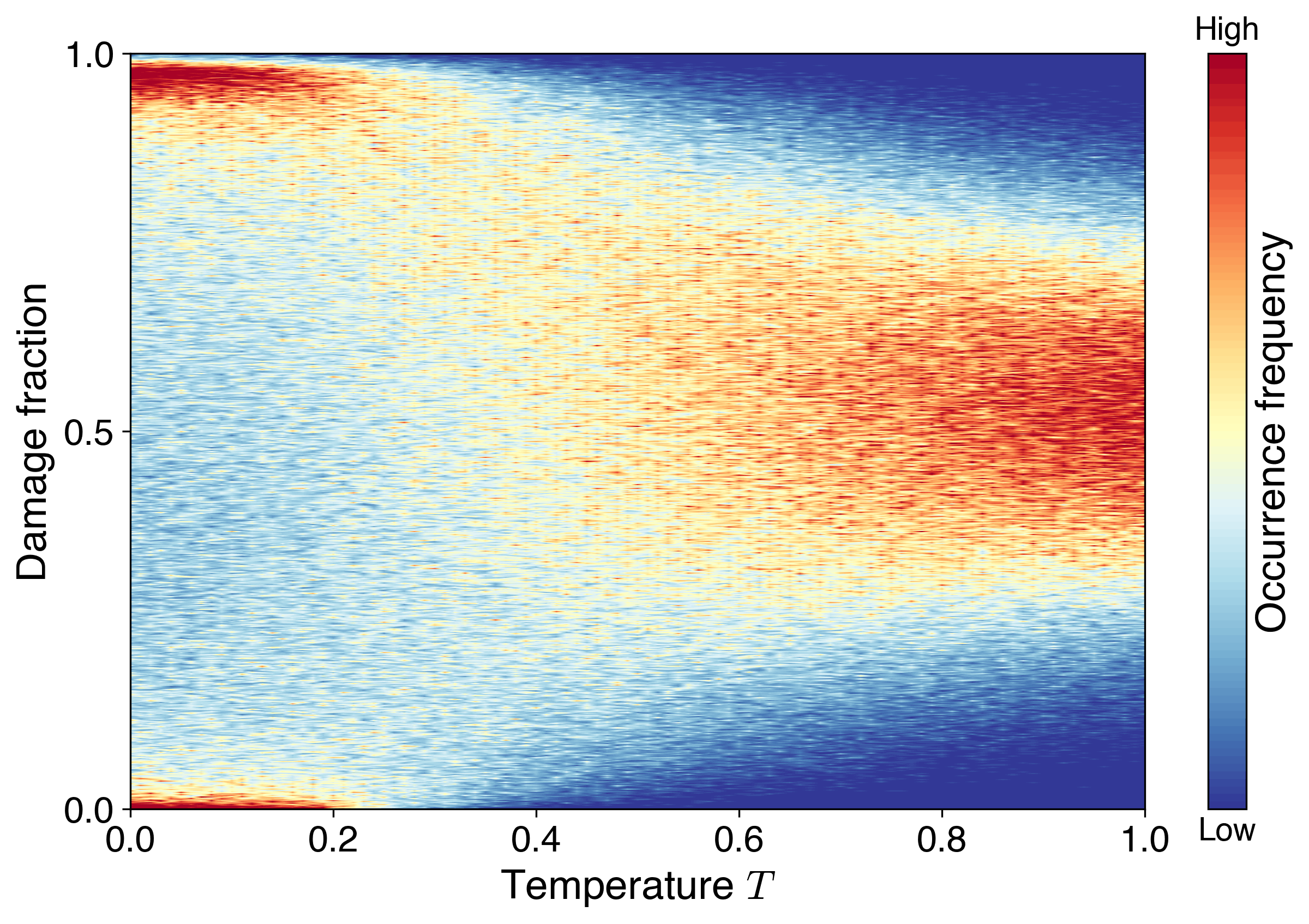}
    \caption{\textbf{Evolution of damage-fraction distribution with temperature in regional simulations.}
    Increasing the temperature-like parameter $T$ progressively smooths the transition between the collectively safe and collectively damaged states, illustrating the analogy of temperature in the random-field Ising model to uncertainty in the hazard--structure interaction model; see Supplementary Method~\ref{sup_method:temp} for implementation details.}
    \label{ext_fig:heatmap_temp}
\end{figure}

\newpage
\begin{figure}[H]
    \centering
    \includegraphics[width=1.0\linewidth]{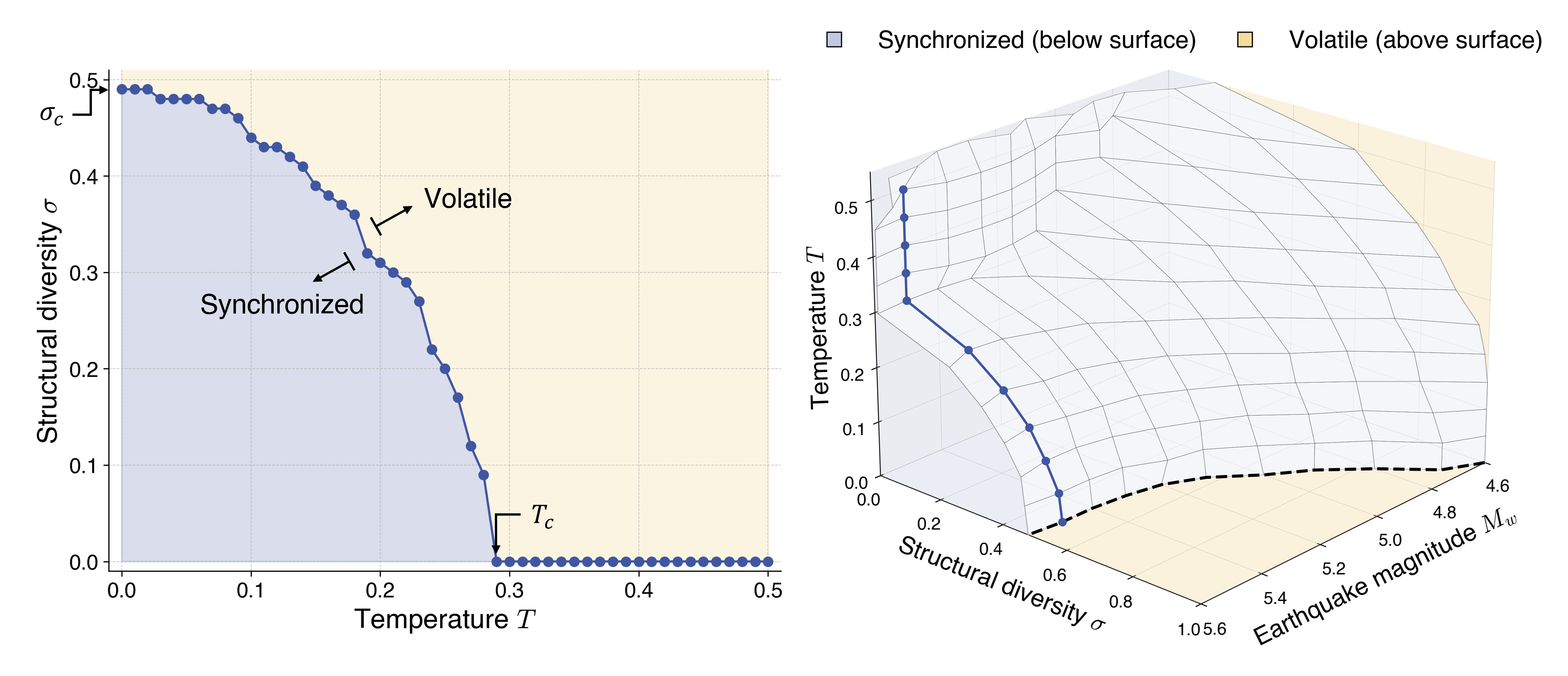}
    \caption{\textbf{Phase diagram for Milpitas, California, with temperature as a control parameter.}
    Collective behavior is shown in structural diversity--temperature space at $M_w=5.5$ (left) and in structural diversity--temperature--magnitude space (right). The left panel shows the decrease in the critical diversity $\sigma_c$ with increasing $T$, which marks the boundary between synchronized regimes (collectively safe or collectively damaged city-scale risk) and a volatile regime with mixed outcomes. The right panel extends this relationship across earthquake magnitudes $M_w$, yielding a continuous phase surface separating synchronized and volatile regions. Increasing $T$ suppresses abrupt transitions, smoothing the boundary and weakening phase separation, consistent with the random-field Ising model~\cite{aharony_tricritical_1978, sebastianes_phase_1987, theodorakis_random-field_2014}; see Supplementary Method~\ref{sup_method:temp} for implementation details.}
    \label{ext_fig:phase_diagram_temperature}
\end{figure}

\newpage
\begin{figure}[H]
    \centering
    \includegraphics[width=0.8\linewidth]{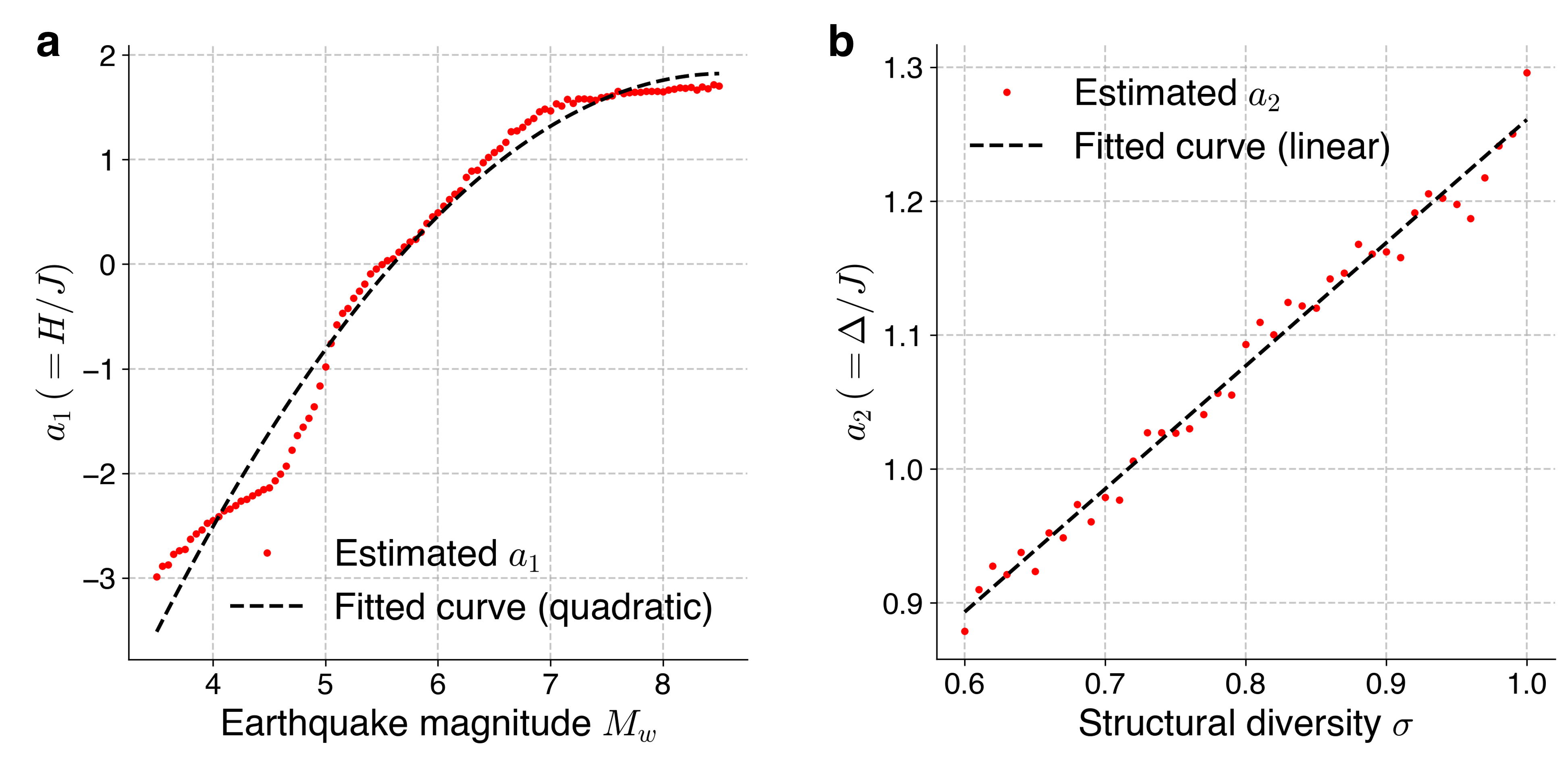}
    \caption{\textbf{Fitted normalized RFIM parameters.} The fitted parameters are shown as the normalized quantities $a_1 \equiv H/J$ and $a_2 \equiv \Delta/J$, with the mean-field coupling strength fixed at $J=1$ throughout the fitting.
    \textbf{a}, Normalized effective field versus earthquake magnitude, with a quadratic fit crossing zero at the inferred critical magnitude.
    \textbf{b}, Normalized effective disorder versus structural diversity, with a non-decreasing linear fit consistent with progressive smoothing of the transition as heterogeneity increases. To stabilize the fit and avoid spurious bistability, only data in the volatile regime ($\sigma>0.6$) were used.}
    \label{ext_fig:rfim_param_est}
\end{figure}

\newpage
\begin{figure}[H]
    \centering
    \includegraphics[width=1.0\linewidth]{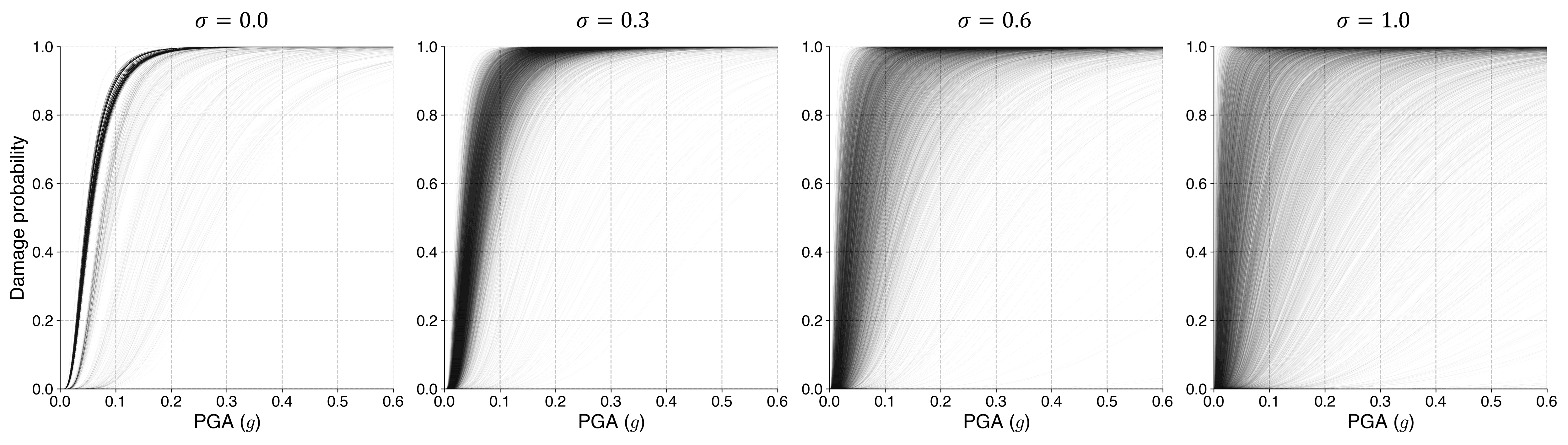}
    \caption{\textbf{Evolution of fragility curves with increasing structural diversity.}
    Fragility curves for the multistory building portfolio in Milpitas, California, shown for four structural diversity levels. Here, $\sigma$ denotes the additional variability introduced to emulate cities with different levels of structural heterogeneity. $\sigma=0.0$ corresponds to the baseline fragility-curve distribution estimated from the IDA for Milpitas, with no added heterogeneity. $g$ denotes gravitational acceleration.}
    \label{ext_fig:frag_evolution}
\end{figure}

\newpage
\begin{figure}[H]
    \centering
    \includegraphics[width=1.0\linewidth]{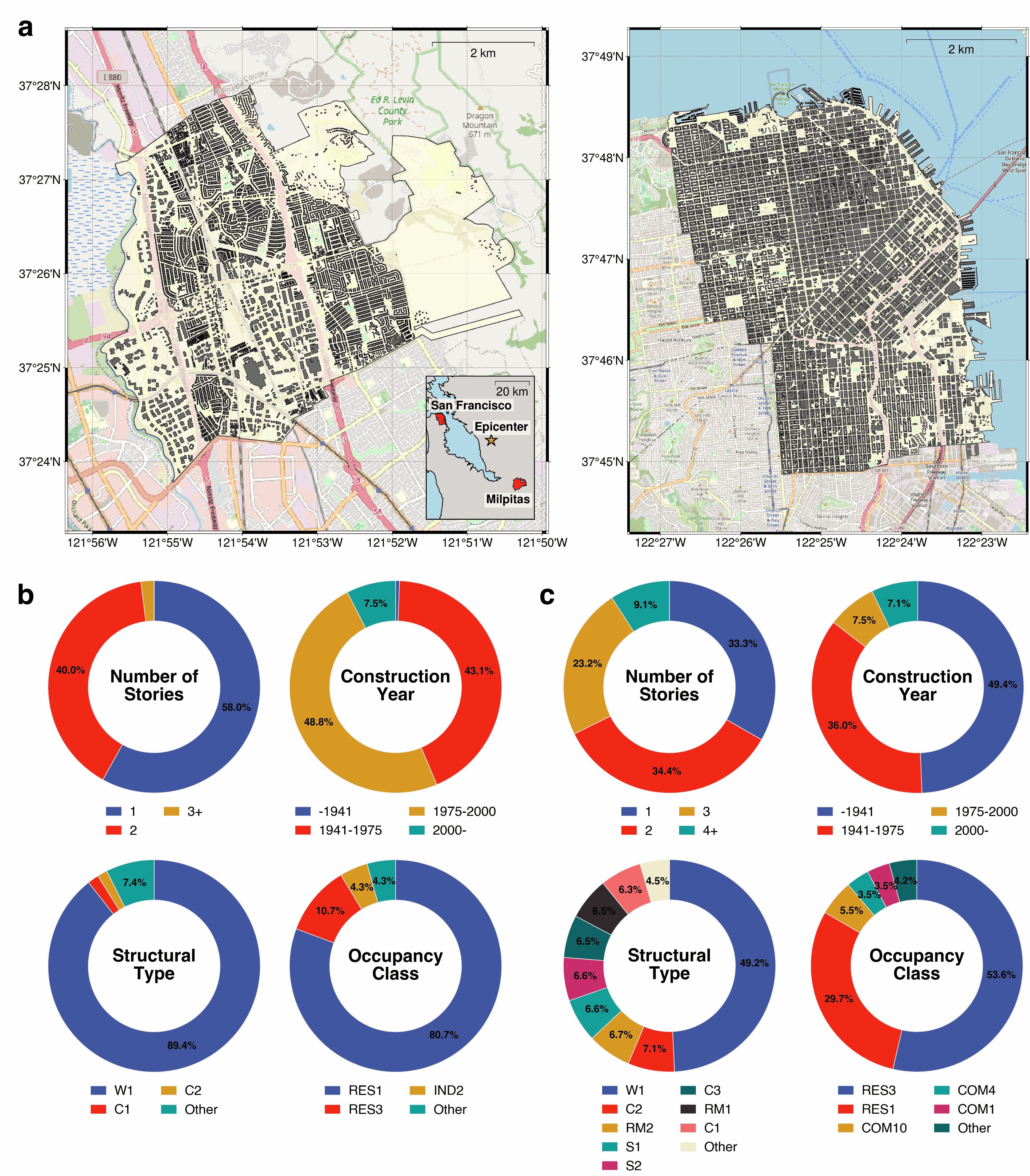}
    \caption{\textbf{Regional building portfolios of Milpitas and San Francisco, California.}
    \textbf{a}, Building footprints of Milpitas (left) and northeastern San Francisco (right), with an inset map showing their relative locations and the earthquake epicenter.
    \textbf{b}, Composition of the Milpitas portfolio by number of stories, year built, structure type, and occupancy class, showing the dominance of low-rise residential buildings.
    \textbf{c}, Same distributions for San Francisco, revealing greater structural and functional diversity across the urban core. Categories for structural type and occupancy class follow Hazus~6.1~\cite{federal_emergency_management_agency_fema_hazus_2024}.}
    \label{ext_fig:region_maps_donuts}
\end{figure}

\newpage
\begin{figure}[H]
    \centering
    \includegraphics[width=0.8\linewidth]{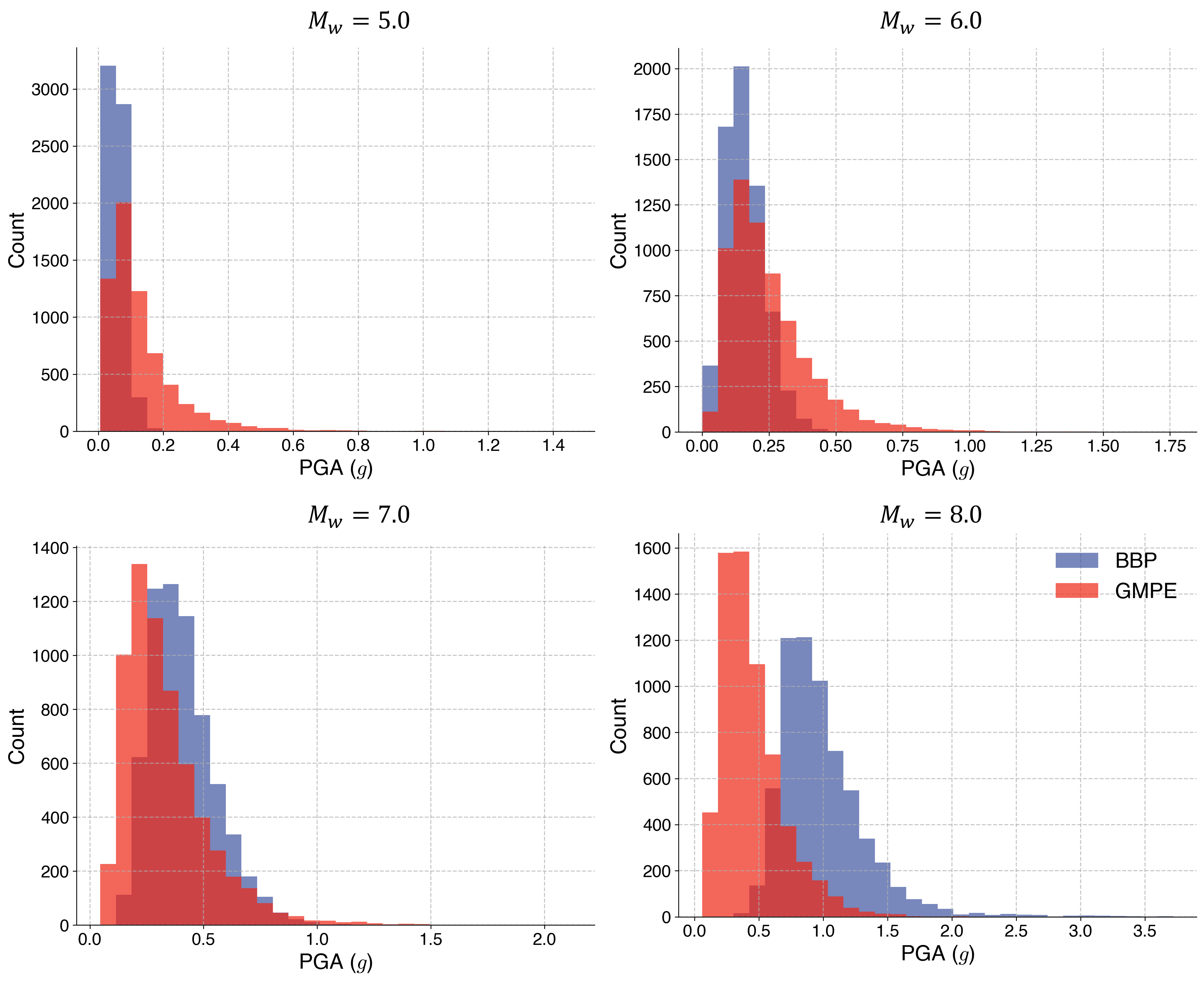}
    \caption{\textbf{Comparison of peak ground acceleration (PGA) distributions between physics-based and empirical models across four earthquake magnitudes.} 
    Histograms show PGA from physics-based SCEC Broadband Platform (BBP) simulations and the empirical Chiou--Youngs ground-motion prediction equation (GMPE)~\cite{chiou_update_2014}. PGA was sampled at 64 recording stations (the same for BBP and GMPE), uniformly distributed within Milpitas. For each method, 100 realizations were generated, yielding $64\times100=6{,}400$ samples per histogram. Supplementary Figs.~\ref{sup_fig:pga_comparison_per_station_Mw5}--\ref{sup_fig:pga_comparison_per_station_Mw8} show the corresponding station-by-station distribution comparisons for the four magnitudes, and Supplementary Fig.~\ref{sup_fig:compare_bbp_gmpe_spatial_correlation} compares the associated log-PGA spatial correlations.}
    \label{ext_fig:pga_comparison}
\end{figure}

\end{bibunit}

%%%%%%%%%%%%%%%%%%%%%%%%%%%%%%%%%%%%%%%%%%%%%%%%%%%%%%
% Supplementary Information
%%%%%%%%%%%%%%%%%%%%%%%%%%%%%%%%%%%%%%%%%%%%%%%%%%%%%%
\clearpage
\begin{bibunit}
\setcounter{tocdepth}{2}

\ResetELSFrontmatter

\begingroup
\makeatletter
\let\MaketitleBox\MaketitleBoxNoRules
\makeatother

\begin{frontmatter}

\title{Supplementary Information for:\\
Phase Transitions in Collective Damage of Civil Structures \\under Natural Hazards}

\author[1]{Sebin Oh}
\author[2]{Jinyan Zhao}
\author[3]{Raul Rincon}
\author[3]{Jamie E. Padgett}
\author[1]{Ziqi Wang\corref{cor1}}
\ead{ziqiwang@berkeley.edu}
\cortext[cor1]{Corresponding author}

\address[1]{Department of Civil and Environmental Engineering, University of California, Berkeley, CA, United States of America}
\address[2]{Department of Mechanical and Civil Engineering, California Institute of Technology, Pasadena, CA, United States of America}
\address[3]{Department of Civil and Environmental Engineering, Rice University, Houston, TX, United States of America}

\end{frontmatter}
\endgroup

\vspace{1.5\baselineskip}
\noindent\textbf{This PDF file includes:}\\[0.5\baselineskip]
\noindent Supplementary Note 1--5\\
Supplementary Methods 1--8\\
Supplementary Figs. 1--30\\
References

\newpage
\renewcommand*\contentsname{Table of Contents}
\tableofcontentsSI

% ---- Prefix numbering in SI ----
\setcounter{figure}{0}
\renewcommand{\figurename}{Fig.}
\renewcommand{\thefootnote}{\fnsymbol{footnote}}

\newpage
\SupplNotesHeading
\suppnote{Probabilistic simulations in regional-scale risk analysis}\label{sup_note:prob_sim}\noindent
Regional-scale risk analysis to natural hazards fundamentally depends on probabilistic simulations. It is infeasible to directly observe or experimentally reproduce the full-scale response of an entire city to a natural hazard. Field surveys and post-disaster reconnaissance provide only a limited, one-shot data point that represents a single realization of complex stochastic processes. To overcome this scarcity of empirical observations, probabilistic simulations serve as an essential tool for exploring the range of possible regional civil structural responses, quantifying uncertainties, and identifying emergent behaviors that cannot be captured from individual events alone. Such simulations enable researchers to represent the inherent randomness in both hazard excitations and structural capacities, forming the foundation for assessing collective risk at the city or regional scale.

\newpage
\suppnote{Interpretation and scope of the temperature-like parameter}\label{sup_note:temp}\noindent
In the random-field Ising model (RFIM), temperature drives thermal fluctuations, allowing spins to probabilistically misalign with their local effective fields. The regional damage simulations in the main text operate in the zero-temperature limit, where a given realization of structural capacity and hazard demand maps deterministically to a binary damage state. Introducing a non-zero temperature replaces this sharp, deterministic threshold with a stochastic damage-assignment rule (Supplementary Method~\ref{sup_method:temp}).

The potential engineering application of introducing this stochasticity is to probe residual, \emph{unstructured} uncertainty on top of the \emph{structured} uncertainty encoded in the capacity and demand models. This includes broad model-form error in the hazard--structure model and omitted effects that cannot be cleanly assigned to a specific building, fragility parameter, or demand model. From a decision-making perspective, reduced confidence in the regional simulation outcomes can likewise be viewed as such residual uncertainty. Therefore, temperature should be interpreted as an effective summary parameter for the net influence of these residual uncertainties, rather than as a uniquely identifiable uncertainty source.

The nonzero-temperature simulations can be interpreted as a complementary sensitivity study, not as a calibration of a literal thermodynamic temperature. Their purpose is to examine how residual, unstructured uncertainty smooths the collective response and shifts the phase boundaries identified in the baseline zero-temperature analysis.

\newpage
\suppnote{Modeled uncertainty, imposed heterogeneity, and RFIM disorder}\label{sup_note:disorder}\noindent
In the RFIM framework, the local random field is characterized by its spatial mean $H$ and disorder strength $\Delta$. Here, $\Delta$ quantifies the spatial dispersion of the local capacity--demand imbalance across the region, shaped jointly by the modeled uncertainty and imposed heterogeneity.

To make this interpretation explicit, we decompose the simulated capacity--demand imbalance into a fixed portfolio-level mean field and conditional fluctuations around that mean field. Under the lognormal capacity and demand models adopted in the main text, the logarithmic capacity--demand imbalance, or the safety-factor field,
\[\vect{X}=\ln \vect{C}-\ln \vect{D}\]
is Gaussian conditional on a fixed perturbed portfolio,
\[\vect{X}\mid\vect{\mu}_{\vect{X}} \sim \mathcal{N}\!\left(\vect{\mu}_{\vect{X}},\,\vect{\Sigma}_{\vect{X}}\right)\,,\]
where $\vect{\mu}_{\vect{X}}$ is the mean safety-factor field of the perturbed portfolio and $\vect{\Sigma}_{\vect{X}}$ is the covariance matrix induced by the capacity and demand models. For each prescribed value of $\sigma$, the additional heterogeneity is imposed by drawing a perturbation to the portfolio-level mean safety-factor field:
\[\vect{\mu}_{\vect{X}} = \vect{\mu}_{\vect{X}_0} + \vect{\varepsilon}_{\sigma}, \qquad \vect{\varepsilon}_{\sigma} \sim \mathcal{N}\!\left(\vect{0},\,\sigma^2\vect{I}\right)\,,\]
where $\vect{\mu}_{\vect{X}_0}$ denotes the inventory-derived baseline mean safety-factor field. Consequently, the unconditional safety-factor field across the entire ensemble is given by:
\[\vect{X} \sim \mathcal{N}\!\left(\vect{\mu}_{\vect{X}_0},\,\vect{\Sigma}_{\vect{X}} + \sigma^2\vect{I}\right)\,.\]

As shown by the unconditional distribution, $\sigma$ adds unstructured variation to scatter the baseline capacities, while $\vect{\Sigma}_{\vect{X}}$ governs the spatially correlated hazard demand and capacity fluctuations. The effective RFIM disorder, $\Delta$, serves as an inferred macroscopic proxy that aggregates the diagonal effects of $\vect{\Sigma}_{\vect{X}} + \sigma^2\vect{I}$. The relative strength of the off-diagonal entries of $\vect{\Sigma}_{\vect{X}}+\sigma^2 \vect I$  compared to its diagonal entries is responsible for the spatial coherence of damage and is effectively captured by the parameter \(\vect J\) in the RFIM.

\newpage
\suppnote{Extension to multiple damage states}
\label{sup_note:multi_damage_states}
\noindent
The main analysis represents each building using a binary damage indicator. Extending the framework to represent several damage states simultaneously would constitute a distinct multi-state problem. Each building would occupy one of several damage states, such as slight, moderate, extensive, or complete damage, and the regional response would be characterized by the fractions of buildings in those states.

One possible statistical-physics formulation is a generalized random-field Potts-type model or another ordinal multi-state spin model. Alternatively, the damage states may be represented through coupled cumulative-exceedance fields. Such formulations may exhibit sequential or overlapping transitions among damage levels, producing a richer phase diagram than the binary case. Their quantitative development requires joint, damage-state-specific fragility and dependence models and lies beyond the scope of the present study.

\newpage
\suppnote{Robustness and interpretation of the IDA-derived capacity-dependence model}
\label{sup_note:capa_dependence}
\noindent
In the regional simulations, each building's capacity is defined as the value of a selected hazard intensity measure at which its structural model first reaches the specified damage threshold, e.g., peak ground acceleration (PGA) for the earthquake application. Preserving the record-wise alignment of the IDA capacity samples retains co-variation associated with record-specific waveform characteristics, while similarities among structural models also contribute to the estimated dependence. We therefore interpret the resulting matrix as an effective capacity-dependence model that may combine shared record-level effects and structural-model similarities, rather than as a uniquely identified matrix of intrinsic pairwise capacity correlations.

Although the matrix dimension equals the number of buildings, the effective model dimensionality is much smaller. In our implementation, each building is represented by a nonlinear multi-degree-of-freedom model parameterized primarily by its Hazus structural type, code level, and number of stories. Given the 16 Hazus structural types, four code levels, and the predominantly low-rise portfolio, the effective dimensionality is on the order of several hundred and is unlikely to substantially exceed 500. Here, effective model dimensionality represents the number of dominant modes in structural responses rather than the algebraic rank of the empirical correlation matrix. This dimensional reduction is not specific to our implementation: real building portfolios are organized around a finite set of structural systems, design eras, and height classes, with continuous differences largely appearing as variations around these archetypes. Thus, although its exact value depends on model resolution, the effective dimensionality should remain far below the total number of buildings.

We therefore extended the IDA analysis from 100 to 500 records and compared the normalized non-zero eigenvalue spectra of the original 100-record estimate, the full 500-record estimate, and random 100-record subsamples drawn from the expanded pool. The spectra are dominated by the first few modes, and the original spectrum lies within the spread of the subsampled spectra (Supplementary Fig.~\ref{sup_fig:capa_eigen_spectrum}). This indicates that the original record set captures the dominant dependence structure within the sampling variability expected for a 100-record analysis.

Because the number of records is smaller than the number of buildings, the empirical capacity-correlation matrix is rank deficient and cannot be directly Cholesky-factorized. It is converted into a capacity covariance matrix and combined with the demand covariance matrix to form the covariance matrix of the logarithmic safety-factor field. A small diagonal jitter, beginning at $10^{-10}$ and increasing when necessary up to $10^{-5}$, is added only to this final covariance matrix to ensure numerical stability of the Cholesky factorization. This constitutes numerical regularization of the covariance used for sampling, rather than statistical shrinkage of the capacity-correlation matrix.

We further compared the IDA-derived model with conditionally independent capacities and a simplified equicorrelation model with $\rho=0.5$, while holding the marginal fragility distributions, building inventory, demand model, and Monte~Carlo settings fixed. The equicorrelation case provides an intermediate-dependence benchmark rather than a calibrated representation of the portfolio. As shown in Supplementary Fig.~\ref{sup_fig:capa_dependence_ablation}, stronger capacity dependence broadens the damage-fraction distribution and makes the bimodal, abrupt transition more pronounced. This comparison illustrates the sensitivity of the collective transition to the assumed strength of capacity dependence.

The sign structure of this dependence is also relevant to the effective RFIM interpretation. The dominant sources of capacity dependence identified above, i.e., shared record-level effects and similarities among structural models sharing a structural type, code level, and height class, tend to shift capacities coherently rather than in opposite directions. The spatially correlated demand field likewise promotes same-sign co-variation in the resulting capacity-demand imbalance. The standard regional simulation pipelines considered in this study do not involve a systematic mechanism by which one building’s capacity increases while another’s decreases. Consistent with this, the estimated capacity correlations are overwhelmingly positive. This sign structure is therefore consistent with a predominantly ferromagnetic effective coupling in the RFIM interpretation, rather than with the frustrated, mixed-sign interactions characteristic of spin-glass systems.

Together, these analyses support using the IDA-derived matrix as an effective portfolio-level dependence model with cooperative, ferromagnetic-like structure, while recognizing that the individual pairwise correlations are not uniquely identified from the available records.

\newpage
\SupplMethodsHeading
\suppmethod{PCA analysis of dominant collective modes}\label{sup_method:pca}\noindent
Principal component analysis (PCA) has been used to identify dominant collective modes and phase-related coordinates from microscopic configurations~\cite{wang_discovering_2016, hu_discovering_2017, bradde_pca_2017}. The use of the regional damage fraction as an order parameter is motivated primarily by its direct relation to the RFIM magnetization, $m=2m_d - 1$. We use PCA as a complementary, minimally parametric analysis to examine whether the simulated damage ensembles are dominated by this uniform aggregate-damage mode or by competing nonuniform modes.

To this end, we employ principal component analysis (PCA) to extract low-dimensional collective modes from ensembles of simulation outcomes. The analyses are conducted for two study regions (Milpitas and northeastern San Francisco) under two distinct simulation scenarios. In the first-order scenario, the earthquake magnitude $M_w$ is varied while structural diversity is fixed at $\sigma=0$ (no additional heterogeneity). In the second-order scenario, the structural diversity $\sigma$ is varied while the magnitude is fixed at the critical value $M_w = M_{w_c}$ defined as the magnitude at which the average damage fraction equals $0.5$. This design yields four analysis cases in total, one for each combination of region and scenario. For each case, PCA is applied to ensembles of realizations sampled across the corresponding control parameter ($M_w$ for the first-order and $\sigma$ for the second-order scenario). 

The resulting explained-variance spectra are shown in Supplementary Fig.~\ref{sup_fig:pca_explained_var}. In all cases, the first principal component dominates the total variance, particularly for the first-order scenarios where the system response is governed by a single collective mode representing the abrupt transition between the safe and damaged states. For the second-order scenarios, the variance explained by the first component is relatively lower, reflecting the distributed, continuous reorganization of the system across a broad range of structural diversity. The explained variance for the San Francisco cases is generally smaller than that of Milpitas, indicating the weak inter-building dependence in the San Francisco region that renders the regional damage-fraction distribution approximately Gaussian.

The equivalence between the first principal component (PC1) and the regional damage fraction is demonstrated in Supplementary Fig.~\ref{sup_fig:pca_order_param_validation}. Across all four analysis cases, PC1 exhibits a nearly one-to-one correspondence with the damage fraction, confirming that the dominant collective mode identified purely from the data coincides with the physically interpretable order parameter of the system. This alignment validates that the macroscopic organization of regional responses can be fully captured by a single scalar quantity, consistent with the analogy between magnetization in the random-field Ising model and the regional damage fraction in city-scale simulations.

Supplementary Fig.~\ref{sup_fig:pca_order_param_validation} shows a near-linear relation between PC1 and the regional damage fraction in all four cases. Such a relation is expected when the leading PCA loading is aligned with the uniform direction, because projection onto this direction is proportional to the centered regional damage fraction. The result therefore indicates that variation in the overall damage level is the dominant linear collective mode in these ensembles, while the remaining components represent residual nonuniform variation.

The visual evolution of ensembles projected onto the PC1–PC2 space is presented in Supplementary Figs.~\ref{sup_fig:pca_1st_Milpitas}–\ref{sup_fig:pca_2nd_SanFrancisco}. Under the first-order scenario (varying $M_w$ at $\sigma=0$), the Milpitas case exhibits an abrupt shift of the distribution along PC1 around $M_w \approx 5.60$, marking a discontinuous transition from the collectively safe to the collectively damaged state. In contrast, the San Francisco case follows a smooth trajectory without a sudden jump, indicating the absence of a sharp first-order transition. Under the second-order scenario (varying $\sigma$ at fixed $M_w = M_{w_c}$), Milpitas shows a gradual merging of distributions along PC1, consistent with continuous, second-order–like behavior. The San Francisco case, by contrast, shows no fundamental reorganization across increasing $\sigma$, confirming that the region is already in a disordered, paramagnetic-like phase. Together, these PCA projections provide direct visual evidence that the nature of collective regional response transitions is governed by the external excitation intensity and by the strength of inter-building coupling.

\newpage
\suppmethod{Non-zero temperature simulation}\label{sup_method:temp}\noindent
The regional simulations described in the main text correspond to the zero-temperature limit; the safety factor (or safety margin) $X_i=\ln C_i-\ln D_i$, where $C_i$ and $D_i$ denote structural capacity and hazard demand of building $i$, respectively, fully determines the damage state. We examine the effect of allowing a \emph{probabilistic margin} to the safety factor. Within the RFIM--regional-simulation analogy, $X$ acts as a local bias between the safe and damaged states. Accordingly, without loss of generality, we write the corresponding energy gap as
\begin{equation*}
    \Delta E = E_{\text{damaged}} - E_{\text{safe}} = X\,.
\end{equation*}
Because only energy differences matter, we may choose $E_{\text{safe}}=0$ and $E_{\text{damaged}}=\Delta E$. The corresponding Boltzmann weights for the two-state system are then
\begin{equation*}
    p_{\text{safe}} \propto \exp(-E_{\text{safe}}/T)=1\,,
    \qquad
    p_{\text{damaged}} \propto \exp(-E_{\text{damaged}}/T)=\exp(-\Delta E/T)\,.
\end{equation*}
Normalizing these probabilities gives
\begin{equation*}
    \Prob{\text{Damage}\mid X}
    =
    \frac{1}{1+\exp\left(X/T\right)}\,,
\end{equation*}
where $T$ is a phenomenological temperature-like parameter that controls the sharpness of the damage threshold. In the limit $T\to0$, this expression reduces to the deterministic rule $\mathds{1}\{X<0\}$, whereas larger $T$ broadens the transition around $X=0$, representing greater residual uncertainty in the damage assignment. The interpretation and scope of $T$ are discussed in Supplementary Note~\ref{sup_note:temp}.

Spatially averaged damage fraction in the magnitude--temperature space is shown in Extended Data Fig.~5, revealing the progressive smoothing of the collective response as temperature increases. Interestingly, the critical structural diversity $\sigma_c$, beyond which the region no longer exhibits a first-order-like abrupt transition, varies systematically with temperature, as illustrated in Extended Data Fig.~6.

We observe that $\sigma_c$ decreases as temperature increases. This suggests that large uncertainty in the hazard--structure model systematically obscures the emergence of collective behavior, smoothing out the distinct phases observed at lower uncertainty levels. Maintaining a consistent phase across analyses therefore becomes crucial for meaningful risk assessment. Conversely, pursuing excessively precise component-level models or datasets may yield limited returns for system-level decision-making, as microscopic refinements are often averaged out when collective dynamics dominate.

\newpage
\suppmethod{Ising model revisited}\label{sup_method:ising}
\subsection*{The classical Ising model}\noindent
The Ising model provides a fundamental framework for describing collective behavior that arises from interactions among many coupled components. In its general form, the system consists of $N$ discrete sites that interact with one another, each represented by a spin variable $s_i \in \{-1,+1\}$ corresponding to one of two possible states. The total energy of a given configuration is expressed by the Hamiltonian
\[
\mathcal{H}(\{s_i\}) = -\sum_{i<j} J_{ij}s_i s_j - H\sum_i s_i\,,
\]
where $J_{ij}$ denotes the coupling strength between sites $i$ and $j$, and $H$ is an external field acting uniformly on all sites. The first term captures the tendency of neighboring spins to align when $J_{ij}>0$, while the second term biases the spins toward the direction of the external field. The balance between these two contributions determines the overall alignment of the system.

At thermal equilibrium, the probability of observing a particular configuration $\{s_i\}$ follows the Boltzmann distribution,
\[
p\left(\{s_i\}\right) = \frac{1}{Z}\exp\left(-\beta \mathcal{H}(\{s_i\})\right)\,,
\]
where $\beta = 1/\left(k_{\mathrm{B}}T\right)$ is the inverse temperature, $k_{\mathrm{B}}$ is the Boltzmann constant, $T$ is the absolute temperature, and $Z$ is the partition function that ensures normalization. The macroscopic state of the system is characterized by the equilibrium magnetization
\[
m^\star \equiv \langle m \rangle = \frac{1}{N}\sum_{i=1}^{N} \langle s_i \rangle\,,
\]
where $\langle s_i \rangle$ denotes the thermal expectation value of spin $s_i$. This quantity serves as the order parameter, distinguishing different phases and capturing critical behavior. When the temperature $T$ is high, thermal fluctuations dominate and the spins are randomly oriented, leading to $m \approx 0$ (a disordered phase). Below a critical temperature $T_c$, the system spontaneously breaks symmetry and develops a nonzero magnetization ($m = \pm m_0$, with $m_0 > 0$), indicating an ordered phase in which most spins align in the same direction. This spontaneous symmetry breaking represents a continuous (second-order) phase transition, providing a minimal mathematical representation of how local interactions can give rise to abrupt macroscopic organization.

A key distinction between the classical Ising model and its regional damage interpretation concerns $\mathbb{Z}_2$ symmetry. The conventional Ising system is symmetric under spin inversion when the external field is zero, whereas civil structures are inherently biased toward the safe state in the absence of hazard excitation. The inherent safety margin of engineered structures acts as an effective inertia opposing hazard excitation, shifting the zero-field condition to the point where this inertia balances the excitation~\cite{oh_longrange_2024}.

\subsection*{Random-field Ising model (RFIM)}\noindent
The RFIM extends the classical Ising framework by introducing spatial disorder through site-dependent local fields. The Hamiltonian of the RFIM is expressed as
\[
\mathcal{H}(\{s_i\}) = -\sum_{i<j} J_{ij}s_i s_j - \sum_i (H + h_i)s_i\,,
\]
where $H$ represents the uniform external field and $h_i$ denotes the local deviation from $H$ at site $i$. The disorder in the random field $h_i$ is quenched, meaning that the local fields are frozen in each realization and remain fixed during thermal averaging. The random fields $\{h_i\}$ are typically drawn from a Gaussian distribution, $h_i \sim \mathcal{N}(0,\Delta^2)$, where $\Delta$ characterizes the strength of disorder. The first term in $\mathcal{H}$ captures the pairwise coupling between sites, while the second term introduces local heterogeneity that competes with the global ordering tendency.

The interplay between the coupling $J_{ij}$ and disorder amplitude $\Delta$ governs the collective behavior of the system. For small $\Delta$, the spins remain strongly correlated, and the system exhibits an abrupt transition between ordered and disordered states as the external field $H$ varies. In contrast, large $\Delta$ suppresses long-range correlations, leading to a smooth evolution of the macroscopic magnetization. This transition from discontinuous to continuous behavior reflects the shift from a second-order phase transition as disorder increases.

The RFIM thus provides a foundational framework for systems in which local variability modulates global organization. In the context of regional-scale structural responses to natural hazards, the disorder strength $\Delta$ represents the degree of heterogeneity in building capacities or subsurface and meteorological conditions, whereas $J_{ij}$ characterizes the effective interaction between structures subjected to correlated hazard excitation. This analogy enables a statistical-physics interpretation of regional risk dynamics through the lens of critical phenomena.

\subsection*{Mean-field approximation and zero-temperature limit of RFIM}\noindent
To obtain a tractable analytical description of the system’s macroscopic behavior, we employ the mean-field approximation, which replaces explicit pairwise interactions with an average coupling to the mean magnetization. In this framework, spatial variations are neglected and the interaction coefficients are replaced by a uniform coupling, $J_{ij} = \tfrac{J}{N}$. Each spin therefore experiences an effective field $h_i^{\mathrm{eff}} = Jm + H + h_i$, combining the mean-field contribution $Jm$, the external field $H$, and the local random perturbation $h_i$.

At zero temperature, thermal fluctuations vanish, and the state of each spin is determined deterministically by the sign of its local effective field:
\[
s_i = \mathrm{sign}\!\left(Jm + H + h_i\right).
\]
The ensemble-averaged magnetization is then obtained as
\[
\langle m \rangle = \frac{1}{N}\sum_{i=1}^{N} \langle s_i \rangle
                  = \int_{-\infty}^{\infty} \mathrm{sign}\!\left(Jm + H + h\right)\,\phi(h)\,\mathrm{d}h\,,
\]
where $\phi(h) = \tfrac{1}{\sqrt{2\pi}\Delta}\exp\!\left(-\tfrac{h^2}{2\Delta^2}\right)$ is the Gaussian distribution of the random field. Exploiting the symmetry of $\phi(h)$, we obtain the self-consistency condition for the magnetization
\begin{align}
m   &= \Prob{Jm + H + h > 0} - \Prob{Jm + H + h < 0} \nonumber \\
    &= 2\Phi\!\left(\frac{Jm + H}{\Delta}\right) - 1 \nonumber \\
    &= \erf\!\left(\frac{Jm + H}{\sqrt{2}\,\Delta}\right)\,, \label{sup_eq:RFIM_meanfield}
\end{align}
where $\Phi(\cdot)$ is the cumulative distribution function of the standard normal random variable. Equation~\eqref{sup_eq:RFIM_meanfield} thus defines the equilibrium magnetization $m^\star\equiv\langle m \rangle$ as the fixed point of the mean-field map.

\subsection*{Landau free-energy formulation}\noindent
The equilibrium states of the mean-field RFIM can be characterized by a phenomenological Landau free-energy potential $F_{\mathrm{MF}}(m)$, whose stationary points reproduce the self-consistent condition in Eq.~\eqref{sup_eq:RFIM_meanfield}. We define its gradient with respect to the order parameter $m$ as
\[
    \frac{\partial F_{\mathrm{MF}}}{\partial m}
    = m - \erf\!\left(\frac{Jm + H}{\sqrt{2}\,\Delta}\right)\,,
\]
so that $\partial F_{\mathrm{MF}}/\partial m = 0$ recovers Eq.~\eqref{sup_eq:RFIM_meanfield}. Integrating with respect to $m$ yields, up to an additive constant $C$,
\begin{align*}
F_{\mathrm{MF}}(m; H, \Delta)
    &= \frac{1}{2} m^2
       - \int \erf\!\left(\frac{Jm + H}{\sqrt{2}\,\Delta}\right)\, \mathrm{d}m + C \\
    &= \frac{1}{2}m^2
       - \left(m + \frac{H}{J}\right)
         \erf\!\left(\frac{Jm + H}{\sqrt{2}\,\Delta}\right)
       - \sqrt{\frac{2}{\pi}}\,
         \frac{\Delta}{J}\,
         \exp\!\left(-\frac{(Jm + H)^2}{2\Delta^2}\right)
       + C\,.
\end{align*}

The shape of $F_{\mathrm{MF}}(m;H,\Delta)$ provides an intuitive picture of how disorder alters collective behavior. For weak disorder ($\Delta \ll J$), the free-energy landscape exhibits two symmetric minima separated by a barrier, corresponding to bistable ordered states with $m = \pm m_0$. As $\Delta$ increases, the barrier gradually flattens and the two minima merge into a single well centered at $m = 0$, exhibiting the disappearance of bistability and the onset of a continuous transition. The evolution of $F_{\mathrm{MF}}(m)$ thus provides an intuitive picture of how local heterogeneity smooths out collective order, linking the microscopic disorder strength to the macroscopic transition type.

In the context of regional-scale structural risk analysis, this free-energy landscape represents the effective potential governing the collective damage state of the built environment. The emergence of bistable minima corresponds to regimes where the region exhibits two distinct macroscopic states, i.e., collectively safe and collectively damaged, while the single well regime reflects a continuous response transition without abrupt shifts in collective condition.

\newpage
\suppmethod{Held-out phase prediction using regional safety-factor statistics}\label{sup_method:heldout}\noindent
To evaluate whether the RFIM can be transferred to regional portfolios not used for its calibration, we reformulated the mapping in terms of regional safety-factor statistics that can be obtained before regional damage simulations are performed. For building $i$, we define the log-safety factor as
\begin{equation}
    X_i = \ln C_i - \ln D_i,
\end{equation}
where $C_i$ and $D_i$ denote structural capacity and hazard demand, respectively. Its mean is
\begin{equation}
    \mu_{X_i} = \E{X_i}
    = \E{\ln C_i - \ln D_i}.
\end{equation}
For a portfolio of $N$ buildings, the regional mean and spatial dispersion of the building-wise mean log-safety factor are defined as
\begin{equation}
    \bar{\mu}_X
    =
    \frac{1}{N}
    \sum_{i=1}^{N}
    \mu_{X_i},
\end{equation}
and
\begin{equation}
    S_{\mu_X}
    =
    \left[
    \frac{1}{N}
    \sum_{i=1}^{N}
    \left(
    \mu_{X_i}-\bar{\mu}_X
    \right)^2
    \right]^{1/2}.
\end{equation}
These quantities are calculated from the building-level mean log-capacity and mean log-demand fields and therefore do not require sampled damage states or regional damage-fraction distributions.

Milpitas was used as the sole calibration portfolio. For the unperturbed portfolio ($\sigma=0$), the regional mean log-safety factor $\bar{\mu}_X$ across earthquake magnitudes was related to the previously inferred normalized RFIM field $a_1=H/J$. Separately, the spatial dispersion $S_{\mu_X}$ was related to the normalized RFIM disorder $a_2=\Delta/J$ using the Milpitas heterogeneity analysis. Linear mappings were fitted as
\begin{equation}
    a_1
    =
    b_{10}
    +
    b_{11}\bar{\mu}_X,
\end{equation}
and
\begin{equation}
    a_2
    =
    b_{20}
    +
    b_{21}S_{\mu_X},
\end{equation}
with the interaction scale fixed at $J=1$. The fitted mappings and their calibration ranges are shown in Supplementary Fig.~\ref{sup_fig:mapping_safety_factor_rfim}. Once estimated, these mappings were fixed for all subsequent held-out analyses.

The fixed mappings were then applied without refitting to the unperturbed ($\sigma=0$) portfolios of northeastern San Francisco, North Berkeley, Belmont, and Livermore over multiple earthquake magnitudes. For each target scenario, $\bar{\mu}_X$ and $S_{\mu_X}$ were calculated solely from its building-level mean log-capacity and mean log-demand fields and transformed into the corresponding $a_1$ and $a_2$. The target-city regional damage simulations were not used in this mapping and were reserved for subsequent evaluation.

Given the mapped $a_1$ and $a_2$, the mean-field RFIM was evaluated using the self-consistency equation given by Eq.~(1) of the main text,
\begin{equation}
    m
    =
    \operatorname{erf}
    \left(
    \frac{m+a_1}
    {\sqrt{2}\,a_2}
    \right).
\end{equation}
The stable solution $m^\star$ corresponds to the equilibrium magnetization and was converted to the predicted most-probable regional damage fraction through
\begin{equation}
    m_d^\star
    =
    \frac{m^\star+1}{2}.
\end{equation}
The corresponding mean-field free-energy landscape provides an effective description of the distributional structure of the regional damage fraction, with its stable minimum determining $m_d^\star$.

For northeastern San Francisco, the forward RFIM estimates were further compared with the held-out regional damage-fraction distributions (Supplementary Fig.~\ref{sup_fig:held_out_sfne_hist}). The simulated distributions were obtained from 10,000 regional realizations at each earthquake magnitude. For visualization of the distributional comparison, the mean-field free energy $F_{\mathrm{MF}}$ was converted to a probability-density profile as
\begin{equation}
    p_{\mathrm{MF}}(m_d)
    \propto
    \exp
    \left[
    -\beta_{\mathrm{eff}}
    F_{\mathrm{MF}}(m_d)
    \right],
\end{equation}
where a positive scenario-specific scale $\beta_{\mathrm{eff}}$ was used to place the RFIM profile on the same probability-density scale as the empirical histogram. This scaling changes the width of the displayed profile but does not alter the location of the free-energy minimum or the predicted mode. The held-out comparison therefore evaluates the transfer of the most-probable collective damage state and its associated phase structure, rather than an independently calibrated prediction of the full damage-fraction distribution.

Finally, the same fixed mappings were evaluated over a grid of $(S_{\mu_X},\bar{\mu}_X)$ to construct the common phase representation shown in Supplementary Fig.~\ref{sup_fig:held_out_phase_diagram}. The calibration and held-out portfolio--hazard scenarios can thereby be located directly in the same phase space using only their capacity and demand models.

\newpage
\suppmethod{Ensemble-based estimation of susceptibility, correlations, and data-driven Landau free energy}\label{sup_method:ensemble_based_calc}\noindent
To quantify collective sensitivity and spatial coherence, we compute the susceptibility $\chi$ and the connected two-point correlation function $C(r)$ from the equilibrium ensemble at fixed $(M_w,\sigma)$ for a fixed portfolio realization (i.e., a fixed set of fragility curves). We repeat the calculation for 50 independent portfolio realizations, all at the same $\sigma$, to quantify realization-to-realization variability and obtain robust distributions of $\chi$ and $\xi$.

\subsection*{Equilibrium ensemble selection}\noindent
We select the narrowest interval of the damage fraction $m_d$, $[m_{d_{\mathrm{lo}}},m_{d_{\mathrm{hi}}}]$, that contains a fixed fraction $f$ of the samples (we use $f=0.10$, i.e., $1{,}000$ of $10{,}000$). Samples with $m_d\in[m_{d_{\mathrm{lo}}},m_{d_{\mathrm{hi}}}]$ form the equilibrium ensemble used to compute the ensemble average $\langle\cdot\rangle$ for $\chi$ and $C(r)$.

\subsection*{Connected correlations}\noindent
For each site $i=1,\cdots,N$, we record $s_i\in\{-1,+1\}$ across $R=10{,}000$ realizations $\alpha=1,\dots,R$ at fixed $(M_w,\sigma)$. Ensemble averages are taken over realizations:
\[
\langle s_i\rangle=\frac{1}{R}\sum_{\alpha=1}^{R}s_i^{(\alpha)}\,,\qquad
\langle s_i s_j\rangle=\frac{1}{R}\sum_{\alpha=1}^{R}s_i^{(\alpha)}s_j^{(\alpha)}\,.
\]
The connected correlation is
\[
C(i,j)=\langle s_i s_j\rangle-\langle s_i\rangle\langle s_j\rangle\,.
\]
To obtain a distance-dependent correlation, we compute pairwise Euclidean distances $r_{ij}=\|\mathbf{x}_i-\mathbf{x}_j\|$ and radially average $C(i,j)$ over all distinct pairs $i<j$ whose distances fall in bins $[r,r+\Delta r)$, yielding $C(r)$. We normalize by the on-site variance,
\[
C(0)=\frac{1}{N}\sum_{i=1}^{N}\Big(\langle s_i^2\rangle-\langle s_i\rangle^2\Big)
     =\frac{1}{N}\sum_{i=1}^{N}\big(1-\langle s_i\rangle^2\big)\,,
\]
and report $C(r)/C(0)$. Correlation lengths $\xi$ are estimated by fitting $C(r)/C(0)\propto e^{-r/\xi}$ over the range shown in Fig.~3d (using only bins with positive $C(r)$).

\subsection*{Susceptibility estimation from a polynomial-fit Landau free energy}\noindent
Susceptibility can also be expressed in terms of the free-energy curvature as
$\chi^{-1}=\left.\tfrac{\partial^2 F}{\partial m^2}\right|_{m=m^\star}$,
where $m^\star$ is the equilibrium order parameter. Here we estimate $\chi$ from an empirical free-energy landscape reconstructed from the simulated distribution of the regional order parameter $m_d$.

For each $(M_w,\sigma)$, we form the empirical distribution $p(m)$ from a histogram of the damage fraction, mapped to $m\in[-1,1]$. Up to an additive constant, we define the empirical Landau free energy as
\[
F_{\mathrm{emp}}(m) = -\ln p(m)\,,
\]
where a small additive offset is applied to $p(m)$ to avoid $\ln 0$ in empty bins.

We then construct a polynomial-fit free energy $F_{\mathrm{poly}}(m)$ by fitting $F_{\mathrm{emp}}(m)$ with a polynomial Landau form,
\[
F_{\mathrm{poly}}(m)=c_0+c_1 m+c_2 m^2+c_4 m^4+c_k m^k\,,
\]
where $k\ge 6$ is an even integer. The linear term $c_1 m$ captures weak asymmetry in the empirical landscape, while the high-order stabilizing term $c_k m^k$ enforces boundedness near $|m|\to 1$ without materially affecting the fit around the minimum (i.e., the equilibrium). We impose $c_k>0$ and fix $k=100$, which yields stable fits and prevents spurious minima near the bounds (Supplementary Fig.~\ref{sup_fig:data_landau_landscape_fitting}).

From the fitted landscape, the equilibrium magnetization $m^\star$ is obtained by the stationary condition $F'_{\mathrm{poly}}(m^\star)=0$. The susceptibility is then estimated from the inverse curvature at the equilibrium magnetization,
\[
\chi^{-1}=\left.\frac{\partial^2 F_{\mathrm{poly}}}{\partial m^2}\right|_{m=m^\star}\,.
\]
This curvature-based definition is consistent with the mean-field linear-response relation $\chi=\partial \langle m\rangle/\partial h$, and provides a robust, quantitative measure of collective sensitivity from a finite ensemble. Supplementary Fig.~\ref{sup_fig:volatile_numerical_investigation} shows the estimated equilibrium order parameter $m^\star$ and susceptibility $\chi$ for three representative values of $\sigma$.

\newpage
\suppmethod{Spatial analysis of observed post-earthquake building damage}\label{sup_method:noto_analy}\noindent
To examine spatial damage organization in an observed event, we analyzed the building-level damage inventory compiled for the 2024 Noto Peninsula earthquake~\cite{vsecovo_2024NotoEq_paper, vescovo_2024NotoEq_dataset}. The dataset provides georeferenced building footprints, binary building-status labels (survived/destroyed), USGS Modified Mercalli Intensity (MMI), and municipality information. Buildings with valid geometries, binary damage labels, and MMI values were retained. The observed damage-state map is shown in Supplementary Fig.~\ref{sup_fig:noto_damage}a.

\subsubsection*{Building-level damage-probability model}\noindent
Let $y_i\in\{0,1\}$ denote the observed status of building~$i$, where $y_i=1$ represents a destroyed building. We first estimated each building's damage probability, $\tilde{p}_i$, by dividing the data into five subsets, fitting the logistic regression to four subsets, and predicting the held-out subset. Each building was therefore assigned an out-of-fold damage probability estimated from a model that excluded that building from training. The predictors included a cubic-spline representation of local USGS MMI, logarithmic building-footprint area, and municipality fixed effects:
\begin{equation}\label{eq:noto_damage_probability}
    \operatorname{logit}(\tilde{p}_i)
    =
    \beta_0
    +
    f_{\mathrm{MMI}}(\mathrm{MMI}_i)
    +
    \beta_A \ln A_i
    +
    \gamma_{G_i},
\end{equation}
where $\operatorname{logit}(p)=\ln[p/(1-p)]$, $\beta_0$ is the intercept, $f_{\mathrm{MMI}}(\cdot)$ is a cubic-spline function representing the nonlinear effect of USGS MMI, $A_i$ is the building-footprint area, and $\gamma_{G_i}$ is the fixed effect associated with the municipality $G_i$ containing building~$i$. The out-of-fold probabilities were adjusted by a common intercept shift to obtain the final probabilities, $\hat{p}_i$, such that their regional mean equaled the observed destroyed fraction. Agreement between the observed and estimated destroyed fractions across MMI intervals is shown in Supplementary Fig.~\ref{sup_fig:noto_damage}b.

\subsubsection*{Damage-state and residual spatial correlations}\noindent
We first calculated the distance-dependent correlation of the observed binary damage states,
\begin{equation}\label{eq:noto_damage_correlation}
    C_y(r) 
    =
    \frac{
    \left\langle
    (y_i-\bar{y})(y_j-\bar{y})
    \right\rangle_r
    }{
    \bar{y}(1-\bar{y})
    },
\end{equation}
where $\langle\cdot\rangle_r$ denotes averaging over building pairs whose separation falls within the distance bin centered at $r$, and $\bar{y}=N^{-1}\sum_{i=1}^{N}y_i$ is the regional mean observed damage state (i.e., the observed destroyed fraction) with $N$ indicating the number of buildings. This damage-state correlation contains both local clustering and correlation induced by broad spatial variation in the marginal damage probability.

To distinguish these contributions, we defined the building-level damage residual as
\begin{equation}\label{eq:noto_damage_residual}
    e_i=y_i-\hat{p}_i
\end{equation}
and calculated the corresponding residual correlation,
\begin{equation}\label{eq:noto_residual_correlation}
    C_e(r)
    =
    \frac{
    \left\langle
    (e_i-\bar{e})(e_j-\bar{e})
    \right\rangle_r
    }{
    N^{-1}\sum_{i=1}^{N}\hat{p}_i(1-\hat{p}_i)
    }.
\end{equation}
The residual correlation measures spatial co-variation remaining after accounting for the available hazard and inventory covariates. It therefore provides a single-event analogue of the connected correlation used in the regional simulations (Supplementary Method~\ref{sup_method:ensemble_based_calc}).

Both $C_y(r)$ and $C_e(r)$ were evaluated using the same sampled building pairs in logarithmically spaced distance bins. Their positive short-range portions were fitted using
\begin{equation}\label{eq:noto_exponential_correlation}
    C(r)=A\exp(-r/\xi),
\end{equation}
where $\xi$ is interpreted as a descriptive correlation length. Supplementary Figs.~\ref{sup_fig:noto_damage}c and \ref{sup_fig:noto_damage}d compare the damage-state and residual correlations, respectively.

\subsubsection*{Conditional randomization}\noindent
We constructed a conditional reference distribution by repeatedly permuting the observed damage labels within groups of buildings belonging to the same municipality and having similar fitted damage probabilities. These randomizations preserve the group-specific number of destroyed buildings and the broad marginal damage pattern while disrupting their fine-scale spatial arrangement. The observed correlation functions were compared with the median and 95\% pointwise envelope obtained from 199 conditional randomizations. Comparison of $C_y(r)$ and $C_e(r)$ distinguishes spatial correlation associated with the fitted marginal damage pattern from spatial coherence remaining after conditioning on the available covariates.

\subsubsection*{Destroyed-building nearest-neighbor clustering}\noindent
Local clustering was assessed independently using the nearest-neighbor distance among destroyed buildings. For each destroyed building~$i$, we calculated
\begin{equation}\label{eq:noto_nearest_neighbor}
    D_{\mathrm{NN},i}
    =
    \min_{\substack{j\neq i\\y_j=1}}
    \left\lVert
    \vect{x}_i-\vect{x}_j
    \right\rVert,    
\end{equation}
where $\vect{x}_i$ denotes the spatial location of building~$i$. The empirical cumulative distribution of $D_{\mathrm{NN}}$ was compared with those obtained from the same conditional randomizations. A shift toward shorter nearest-neighbor distances indicates that destroyed buildings are more tightly clustered than expected from the fitted marginal damage pattern alone (Supplementary Fig.~\ref{sup_fig:noto_damage}e).

This analysis characterizes event-conditioned spatial dependence and clustering within a single observed damage realization. It does not directly assess ensemble bimodality or regional loss tails. Moreover, the damage classification used in this dataset differs from the slight-damage threshold used in the regional simulations; comparisons of their correlation lengths are therefore interpreted as order-of-magnitude consistency rather than damage-state-specific calibration.

\newpage
\suppmethod{Simulation framework for regional-scale structural responses}\label{sup_method:sim_framework}\noindent
For a region consisting of $n$ structures, consider a safety factor vector $\vect{X} \in \mathbb{R}^n$, where each element $X_i$ denotes the safety factor of structure $i$. The damage state of each structure is determined based on its safety factor. For instance, if a binary damage state is of interest, the damage state $\mathrm{DS}_i$ of structure $i$ is defined as
\[
\mathrm{DS}_i =
\begin{cases}
    1, & \text{if } X_i \leq 0\,, \\
    0, & \text{if } X_i > 0\,,
\end{cases}
\]
where $\mathrm{DS}_i$ indicates whether structure $i$ has damaged ($\mathrm{DS}_i = 1$) or not ($\mathrm{DS}_i = 0$). To capture the uncertainties inherent in hazard events and structural responses, each safety factor $X_i$ is modeled as a random variable. The failure probability of structure $i$, denoted by $P_{f_i}$, is then given by
\[
P_{f_i} = \Prob{X_i \leq 0}\,,
\]
which depends on the probability distribution of $X_i$. Since our interest lies in the collective behavior of multiple structures, the \textit{joint} probability distribution of the vector $\vect{X}$ becomes central. In other words, regional-scale risk analyses and simulations focus on characterizing the joint distribution of $\vect{X}$, including its mean vector, covariance matrix, and distribution type. This perspective encompasses both stochastic and physics-based approaches, where the resulting distribution may not belong to any standard family of probability distributions. In such cases, the distribution can be developed through repeated simulations under varying sources of uncertainty.

In structural engineering practice, the first- and second-order central and cross moments of $\vect{X}$ are often available; the reliability of individual structures typically needs to be assessed during the design phase, which requires the mean and variance of the safety factor. These moments can be obtained using methods ranging from full-scale experiments to high-fidelity finite element analyses with accurate physical parameters, while inter-building correlation has only recently begun to receive attention~\cite{xiang_structure--structure_2024, you_spatial_2024, zhong_regional_2024, Kang2021EvaluationAnalysis, heresi_structure--structure_2022}. \citet{oh_longrange_2024} presented an unbiased maximum-entropy approach for constructing a joint probability distribution of $\vect{X}$ when the first- and second-order central moments are known.

\newpage
\suppmethod{Implementation of regional-scale structural simulations}\label{sup_method:sim_implem}
\subsection*{Regional data retrieval}\noindent 
All data used in this study were obtained from publicly accessible repositories and open-source libraries to ensure reproducibility. The datasets encompass building inventories, ground-motion records, and administrative boundaries, which together form the basis of the regional simulations.

Building-level attributes were retrieved from the \textit{NHERI SimCenter Earthquake Testbed for the San Francisco Bay Area}~\cite{mckenna_nheri-simcenterr2dtool_2025}. The database provides detailed information on the number of stories, construction year, structural type, occupancy class, plan area, repair cost, and replacement cost for more than 200,000 buildings in the region. Two subsets were used in this study: (1) the city of Milpitas, including 5,943 multi-story buildings, and (2) the northeastern districts of San Francisco, comprising 6,323 buildings across mixed structural types. The data were accessed through the DesignSafe DataDepot (PRJ-3880~\cite{zsarnoczay_simcenter_2023}) and formatted as geospatial layers to enable spatial mapping and structural model generation.

A total of 100 empirical ground-motion acceleration records were obtained from the \textit{Enhancement of Next Generation Attenuation Relationships for Western US} (NGA-West2) database~\cite{ancheta_nga-west2_2014, bozorgnia_nga-west2_2014}. The NGA‐West2 database provides a comprehensive collection of uniformly processed strong‐motion records from shallow‐crustal earthquakes in active tectonic regions. The retrieved 100 time‐history records were used to perform incremental dynamic analyses for estimating structural capacities. These motions represent crustal earthquakes recorded in California and were scaled to varying intensity levels following the procedure described in the Methods.

City boundaries and building footprints were generated using the open-source \textit{OSMnx} library~\cite{boeing_modeling_2025} with base map data from OpenStreetMap~\cite{OpenStreetMap}. Neighborhood boundaries for San Francisco were obtained from the \textit{Analysis Neighborhoods} dataset on the San Francisco Open Data Portal~\cite{sfgov_neighborhoods}. All retrieved polygons were projected to the EPSG:3857 coordinate system and used to clip and visualize the regional building portfolios.

\subsection*{Structural capacity modeling}\noindent
We adopt the peak inter‐story drift ratio (IDR) as the engineering demand parameter (EDP). For each structural type and design era, the \textit{Hazus Earthquake Model Technical Manual} (Version 6.1; FEMA, 2024) defines threshold drift ratios at which a building is considered to have reached the slight‐damage state~\cite{federal_emergency_management_agency_fema_hazus_2024}. Building capacity is defined as the ground‐motion intensity measure (IM) at which the model first reaches this slight‐damage EDP under record scaling. The peak ground acceleration (PGA) is adopted as the intensity measure, expressed in units of $g$, the gravitational acceleration.

Following the baseline approach of \citet{lu_open-source_2020}, as adopted in the SimCenter R2D Tool~\cite{mckenna_nheri-simcenterr2dtool_2025}, we developed a multi–degree‐of‐freedom (MDOF) shear model for each building using the Hazus~6.1 tabulations~\cite{federal_emergency_management_agency_fema_hazus_2024}, characterized by fundamental building attributes such as the number of stories, structural material, construction year, and plan area. Each model is a planar system with one lateral degree of freedom per story (rigid diaphragm action at floor levels). Story response is represented by zero-length elements with a uniaxial hysteretic material capturing yielding, cyclic degradation, and pinching. Models are implemented in OpenSeesPy~\cite{mckenna_opensees_2011,zhu_openseespy_2018}.

Incremental dynamic analyses (IDA)~\cite{vamvatsikos_incremental_2002} were performed with 100 empirical NGA-West2 ground-motion records. Each record was scaled until the structure reached its slight-damage drift-ratio limit, producing one threshold IM (here, PGA) per record. The resulting 100 threshold PGAs per building constitute a capacity sample in terms of PGA, from which the building-level fragility curve is derived. The IDA-derived capacity samples were converted into probabilistic models at the building level to propagate uncertainty in regional simulations. 

In earthquake engineering practice, building-level fragility curves are most commonly modeled as lognormal, and this convention is adopted widely in guidelines and applications (e.g., Hazus~6.1~\cite{federal_emergency_management_agency_fema_hazus_2024}; FEMA P-58~\cite{fema_p58_1_2018}). Accordingly, for each building $i$ with structural capacity $C_i$, we model $\ln C_i \sim \mathcal{N}(\mu_i,\sigma_i^2)$ with $(\mu_i,\sigma_i)$ calibrated from the 100 samples. The resulting fragility (probability of damage at a given $\mathrm{IM}$) is
\[
\Prob{\text{Damage}\ge\text{Slight}~|~\mathrm{IM}} \;=\; \Phi\!\left(\frac{\ln \mathrm{IM}-\mu_i}{\sigma_i}\right),
\]
where $\Phi$ is the standard normal cumulative distribution function (CDF).

\begin{figure}[H]
    \centering
    \includegraphics[width=0.7\linewidth]{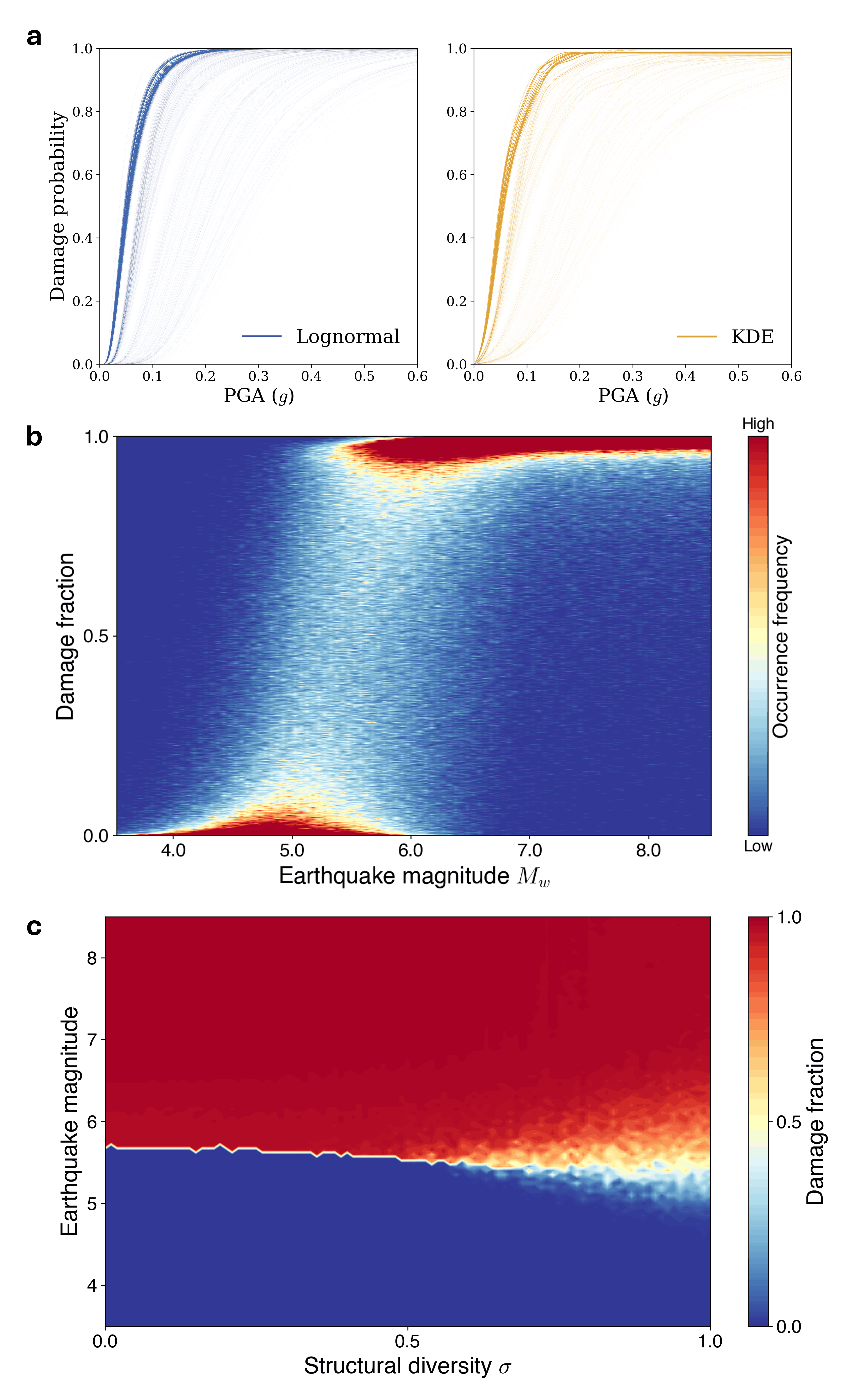}
    \caption{\textbf{Sensitivity analysis: kernel density estimation (KDE)-based fragility curves.}
    \textbf{a}, Comparison of fragility curves for the Milpitas portfolio constructed using the conventional lognormal form (left) and the KDE-based non-parametric form (right).
    \textbf{b}, Evolution of regional damage-fraction distributions with increasing earthquake magnitude when the KDE-based fragility curves are used.
    \textbf{c}, Empirical phase diagram of the Milpitas portfolio obtained using the KDE-based fragility curves. It preserves the key features of the phase diagram shown in Fig.~3a of the main text, which is constructed using lognormal fragility curves.}
    \addcontentsline{sitoc}{subsection}{Supplementary Fig.~\arabic{figure}. Sensitivity analysis: kernel density estimation (KDE)-based fragility curves}
    \label{sup_fig:sensitivity_frag_kde}
\end{figure}

To evaluate the impact of distributional assumptions, we also construct building-level fragility curves directly from the capacity samples using Gaussian kernel density estimation (KDE) on $\ln C$, without imposing a parametric form. The KDE-based density is numerically integrated to obtain a CDF, which represents the individual fragility curves for buildings~\cite{cao_kde-based_2023, mai_seismic_2017}. Supplementary Figs.~\ref{sup_fig:sensitivity_frag_kde} present the resulting fragility curves and the corresponding regional simulation outcomes.

We preserve inter-building dependence in structural capacities by estimating a correlation matrix from the aligned sets of 100 IDA-derived capacity samples and using this matrix when sampling capacities in the regional analyses. Preserving the record-wise alignment of the capacity samples across buildings retains co-variation induced by record-specific waveform characteristics, while similarities among structural models also contribute to the estimated dependence. Given the relatively compact geographic extent of the portfolios considered in this study, this serves as an effective representation of common record-level influences on structural response within the region, whereas location-specific variability in scenario PGA is modeled separately through the spatially correlated demand field. We therefore interpret the estimated matrix as an effective capacity-dependence model that may combine shared record-level effects with similarities among structural models, rather than as a uniquely identified matrix of intrinsic pairwise capacity correlations. Additional analyses of the sampling stability and numerical treatment of this dependence model, together with sensitivity analyses using alternative dependence representations, are provided in Supplementary Note~\ref{sup_note:capa_dependence}. For the KDE-based fragility curves, we use a Gaussian copula with the same latent Gaussian correlation matrix to impose a consistent dependence structure on the non-parametric marginals, thereby maintaining comparable inter-building dependence in the parametric and non-parametric formulations.

\subsection*{Hazard demand modeling}\noindent
We adopt the peak ground acceleration (PGA) as the ground‐motion intensity measure at each building location, representing the maximum horizontal acceleration experienced during an earthquake. For each earthquake scenario, spatial PGA fields are computed using empirical ground‐motion prediction equations (GMPEs) based on predefined earthquake scenarios that are physically plausible for the study regions. To ensure both statistical realism and physical consistency with regional seismotectonic conditions, physics‐based ground‐motion simulations are also performed for selected scenarios to cross‐validate the PGA fields derived from the GMPEs.

The earthquake scenarios were defined by varying the moment magnitude $M_w$ from 3.5 to 8.5 in increments of 0.05, in which reliability of the adopted GMPEs is ensured, covering the sufficient range from small to large regional events. All scenarios share a common epicenter located at latitude 37.666° N and longitude 122.076° W, approximately along the central segment of the Hayward fault, which poses one of the most significant seismic threats to the San Francisco Bay Area. Since the Hayward fault is a right-lateral strike-slip fault, the fault geometry is defined by a strike of 325°, dip of 90°, and rake of 180°, representing a near‐vertical, right‐lateral strike‐slip mechanism typical of the Hayward fault system. The depth to the top of the rupture plane is set to 3 km, consistent with empirical observations of shallow crustal earthquakes in California. These configurations ensure that the generated ground‐motion fields are both realistic and representative of the dominant regional seismotectonic conditions.

For each scenario, we compute spatial fields of the logarithmic PGA mean and standard deviation at every building location using the \textit{Enhancement of Next Generation Attenuation Relationships for Western U.S.} (NGA-West2) GMPE of~\citet{chiou_update_2014} (CY14). The selected GMPE is implemented through the \texttt{pygmm} Python library~\cite{kottke_arkottkepygmm_2025}. The rupture geometry defined above is employed, while the $V_{s30}$ value at each location is retrieved from~\citet{thompson_updated_2018} (Supplementary Fig.~\ref{sup_fig:vs30_milpitas_sf}) and the rupture length $L$ and width $W$ are estimated from the moment magnitude $M_w$ using the empirical regressions of ~\citet{wells_new_1994}. Spatial correlation of intra-event residuals was modeled following~\citet{Jayaram2009CorrelationIntensities} with an exponential decay function, ensuring realistic spatial variability in ground‐motion intensity. 

\begin{figure}[H]
    \centering
    \includegraphics[width=1.0\linewidth]{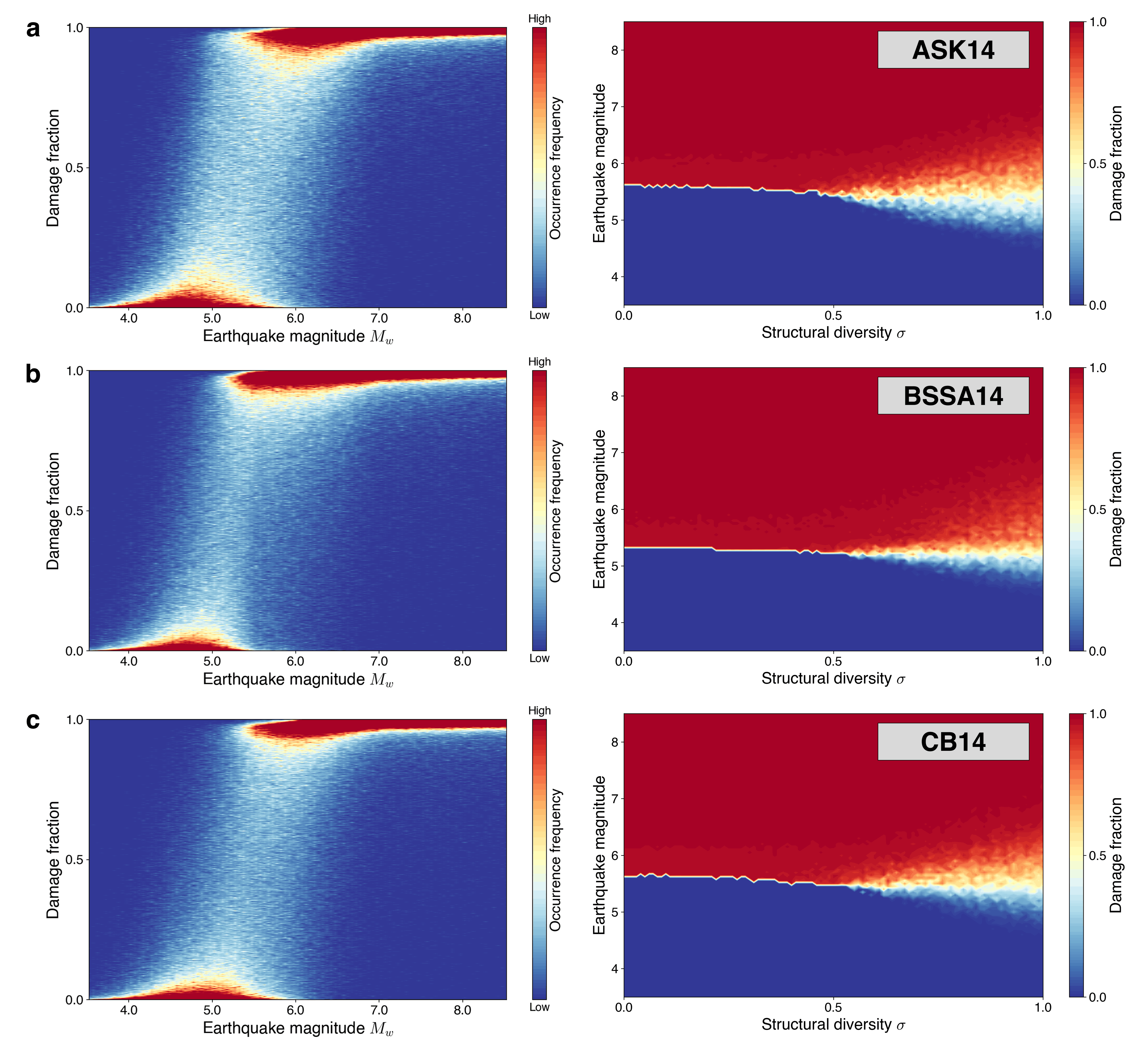}
    \caption{\textbf{Sensitivity analysis: empirical ground‐motion prediction equations (GMPEs).}
    The benchmark Milpitas portfolio is evaluated using three NGA‐West2 GMPEs. For each model, the evolution of regional damage‐fraction distributions with earthquake magnitude (left) and the corresponding empirical phase diagram (right) are shown.
    \textbf{a}, Abrahamson–Silva–Kamai 2014 (ASK14)~\cite{abrahamson_summary_2014}.
    \textbf{b}, Boore–Stewart–Seyhan–Atkinson 2014 (BSSA14)~\cite{boore_nga-west2_2014}.
    \textbf{c}, Campbell–Bozorgnia 2014 (CB14)~\cite{campbell_nga-west2_2014}.
    All three preserve the key features of the phase diagram in Fig.~3a of the main text, which is constructed using the Chiou–Youngs 2014 (CY14) model~\cite{chiou_update_2014}.}
    \addcontentsline{sitoc}{subsection}{Supplementary Fig.~\arabic{figure}. Sensitivity analysis: empirical ground‐motion prediction equations (GMPEs)}
    \label{sup_fig:sensitivity_gmpes}
\end{figure}

Different empirical GMPEs are applied to the benchmark Milpitas example, including the Abrahamson–Silva–Kamai 2014 (ASK14) GMPE~\cite{abrahamson_summary_2014}, the Boore–Stewart–Seyhan–Atkinson 2014 (BSSA14) GMPE~\cite{boore_nga-west2_2014}, and the Campbell–Bozorgnia 2014 (CB14) GMPE~\cite{campbell_nga-west2_2014}, all of which are suitable for active tectonic regions such as California. Supplementary Figs.~\ref{sup_fig:sensitivity_gmpes} present the sensitivity analysis comparing the regional simulation outcomes obtained using these different GMPEs.

To complement the empirical models, we employ the \textit{Southern California Earthquake Center Broadband Platform} (SCEC-BBP) to simulate three-dimensional broadband ground motions. For the same earthquake scenarios and input parameters used in the GMPE analyses, time histories are generated on an $8\time8$ grid (64 recording stations) within the study region, and the corresponding PGA values are extracted. These simulations capture long-period basin effects and rupture directivity patterns that empirical models cannot represent. The simulated and GMPE-based PGA maps are cross-validated to confirm the realism of the generated ground-motion intensity fields, which are presented in Extended Data Fig.~10 and Supplementary Figs.~\ref{sup_fig:pga_comparison_per_station_Mw5} to~\ref{sup_fig:pga_comparison_per_station_Mw8}.

\newpage
\SupplFiguresHeading
\renewcommand{\figurename}{Supplementary Fig.}

\newpage

\newpage
\begin{figure}[H]
    \centering
    \includegraphics[width=1.0\linewidth]{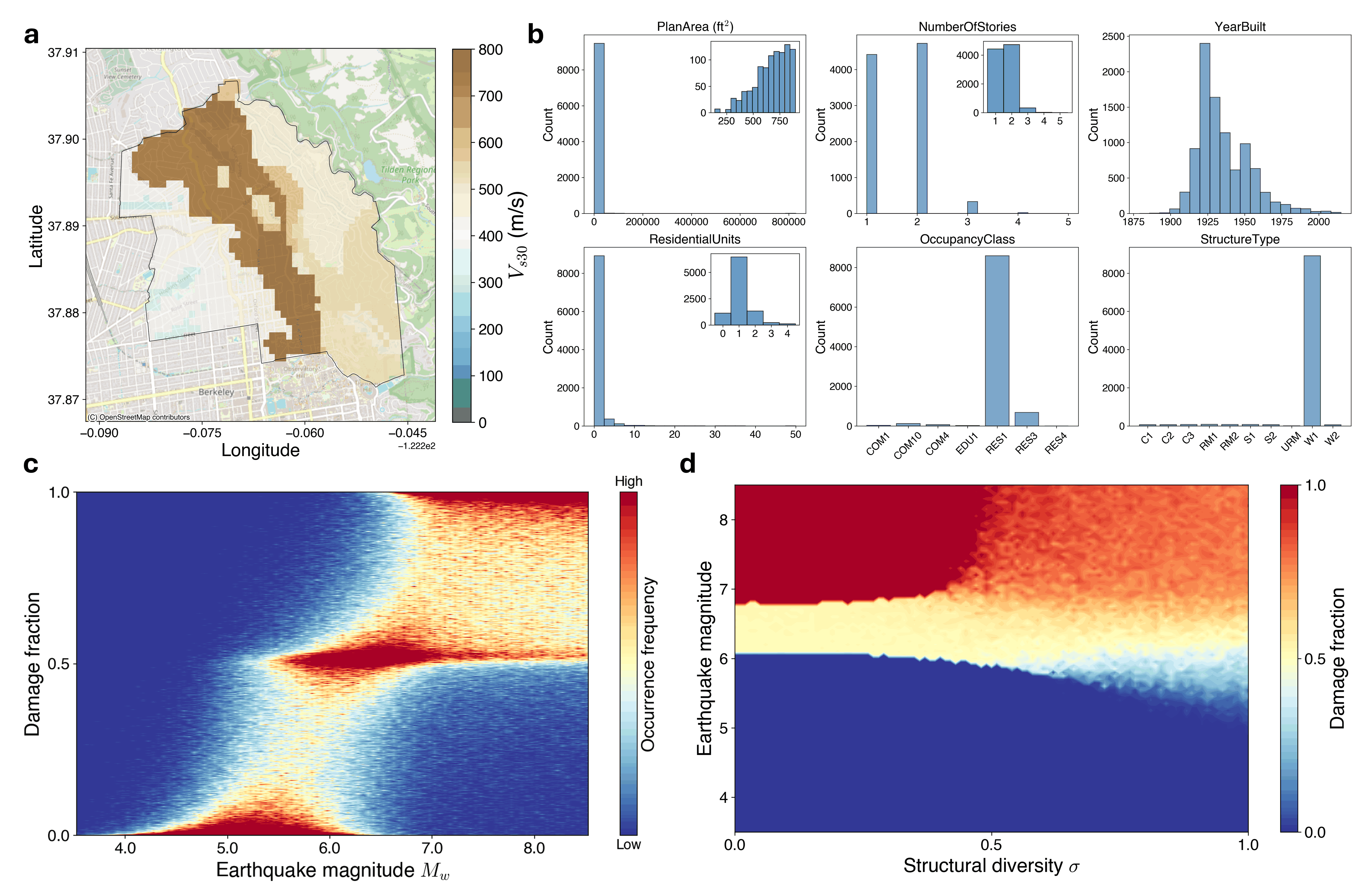}
    \caption{\textbf{Sensitivity analysis: the North Berkeley building portfolio.}
    \textbf{a}, Map of the near-surface shear-wave velocity $V_{s30}$ across the North Berkeley region.
    \textbf{b}, Distributions of key building attributes, including plan area, number of stories, construction year, residential units, occupancy class, and structural type.
    \textbf{c}, Evolution of regional damage-fraction distributions with increasing earthquake magnitude for the North Berkeley portfolio.
    \textbf{d}, Empirical phase diagram showing the collective structural response as a function of structural diversity $\sigma$ and earthquake magnitude.
    Since the numbers of single-story and multi-story buildings are comparable, while other factors that strongly influence seismic response are relatively uniform across the region, the North Berkeley portfolio exhibits three collective states with two transition boundaries.}
    \addcontentsline{sitoc}{subsection}{Supplementary Fig.~\arabic{figure}. Sensitivity analysis: the North Berkeley building portfolio}
    \label{sup_fig:sensitivity_city_northberkeley}
\end{figure}

\newpage
\begin{figure}[H]
    \centering
    \includegraphics[width=1.0\linewidth]{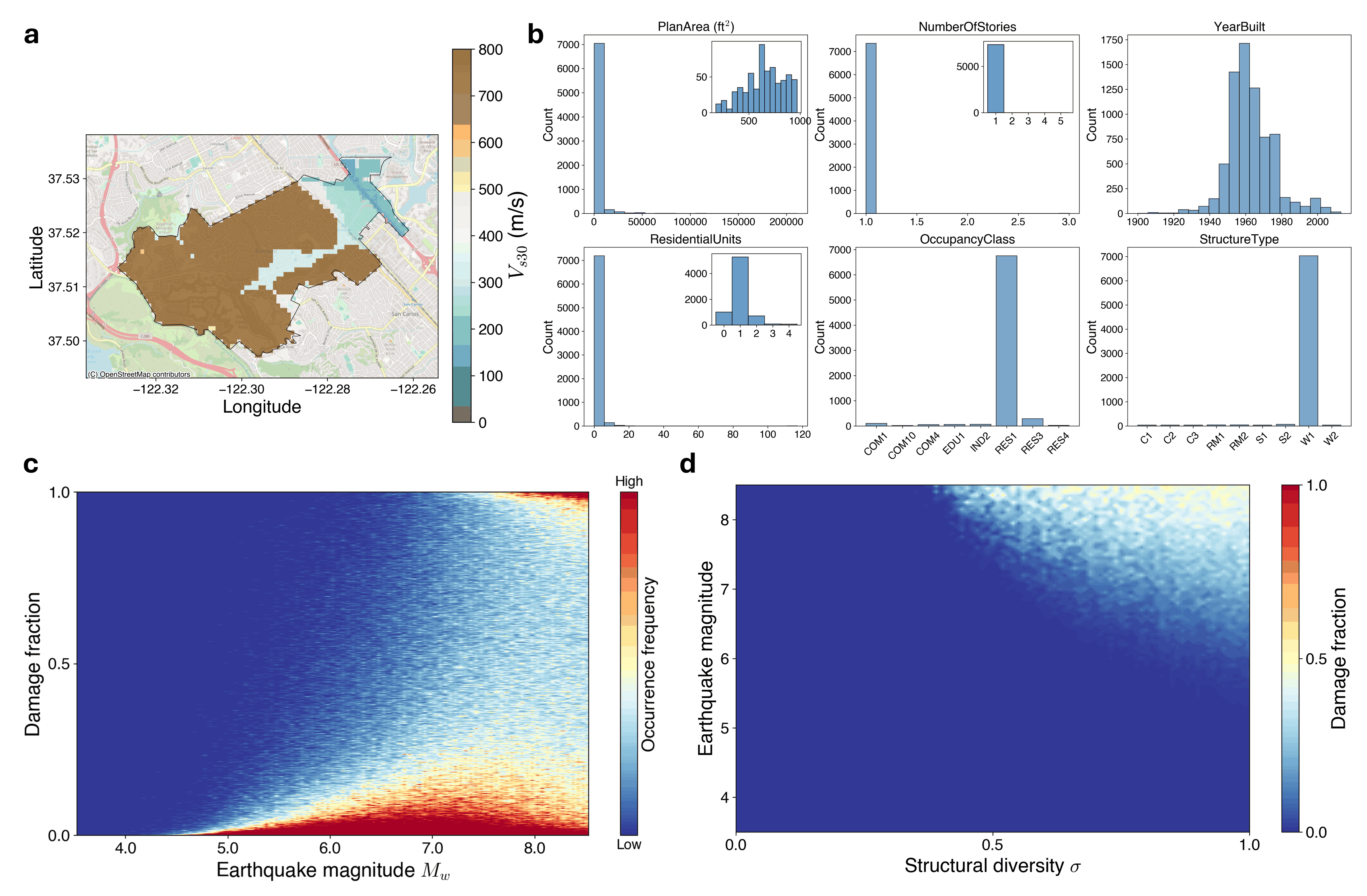}
    \caption{\textbf{Sensitivity analysis: the Belmont building portfolio.}
    \textbf{a}, Map of the near-surface shear-wave velocity $V_{s30}$ across the Belmont region.
    \textbf{b}, Distributions of key building attributes, including plan area, number of stories, construction year, residential units, occupancy class, and structural type.
    \textbf{c}, Evolution of regional damage-fraction distributions with increasing earthquake magnitude for the Belmont portfolio.
    \textbf{d}, Empirical phase diagram showing the collective structural response as a function of structural diversity $\sigma$ and earthquake magnitude.
    Due to the single-story–dominant building portfolio and the high $V_{s30}$ values across the region, Belmont shows little transition behavior and remains predominantly in the collectively safe state throughout the examined parameter range.}
    \addcontentsline{sitoc}{subsection}{Supplementary Fig.~\arabic{figure}. Sensitivity analysis: the Belmont building portfolio}
    \label{sup_fig:sensitivity_city_Belmont}
\end{figure}

\newpage
\begin{figure}[H]
    \centering
    \includegraphics[width=1.0\linewidth]{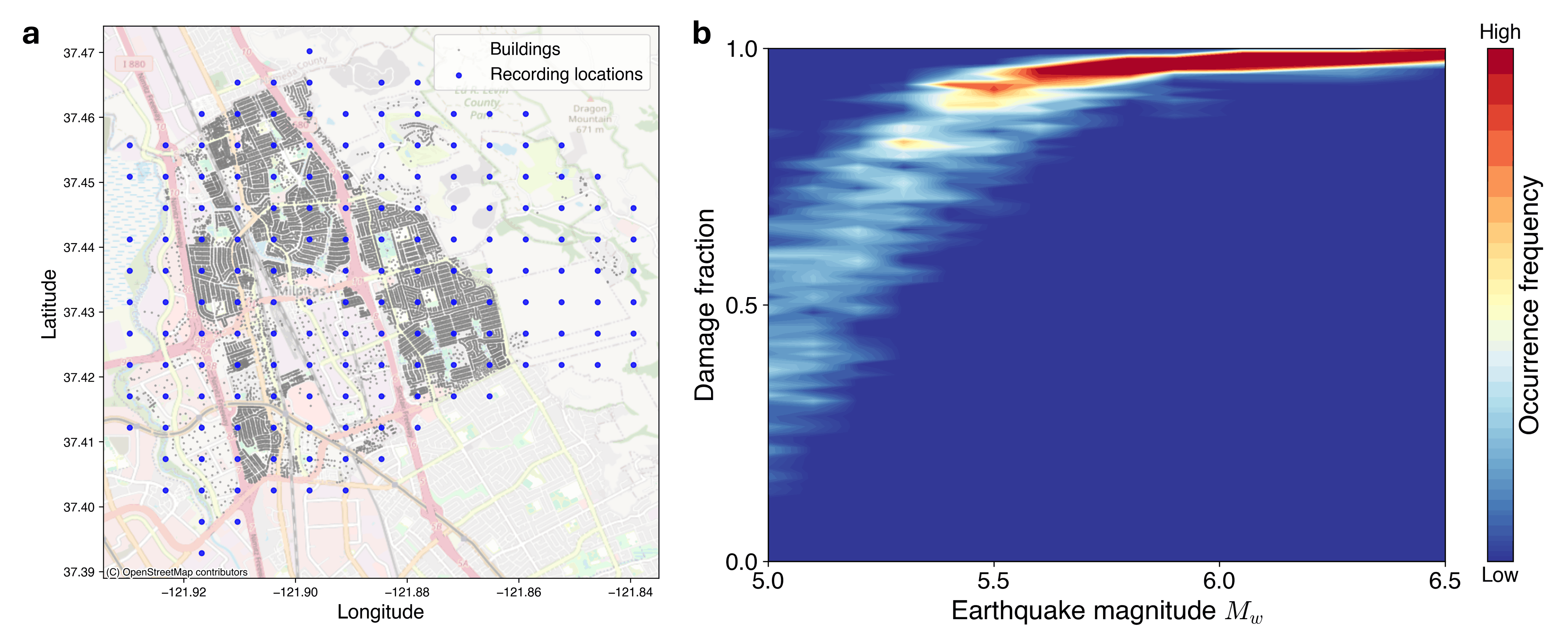}
    \caption{\textbf{Sensitivity analysis: physics-based regional damage simulation.}
    \textbf{a,} Spatial layout of the benchmark Milpitas portfolio and the prescribed SCEC Broadband Platform (SCEC-BBP) recording grid. Gray points denote analyzed buildings, and blue points denote synthetic recording locations. \textbf{b,} Occurrence-frequency heatmap of the regional damage fraction obtained from the physics-based workflow. For each earthquake magnitude, 100 SCEC-BBP ground-motion realizations are generated at 168 prescribed recording locations. Following the baseline approach~\cite{mckenna_nheri-simcenterr2dtool_2025}, each building is assigned one acceleration time history sampled from one of its four nearest recording locations, and nonlinear time-history analysis is performed using the OpenSeesPy-based multi-degree-of-freedom structural model to evaluate building-level damage. Since the physics-based scenarios begin at $M_w=5.0$, the low-magnitude collectively safe state is not sampled enough. Compared with the corresponding fragility curve- and GMPE-based result for the same Milpitas portfolio in the main text, the bistable phase-coexistence band near $M_w \approx 5.2$--$5.8$ is substantially weakened. This difference may arise because the SCEC-BBP simulations explicitly resolve source-, path-, site-, and station-dependent spatial variations in seismic demand that are represented more coarsely, or through empirical median trends and stochastic residual fields, in the GMPE-based workflow. In the RFIM-style interpretation, this explicitly represented spatial variability in the capacity--demand imbalance can act like a larger effective disorder, qualitatively similar to increasing $\sigma$, and can therefore smooth the synchronized transition. The comparison is qualitative: no calibrated $\sigma$ is assigned to the physics-based workflow, and the result does not imply uniformly lower total uncertainty. Instead, it demonstrates that different engineering representations can diagnose the same portfolio as distinct collective-response regimes.}
    \addcontentsline{sitoc}{subsection}{Supplementary Fig.~\arabic{figure}. Sensitivity analysis: physics-based regional damage simulation}
    \label{sup_fig:sensitivity_phy_Milpitas}
\end{figure}

\newpage
\begin{figure}[H]
    \centering
    \includegraphics[width=1.0\linewidth]{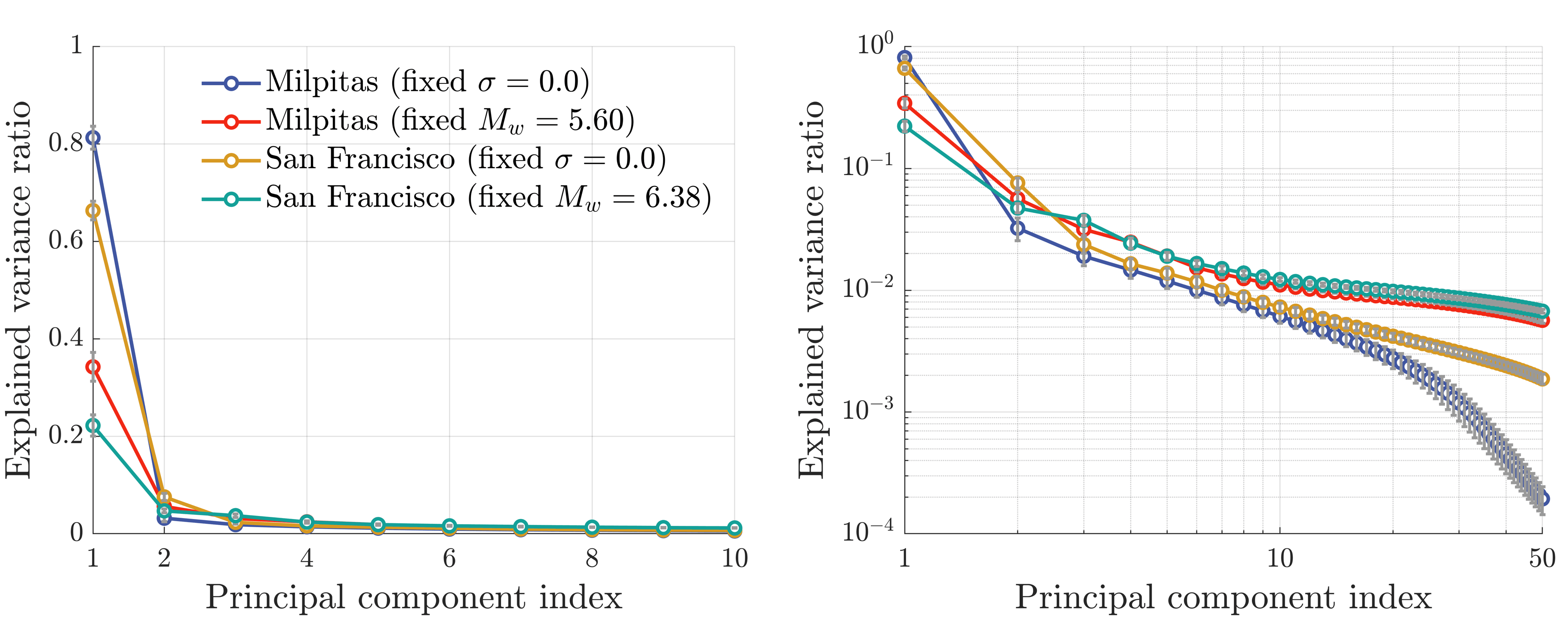}
    \caption{\textbf{Explained-variance spectra from principal component analysis (PCA).}
    Explained variance ratios of successive principal components are shown on linear (left) and log–log (right) scales for four analysis cases: Milpitas and northeastern San Francisco under first-order (fixed $\sigma = 0$) and second-order (fixed $M_w = M_{w_c}$) conditions. The first principal component dominates the variance, particularly for the first-order scenarios, indicating that a single linear mode accounts for a large fraction of the ensemble variance. For the second-order scenarios, the variance is distributed across multiple components, consistent with the gradual and correlated reorganization of the system as structural diversity increases. The overall lower explained variance in the San Francisco cases reflects weaker inter-building dependence and a more Gaussian-like distribution of regional damage fractions.}
    \addcontentsline{sitoc}{subsection}{Supplementary Fig.~\arabic{figure}. Explained-variance spectra from principal component analysis (PCA)}
    \label{sup_fig:pca_explained_var}
\end{figure}

\newpage
\begin{figure}[H]
    \centering
    \includegraphics[width=1.0\linewidth]{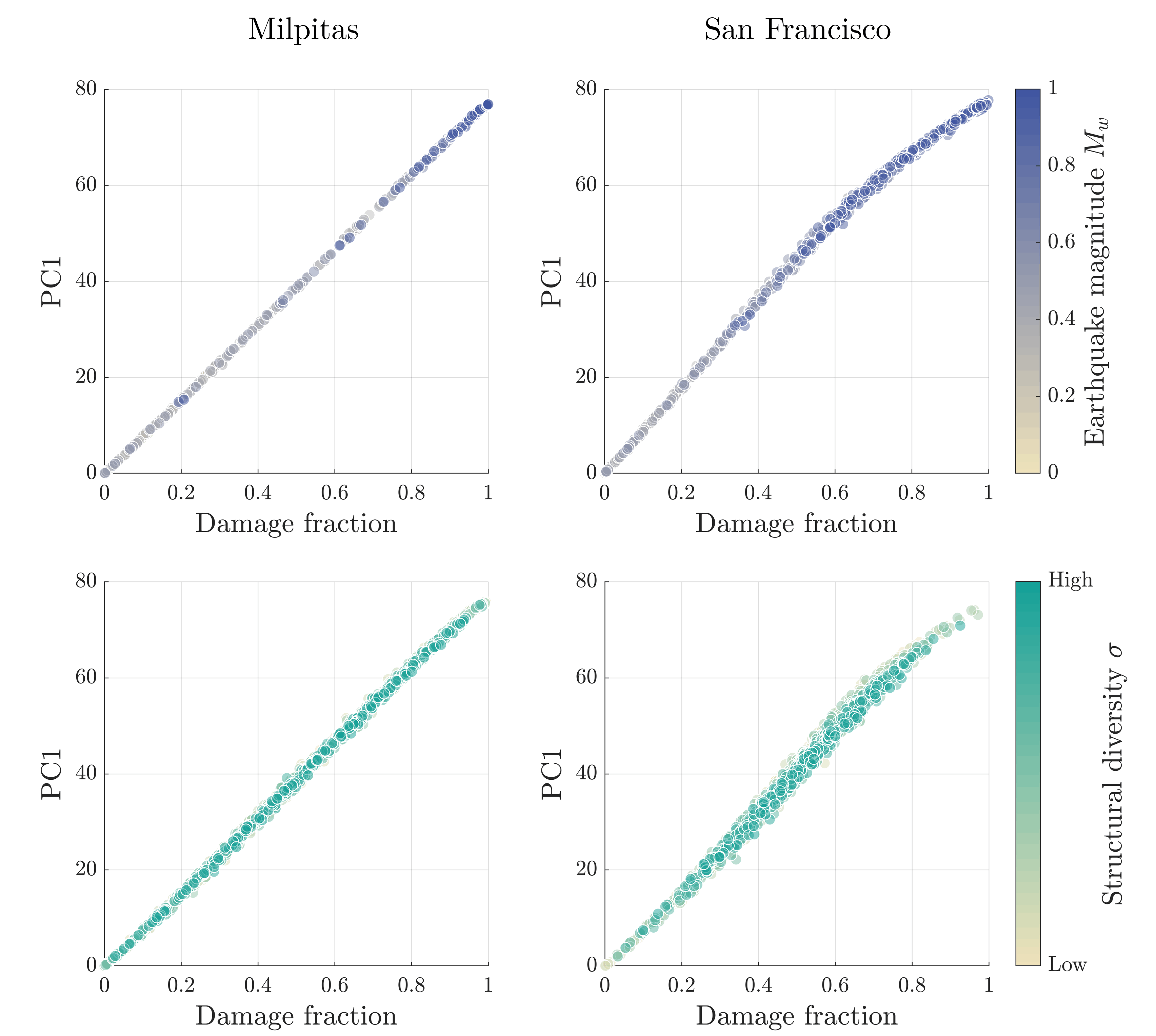}
    \caption{\textbf{Alignment of the first principal component with the regional damage fraction.}
    The first principal component (PC1) obtained from principal component analysis (PCA) is plotted against the regional damage fraction for two study regions (Milpitas and northeastern San Francisco) under two simulation scenarios. The top row shows the first-order scenario, where the earthquake magnitude $M_w$ is varied at fixed structural diversity $\sigma=0$, and the bottom row shows the second-order scenario, where $\sigma$ is varied at fixed $M_w=M_{w_c}$. Ten realizations are shown for each control parameter ($M_w$ or $\sigma$), where each point represents a single simulation realization, colored by the corresponding control parameter. In all cases, PC1 exhibits a near one-to-one correspondence with the damage fraction, demonstrating that the dominant mode identified purely from data coincides with the physically interpretable order parameter of the system.}
    \addcontentsline{sitoc}{subsection}{Supplementary Fig.~\arabic{figure}. Alignment of the first principal component with the regional damage fraction}
    \label{sup_fig:pca_order_param_validation}
\end{figure}

\newpage
\begin{figure}[H]
    \centering
    \includegraphics[width=1.0\linewidth]{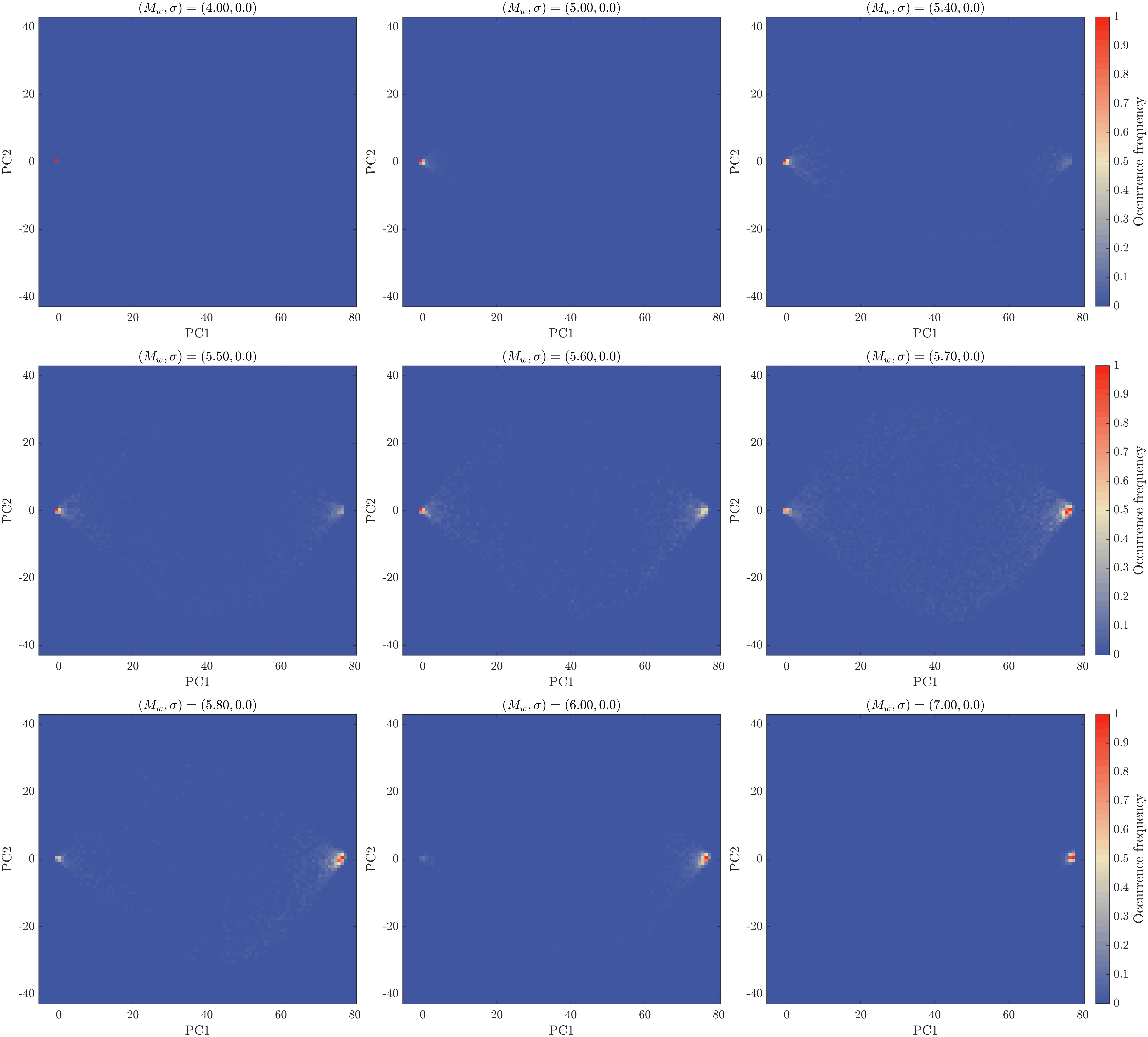}
    \caption{\textbf{Evolution of the projected PCA space with increasing earthquake magnitude for Milpitas.}
    Simulation realizations are projected into the space spanned by the first and second principal components (PC1 and PC2) at fixed structural diversity $\sigma = 0.0$ as the earthquake magnitude $M_w$ increases. Each panel corresponds to a different $(M_w, \sigma)$ condition, with color indicating the normalized occurrence frequency from 5,000 realizations.
    The distribution exhibits an abrupt shift from the left to the right cluster in PC1 around $M_w \approx 5.60$, marking a sudden transition from the collectively safe to the collectively damaged state, characteristic of a first-order-like phase transition.}
    \addcontentsline{sitoc}{subsection}{Supplementary Fig.~\arabic{figure}. Evolution of the projected PCA space with increasing earthquake magnitude for Milpitas}
    \label{sup_fig:pca_1st_Milpitas}
\end{figure}

\newpage
\begin{figure}[H]
    \centering
    \includegraphics[width=1.0\linewidth]{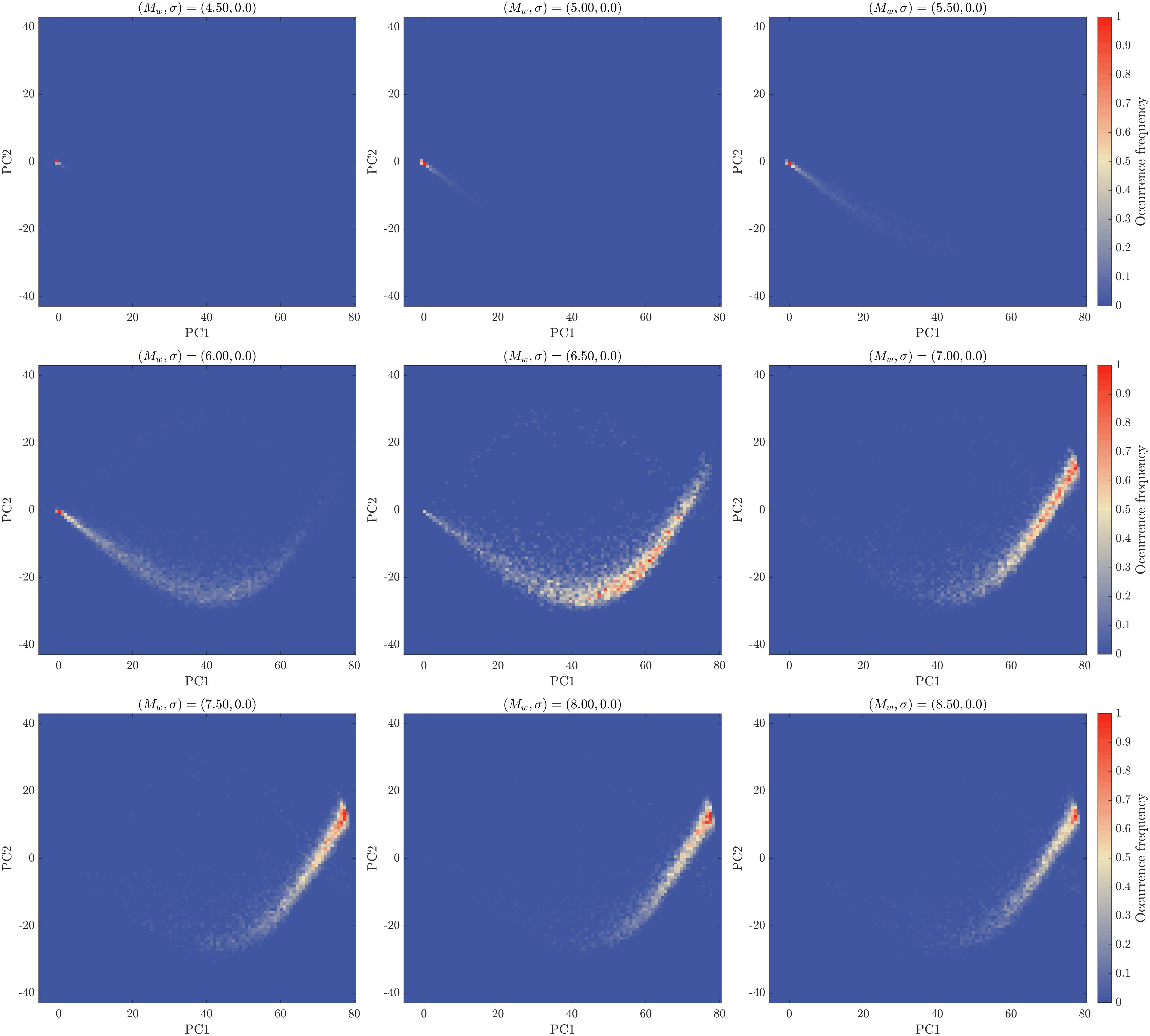}
    \caption{\textbf{Evolution of the projected PCA space with increasing earthquake magnitude for northeastern San Francisco.}
    Same analysis setup as in Supplementary Fig.~\ref{sup_fig:pca_1st_Milpitas}, performed for San Francisco at fixed structural diversity $\sigma = 0.0$. As $M_w$ increases, the distribution shifts gradually and forms a smooth curved trajectory along PC1–PC2 without an abrupt jump, indicating the absence of a sharp first-order transition. This continuous evolution reflects weaker inter-building coupling and higher intrinsic diversity in the San Francisco region.}
    \addcontentsline{sitoc}{subsection}{Supplementary Fig.~\arabic{figure}. Evolution of the projected PCA space with increasing earthquake magnitude for northeastern San Francisco}
    \label{sup_fig:pca_1st_SanFrancisco}
\end{figure}

\newpage
\begin{figure}[H]
    \centering
    \includegraphics[width=1.0\linewidth]{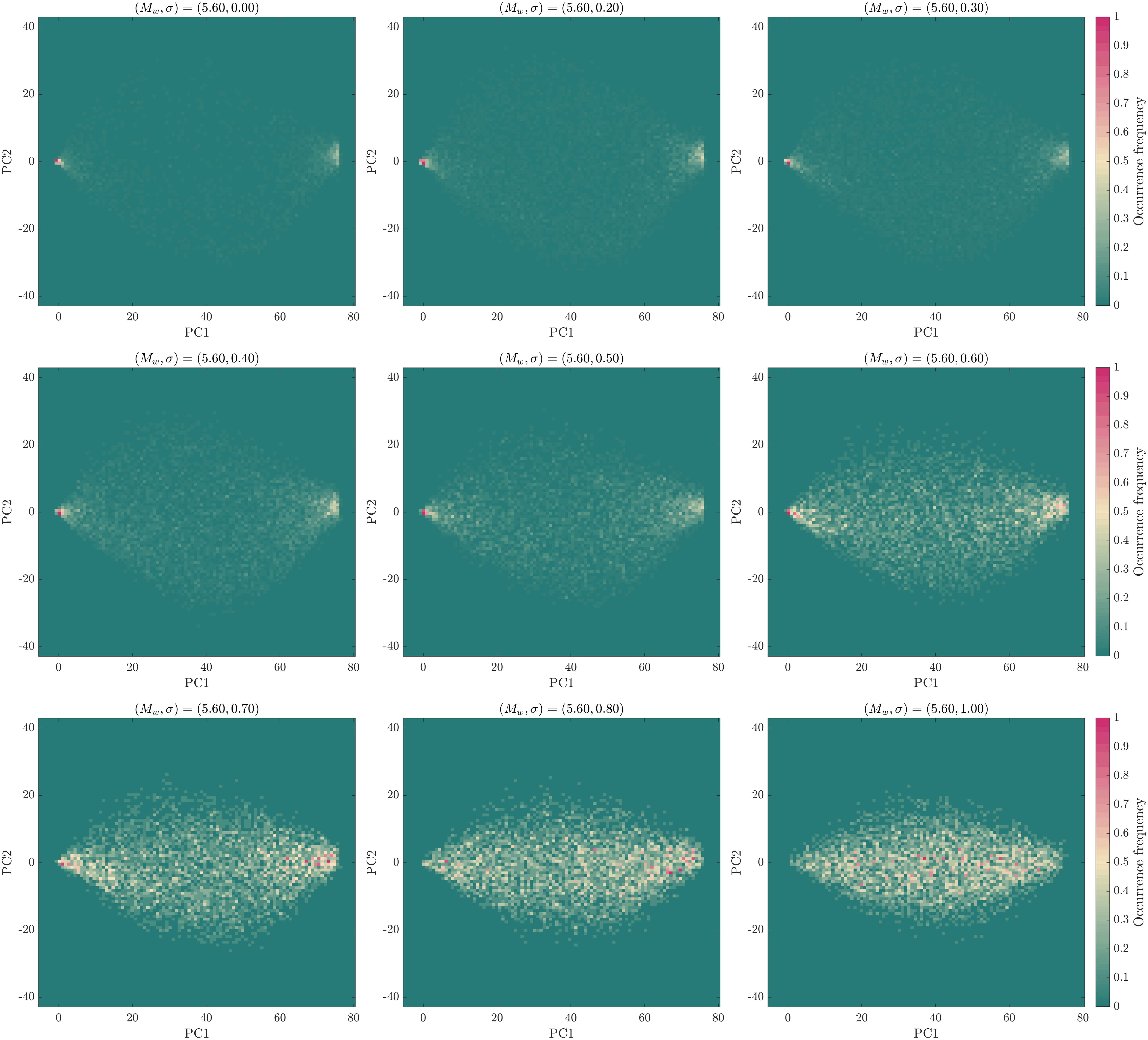}
    \caption{\textbf{Evolution of the projected PCA space with increasing structural diversity for Milpitas.}
    Simulation realizations are projected into the space spanned by PC1 and PC2 for Milpitas at fixed earthquake magnitude $M_w = 5.60$ under increasing structural diversity $\sigma$.
    Each panel shows the normalized occurrence frequency computed from 5,000 realizations.
    With increasing $\sigma$, the initially bimodal distribution along PC1 gradually merges into a single broadened cluster, reflecting a continuous transition from an ordered to a disordered regime, consistent with second-order–like behavior.}
    \addcontentsline{sitoc}{subsection}{Supplementary Fig.~\arabic{figure}. Evolution of the projected PCA space with increasing structural diversity for Milpitas}
    \label{sup_fig:pca_2nd_Milpitas}
\end{figure}

\newpage
\begin{figure}[H]
    \centering
    \includegraphics[width=1.0\linewidth]{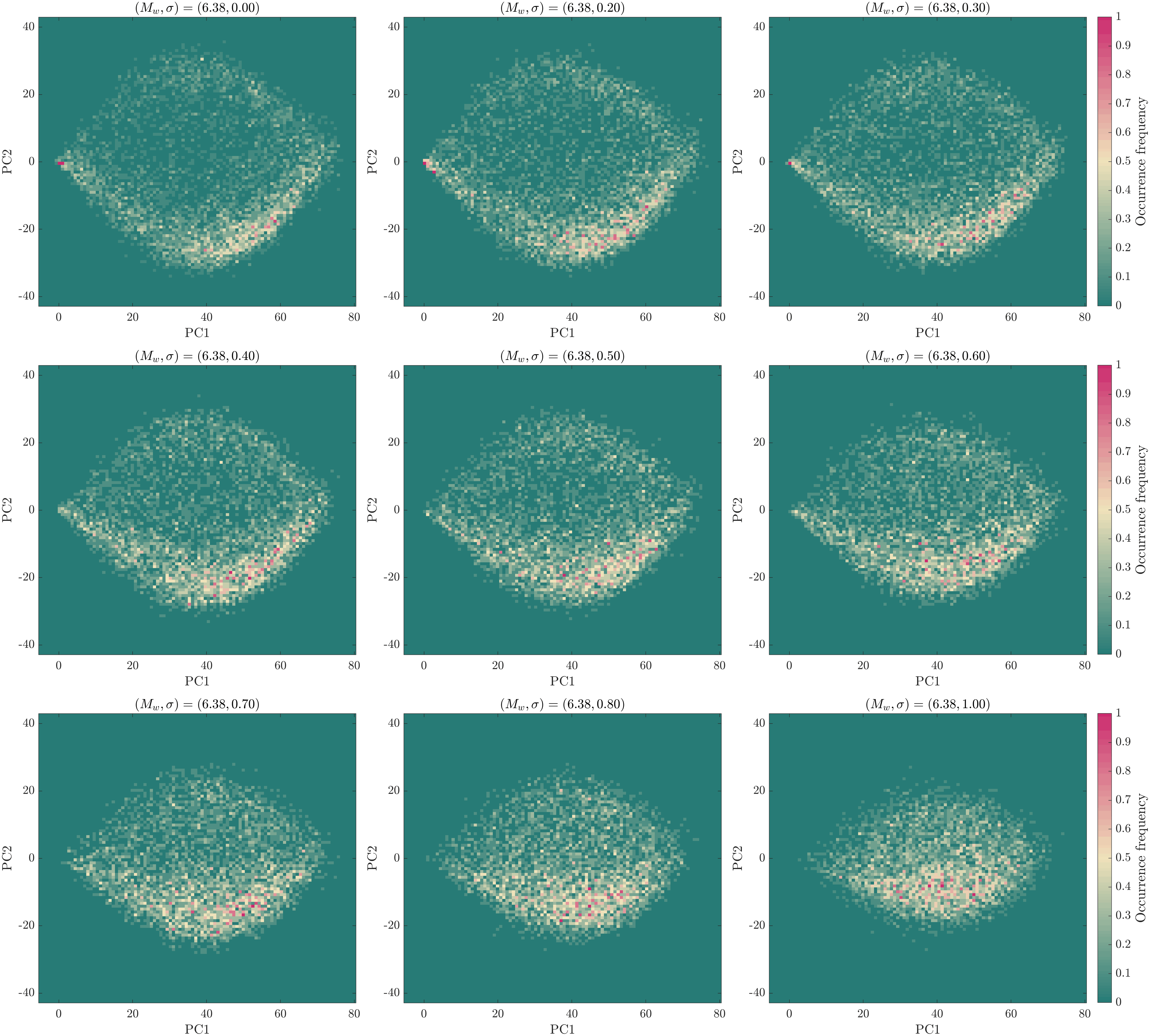}
    \caption{\textbf{Evolution of the projected PCA space with increasing structural diversity for northeastern San Francisco.}
    Same analysis setup as in Supplementary Fig.~\ref{sup_fig:pca_2nd_Milpitas}, performed for San Francisco at fixed earthquake magnitude $M_w = 6.38$. As $\sigma$ increases, the distribution contracts slightly toward the central cluster but does not exhibit any fundamental reorganization across structural diversity. This behavior visually confirms that the region remains in a disordered, paramagnetic-like phase without undergoing a phase transition.}
    \addcontentsline{sitoc}{subsection}{Supplementary Fig.~\arabic{figure}. Evolution of the projected PCA space with increasing structural diversity for northeastern San Francisco}
    \label{sup_fig:pca_2nd_SanFrancisco}
\end{figure}

\newpage
\begin{figure}[H]
    \centering
    \includegraphics[width=1.0\linewidth]{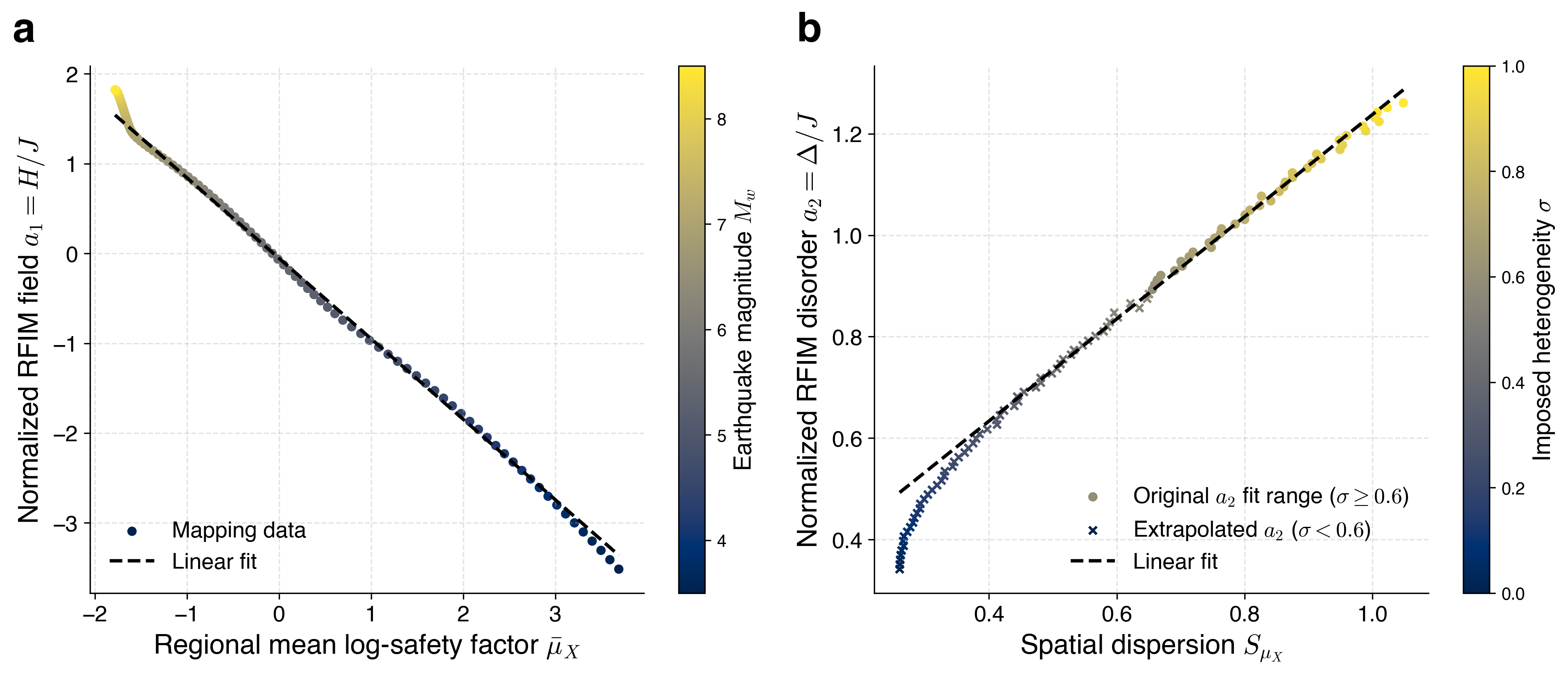}
    \caption{\textbf{Mappings from regional safety-factor statistics to normalized random-field Ising model parameters.}    
    \textbf{a}, Normalized RFIM external field $a_1=H/J$ versus the regional mean log-safety factor $\bar{\mu}_{X}$ for the Milpitas scenarios at $\sigma=0$. Points are colored by earthquake magnitude $M_w$; the dashed line shows the linear map fitted over $-1 \leq \bar{\mu}_{X} \leq 3$.
    \textbf{b}, Normalized RFIM disorder $a_2=\Delta/J$ versus the spatial dispersion $S_{\mu_{X}}$ of the building-wise mean log-safety factor for Milpitas at $M_w=5.6$ while $\sigma$ varies. Points are colored by $\sigma$. Circles denote values within the original $a_2(\sigma)$ fit range ($\sigma\geq0.6$), whereas crosses denote values extrapolated below that range. The dashed line shows the linear map fitted over $0.4 \geq S_{\mu_{X}} \geq 1.0$. The interaction scale is fixed at $J=1$, and no held-out-city data enter either mapping.}
    \addcontentsline{sitoc}{subsection}{Supplementary Fig.~\arabic{figure}. Mappings from regional safety-factor statistics to normalized random-field Ising model parameters}
    \label{sup_fig:mapping_safety_factor_rfim}
\end{figure}

\newpage
\begin{figure}[H]
    \centering
    \includegraphics[width=1.0\linewidth]{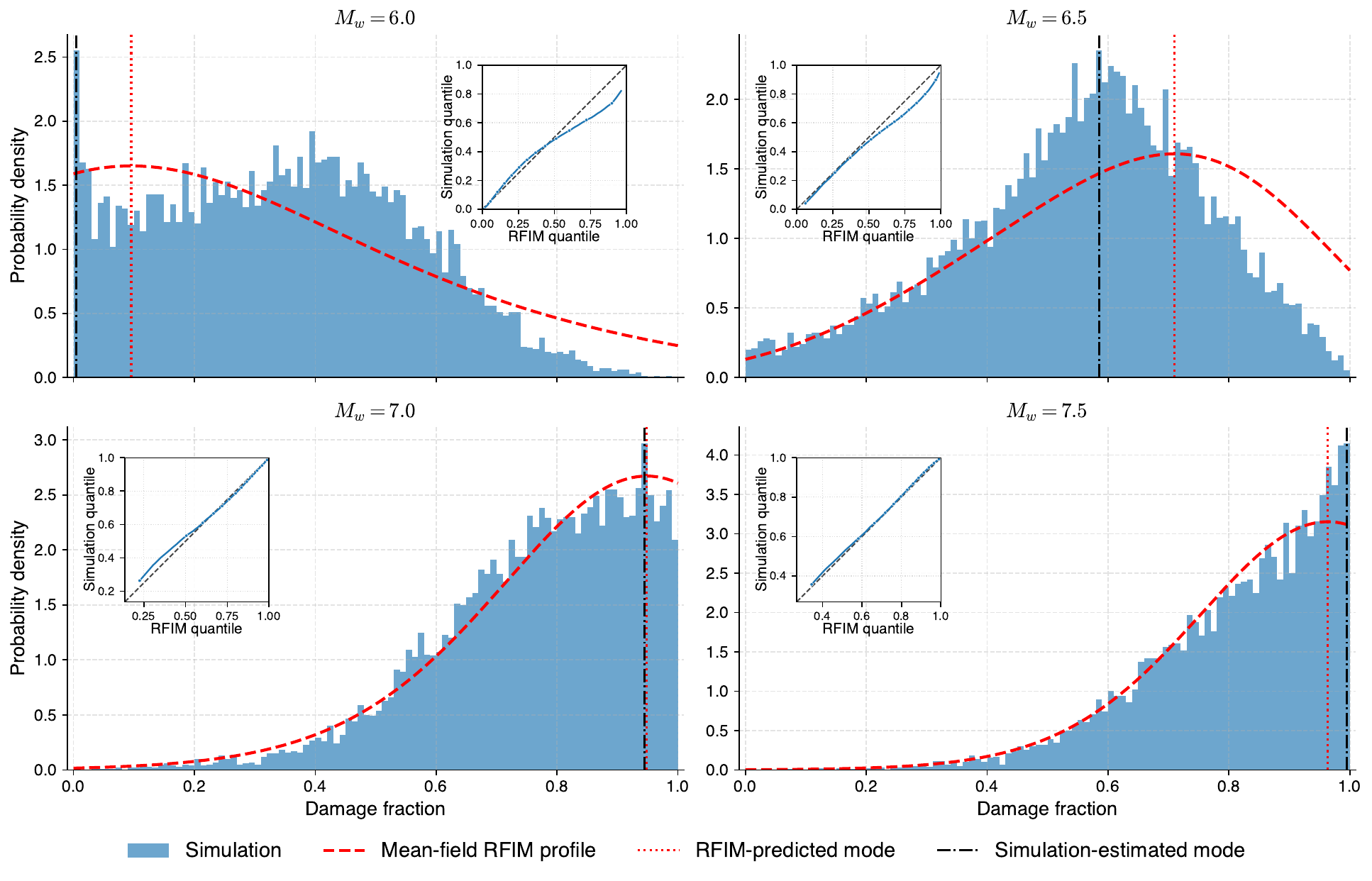}
    \caption{\textbf{Held-out comparison of predicted and simulated damage-fraction distributions for northeastern San Francisco.}
    Blue histograms show the damage-fraction distributions of the unperturbed northeastern San Francisco portfolio at $M_w=6.0$,6.5,7.0, and 7.5, based on 10,000 regional simulations per magnitude. These target-city outcomes were not used to fit the mappings in Supplementary Fig.~\ref{sup_fig:mapping_safety_factor_rfim} or infer $a_1$ and $a_2$. Red dashed curves show profiles derived from the mean-field RFIM free-energy landscapes obtained with the fixed Milpitas mappings. Red dotted and black dash-dotted lines mark the RFIM-predicted and simulation-estimated modes, respectively. Insets show quantile–quantile plots of the simulated damage fractions (vertical axis) against the corresponding mean-field RFIM profile (horizontal axis), with the black dashed line indicating the 1:1 reference. The quantiles track the reference line closely at $M_w=7.0$ and 7.5; at $M_w=6.0$ and 6.5 they fall below it in the upper range, indicating that the RFIM profile assigns more weight to high damage fractions than the simulations do.}
    \addcontentsline{sitoc}{subsection}{Supplementary Fig.~\arabic{figure}. Held-out comparison of predicted and simulated damage-fraction distributions for northeastern San Francisco}
    \label{sup_fig:held_out_sfne_hist}
\end{figure}

\newpage
\begin{figure}[H]
    \centering
    \includegraphics[width=0.85\linewidth]{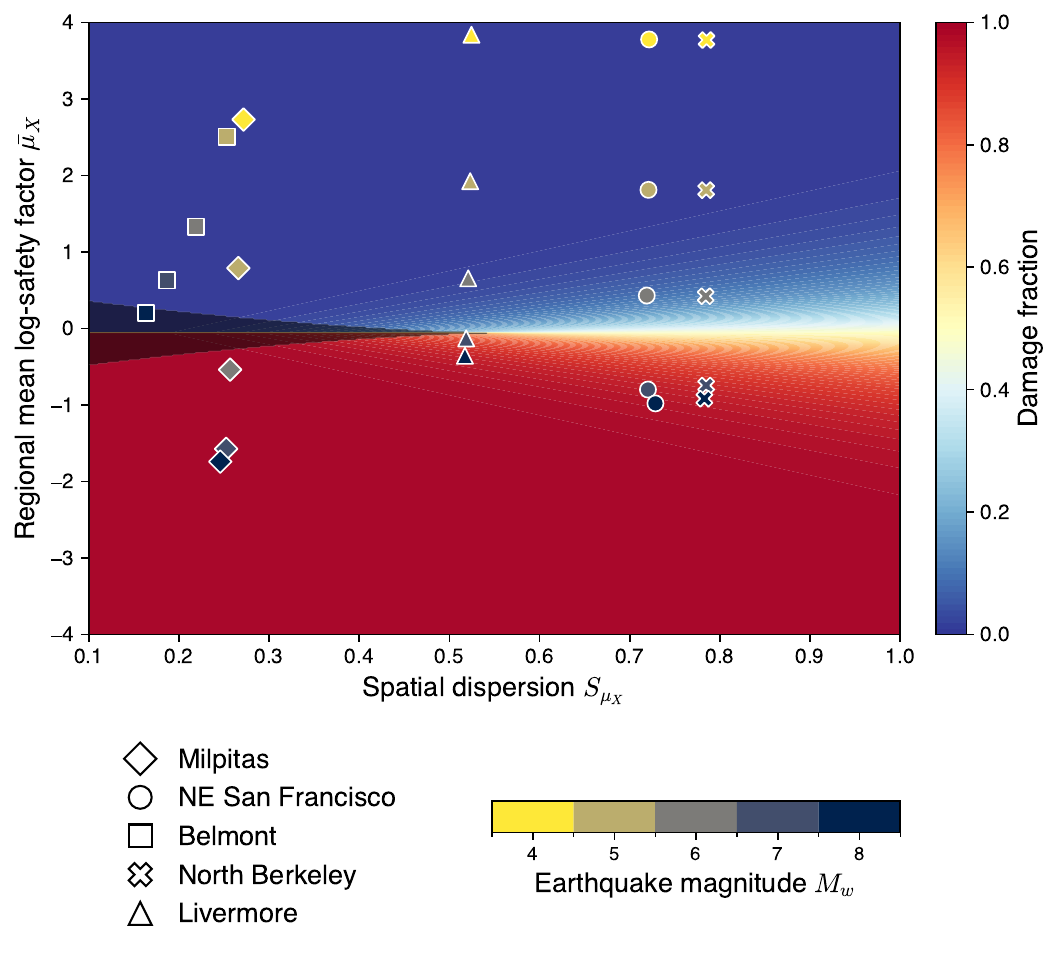}
    \caption{\textbf{Held-out city portfolios on a phase diagram.}
    Background colors show the mean-field RFIM most-probable damage fraction $m_d^\star$ in $(S_{\mu_{X}}, \bar{\mu}_{X})$ coordinates, constructed from the fixed Milpitas mappings in Supplementary Fig.~\ref{sup_fig:mapping_safety_factor_rfim}. Symbols locate the unperturbed ($\sigma=0$) portfolios across earthquake magnitudes; marker shape identifies the portfolio and marker color denotes $M_w$. Milpitas is the calibration portfolio, whereas northeastern San Francisco, Belmont, North Berkeley, and Livermore are held out. Each held-out point is determined only from the target portfolio's building-level mean log-capacity and mean log-demand fields, without using its regional damage simulations.}
    \addcontentsline{sitoc}{subsection}{Supplementary Fig.~\arabic{figure}. Held-out city portfolios on a phase diagram}
    \label{sup_fig:held_out_phase_diagram}
\end{figure}

\begin{figure}[H]
    \centering
    \includegraphics[width=1.0\linewidth]{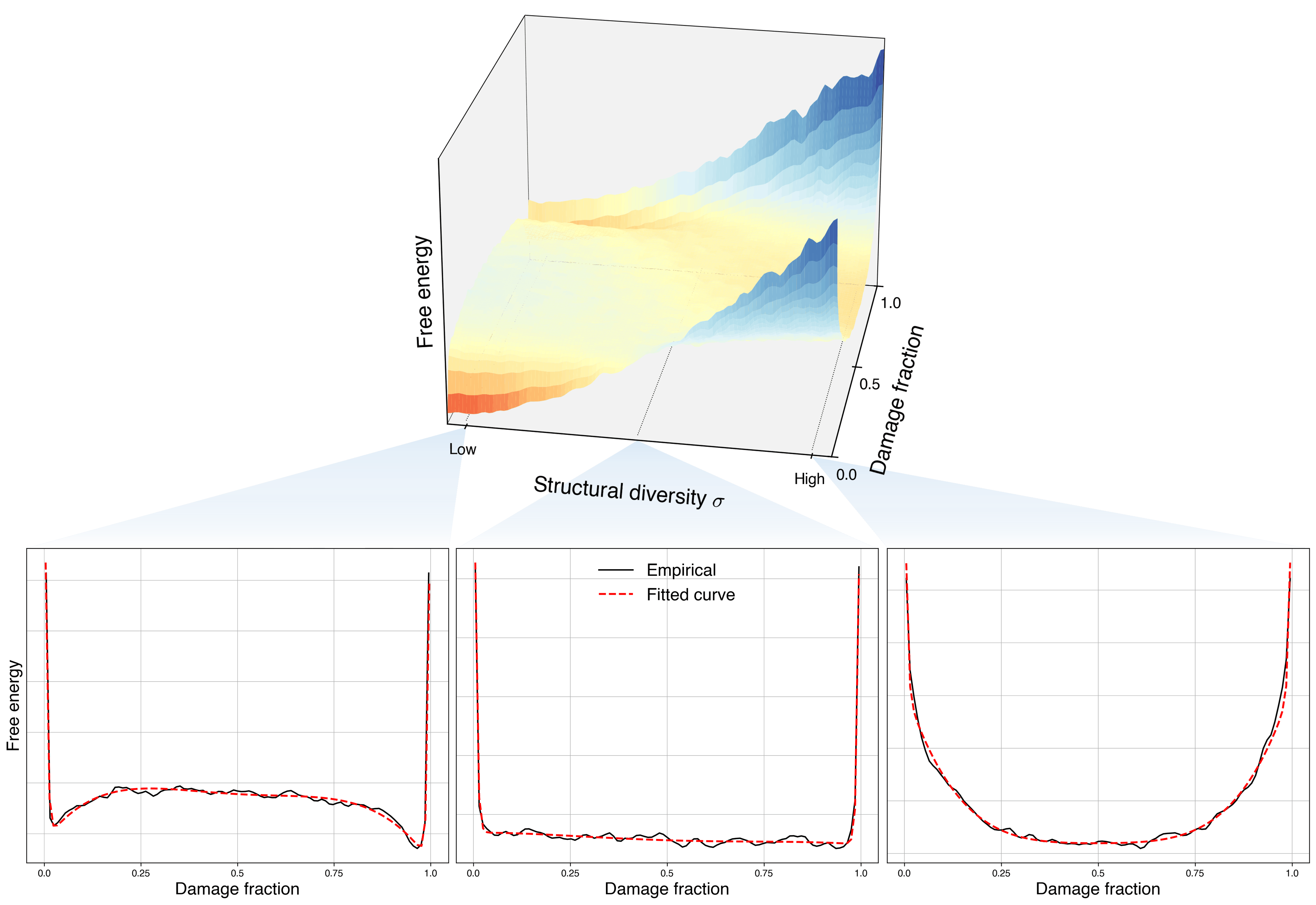}
    \caption{\textbf{Data-driven Landau free-energy landscape.} 
    For each $(M_w,\sigma)$, we form the empirical distribution $p(m)$ by histogramming the simulated damage fraction $m_d$ after mapping to $m\in[-1,1]$ (with $m_d=(m+1)/2$), and define the empirical free energy $F_{\mathrm{emp}}(m)\propto-\ln p(m)$. We fit $F_{\mathrm{emp}}(m)$ with a polynomial Landau form $F_{\mathrm{poly}}(m)=c_0+c_1m+c_2m^2+c_4m^4+c_k m^k$, using $k=100$ for the stabilizing high-order term. The 3D surface (top) shows $F_{\mathrm{emp}}(m)$, and the 2D projections (bottom) show cross-sections with the fitted $F_{\mathrm{poly}}(m)$ for three representative cases: $\sigma<\sigma_c$ (left), $\sigma\approx\sigma_c$ (center), and $\sigma>\sigma_c$ (right). This polynomial-fit Landau free energy enables curvature-based susceptibility estimates (Supplementary Fig.~\ref{sup_fig:volatile_numerical_investigation}).}
    \addcontentsline{sitoc}{subsection}{Supplementary Fig.~\arabic{figure}. Data-driven Landau free-energy landscape}
    \label{sup_fig:data_landau_landscape_fitting}
\end{figure}

\newpage
\begin{figure}[H]
    \centering
    \includegraphics[width=1.0\linewidth]{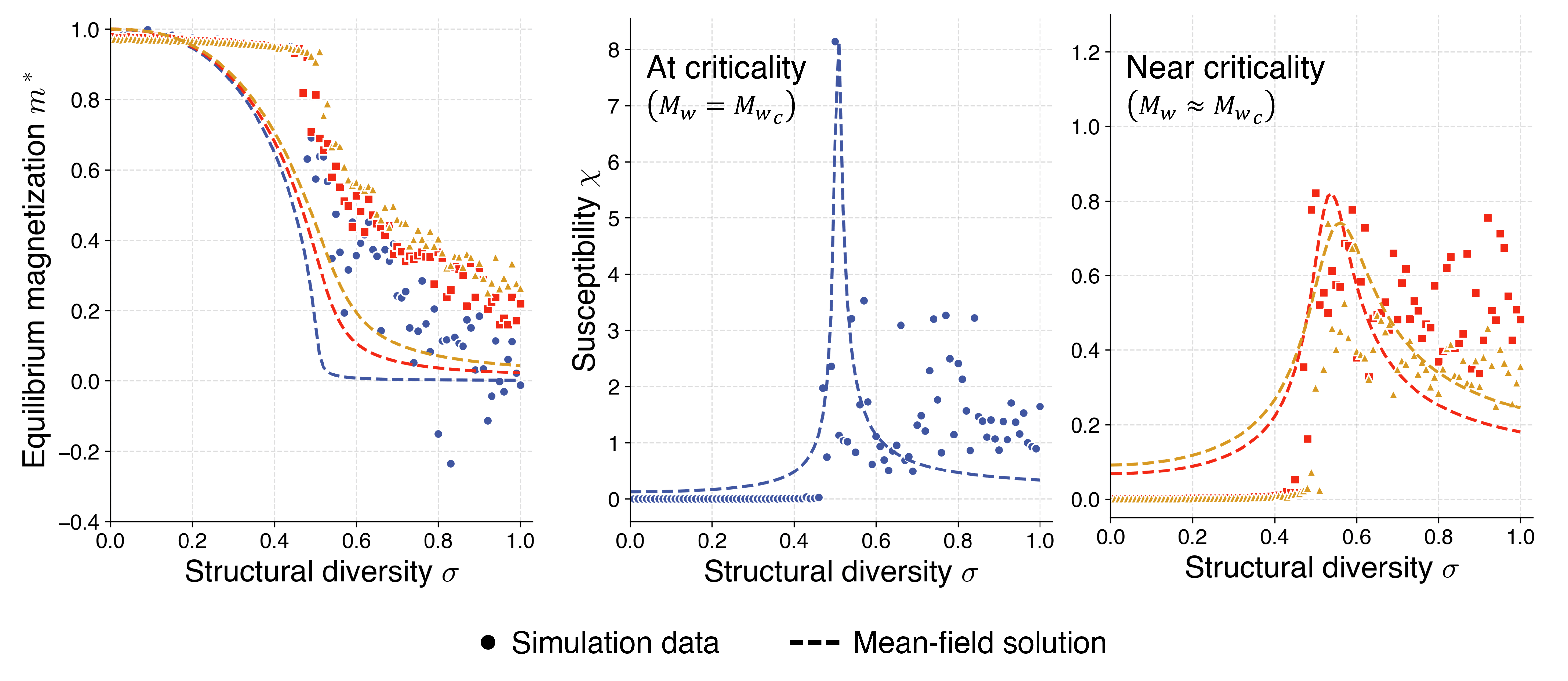}
    \caption{\textbf{Equilibrium magnetization and susceptibility from the data-driven Landau free-energy landscape.}
    Equilibrium magnetization $m^\star$ (left) and curvature-based susceptibility $\chi$ inferred from the polynomial-fit empirical free energy $F_{\mathrm{poly}}(m)$ (Supplementary Method~\ref{sup_method:ensemble_based_calc}; Supplementary Fig.~\ref{sup_fig:data_landau_landscape_fitting}). Blue, red, and yellow indicate $M_w=5.55, 5.60,$ and $5.65$ cases. For each $(M_w,\sigma)$, $m^\star$ is obtained from the minimum of $F_{\mathrm{poly}}(m)$ and $\chi$ from the inverse curvature at $m^\star$, $\chi^{-1}=\left.\partial^2 F_{\mathrm{poly}}/\partial m^2\right|_{m=m^\star}$. The center and right panels report $\chi$ at criticality ($M_w=M_{w_c}$) and near criticality ($M_w\approx M_{w_c}$), respectively. Markers show simulation-based estimates from a single portfolio realization and dashed lines indicate the mean-field prediction. The lingering magnetization and slowly decaying susceptibility for $\sigma>\sigma_c$ are consistent with the Griffiths-like behavior described in Figs.~3c,d}
    \addcontentsline{sitoc}{subsection}{Supplementary Fig.~\arabic{figure}. Equilibrium magnetization and susceptibility from the data-driven Landau free-energy landscape}
    \label{sup_fig:volatile_numerical_investigation}
\end{figure}

\newpage
\begin{figure}[H]
    \centering
    \includegraphics[width=1.0\linewidth]{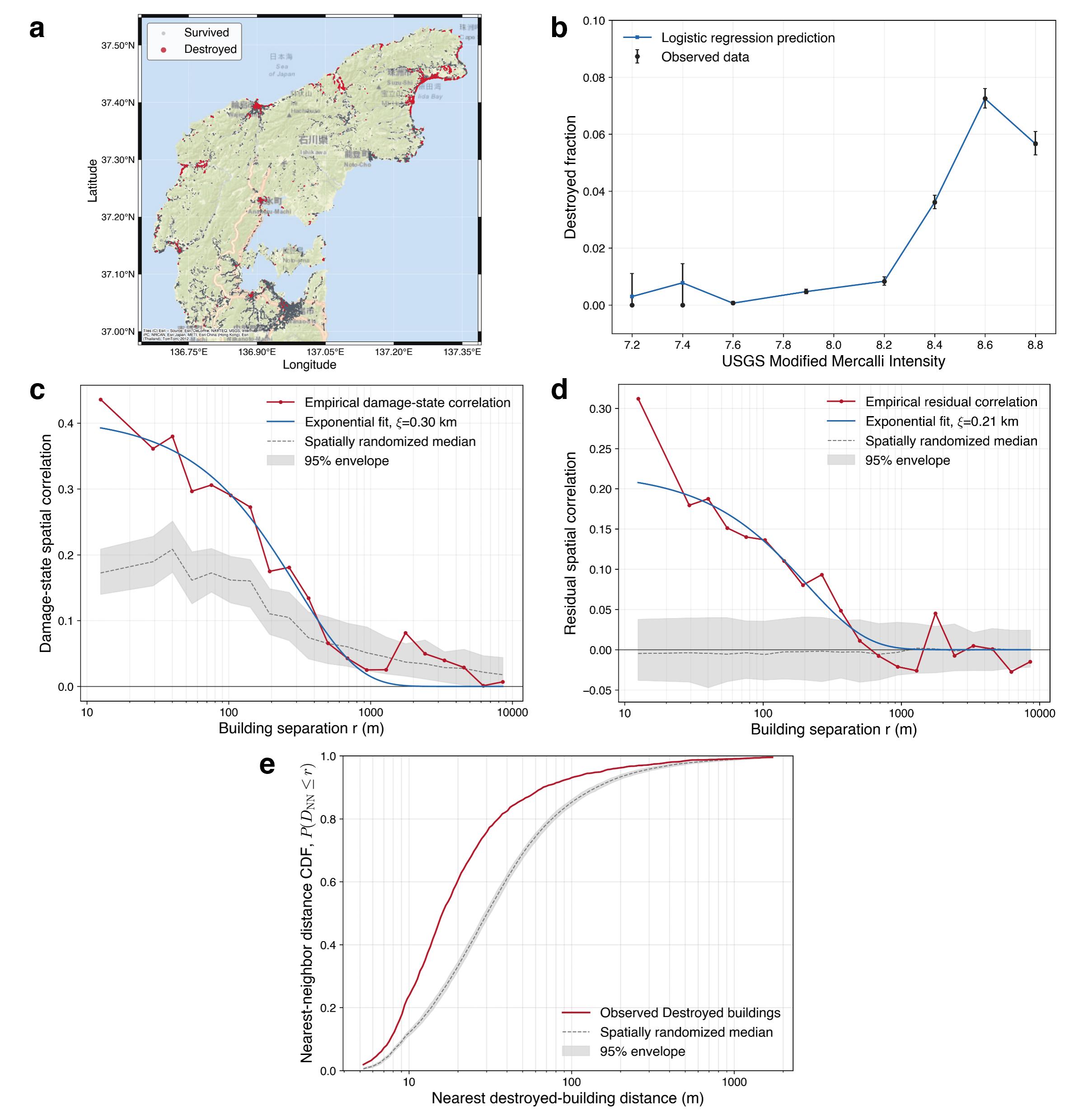}
    \caption{\textbf{Spatial organization of building damage in the 2024 Noto Peninsula earthquake.}
    \textbf{a}, Spatial distribution of buildings classified as survived or destroyed in the post-earthquake damage inventory.
    \textbf{b}, Observed and fitted logistic-regression estimates of the destroyed fraction across USGS MMI intervals. Error bars denote 95\% intervals for the observed fractions.
    \textbf{c}, Distance-dependent correlation of the observed binary damage states. The positive randomized correlation reflects the broad spatial variation in the fitted marginal damage pattern that is retained by the conditional randomization. The empirical correlation yields a descriptive correlation length of $\xi=0.30$ km.
    \textbf{d}, Spatial correlation of the damage residuals after accounting for USGS MMI, building-footprint area, and municipality. Compared to \textbf{c}, residualization substantially reduces the broad covariate-associated correlation, while a positive short-range correlation remains with a correlation length of $\xi=0.21$ km. In \textbf{c} and \textbf{d}, dashed lines and shaded regions denote the median and 95\% pointwise envelope from conditional randomizations, respectively.
    \textbf{e}, Cumulative distribution of the nearest-destroyed-building distance. The observed distribution is shifted toward shorter distances than the conditional-randomization reference, indicating tighter local clustering of destroyed buildings.}
    \addcontentsline{sitoc}{subsection}{Supplementary Fig.~\arabic{figure}. Spatial organization of building damage in the 2024 Noto Peninsula earthquake}
    \label{sup_fig:noto_damage}
\end{figure}

\newpage
\begin{figure}[H]
    \centering
    \includegraphics[width=1.0\linewidth]{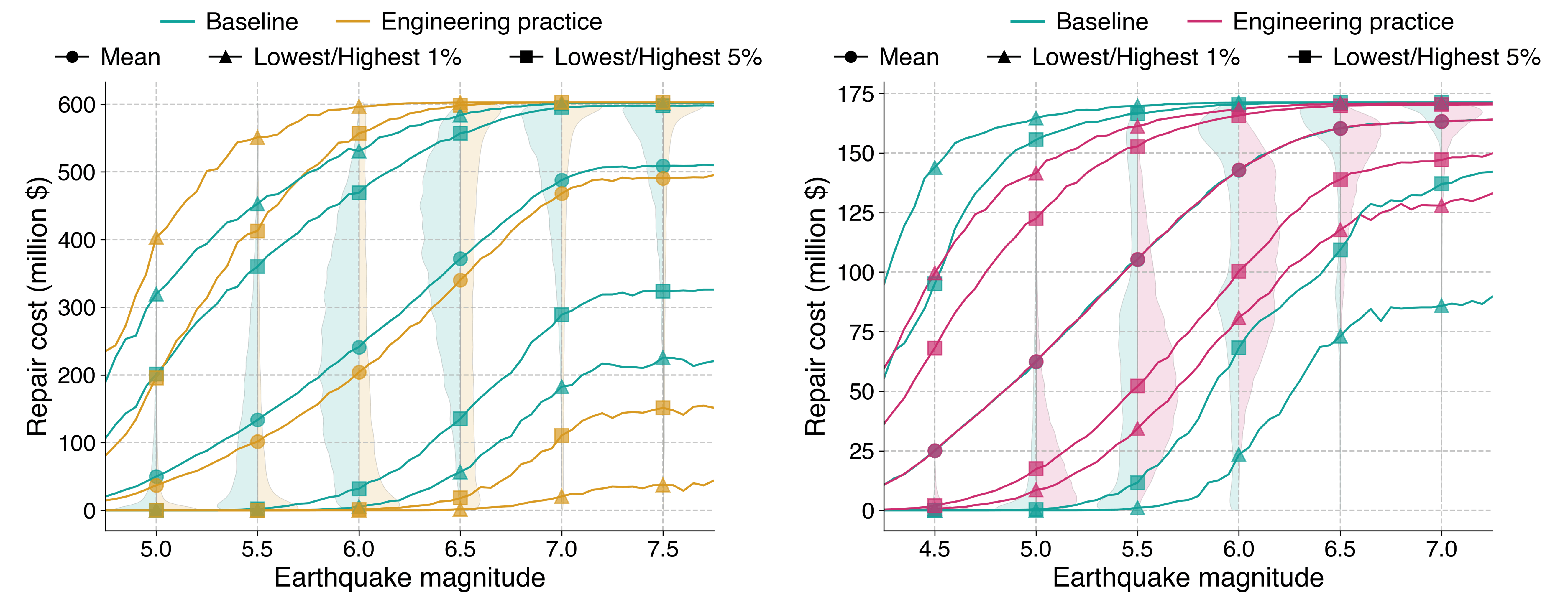}
    \caption{\textbf{Mean and quantile values of regional repair-cost distributions computed with and without engineering simplifications.}
    Left: northeastern San Francisco portfolio; right: Milpitas portfolio. The background violins show the full repair-cost distributions at representative earthquake magnitudes. Mean repair costs are largely insensitive to engineering simplifications, whereas the lower and upper 1\% and 5\% quantiles differ substantially. In the northeastern San Francisco case, structural categorization pushes the lower and upper quantile curves farther apart, overestimating upper-tail losses and underestimating lower-tail losses. This lower-tail bias is particularly pronounced at $M_w=6.5$, where the 1\% quantile decreases from US\$56.66 million in the baseline analysis to US\$1.72 million under structural categorization, corresponding to an about 33-fold underestimation. In Milpitas, the conditional-independence assumption has the opposite effect, drawing the two quantile curves closer together by reducing upper-tail losses and inflating lower-tail estimates; at $M_w=5.5$, it increases the 1\% quantile from US\$1.14 million to US\$34.31 million, corresponding to an about 30-fold overestimation.}
    \addcontentsline{sitoc}{subsection}{Supplementary Fig.~\arabic{figure}. Mean and quantile values of regional repair-cost distributions computed with and without engineering simplifications}
    \label{sup_fig:eng_implications_curve}
\end{figure}

\newpage
\begin{figure}[H]
    \centering
    \includegraphics[width=1.0\linewidth]{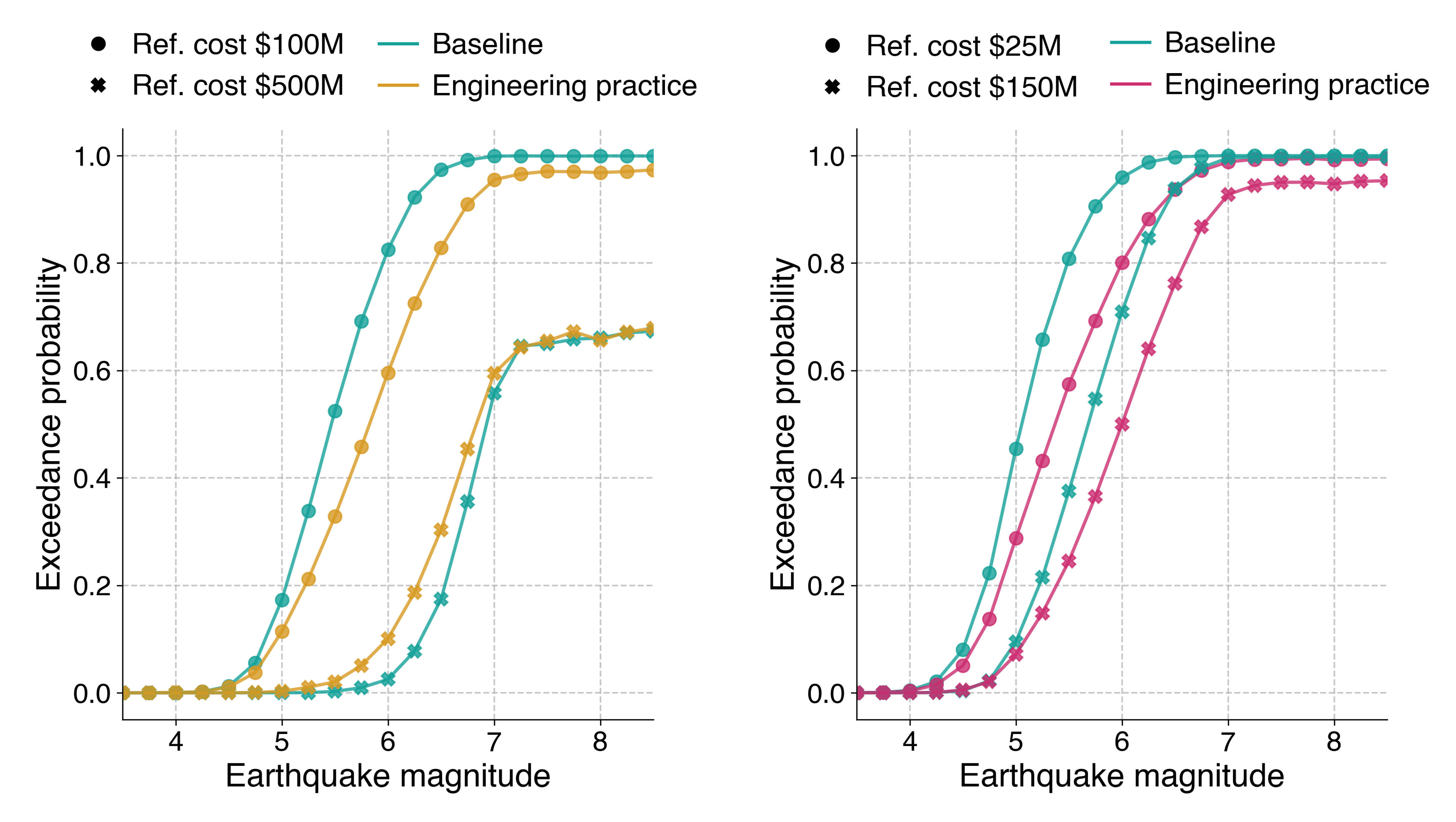}
    \caption{\textbf{Exceedance probabilities for two reference repair costs.}
    Left: northeastern San Francisco portfolio; right: Milpitas portfolio. Curves show the probability that regional repair costs exceed two reference thresholds (circles: lower threshold; crosses: higher threshold) with and without engineering simplifications. In both regions, exceedance probabilities shift noticeably when engineering simplifications are applied. These simplifications introduce systematic bias in the exceedance probability over $M_w \in [5.0,6.0]$, with relative errors generally on the order of 50\% across this magnitude range in both the Milpitas and northeastern San Francisco cases.}
    \addcontentsline{sitoc}{subsection}{Supplementary Fig.~\arabic{figure}. Exceedance probabilities for two reference repair costs}
    \label{sup_fig:eng_implications_cdf}
\end{figure}

\newpage
\begin{figure}[H]
    \centering
    \includegraphics[width=1.0\linewidth]{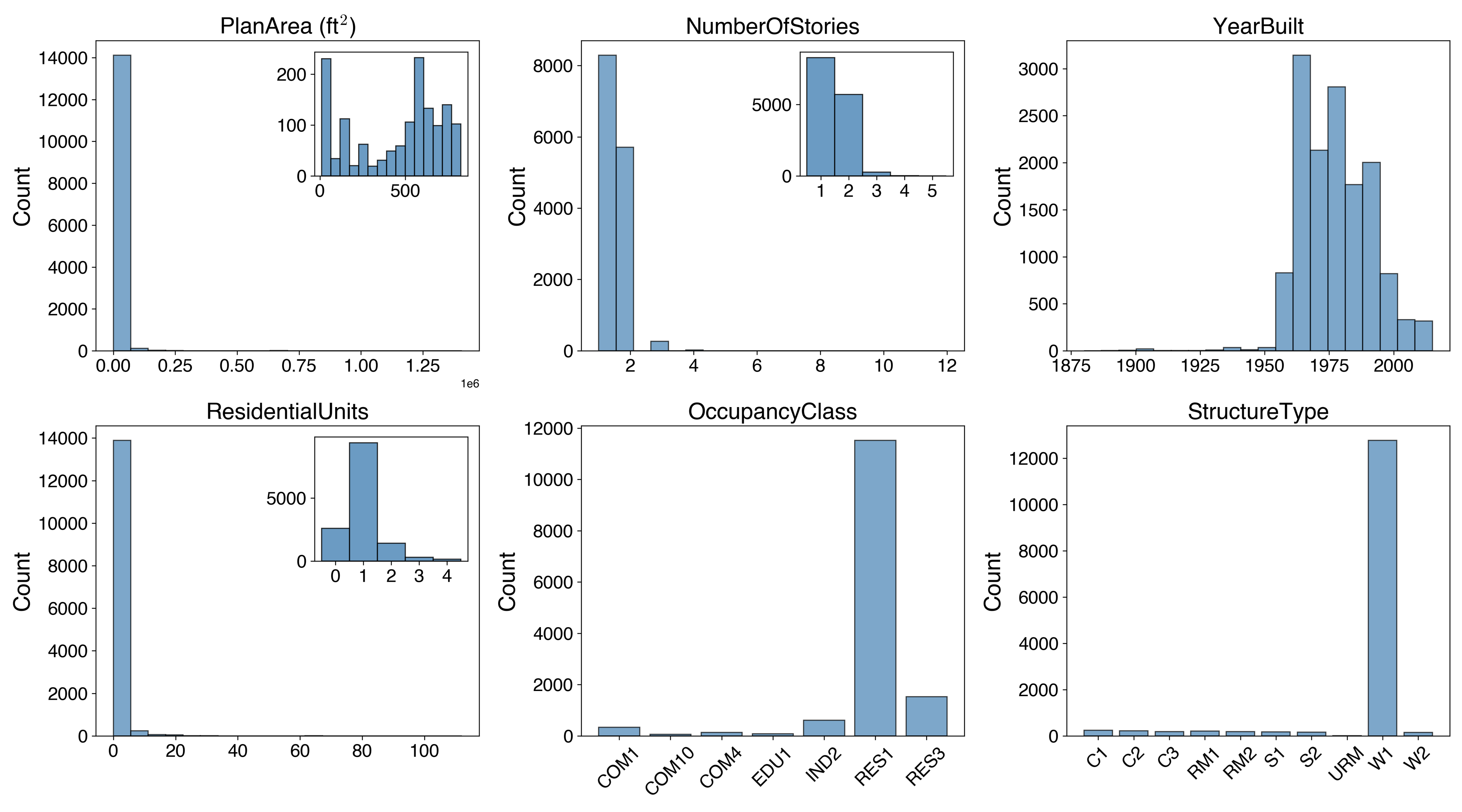}
    \caption{\textbf{Building attribute distributions for the Milpitas portfolio.} 
    Histograms show the distributions of key building attributes, including plan area, number of stories, construction year, residential units, occupancy class, and structure type. Insets magnify the histograms near small values.}
    \addcontentsline{sitoc}{subsection}{Supplementary Fig.~\arabic{figure}. Building attribute distributions for the Milpitas portfolio}
    \label{sup_fig:milpitas_attributes_histograms}
\end{figure}

\newpage
\begin{figure}[H]
    \centering
    \includegraphics[width=0.6\linewidth]{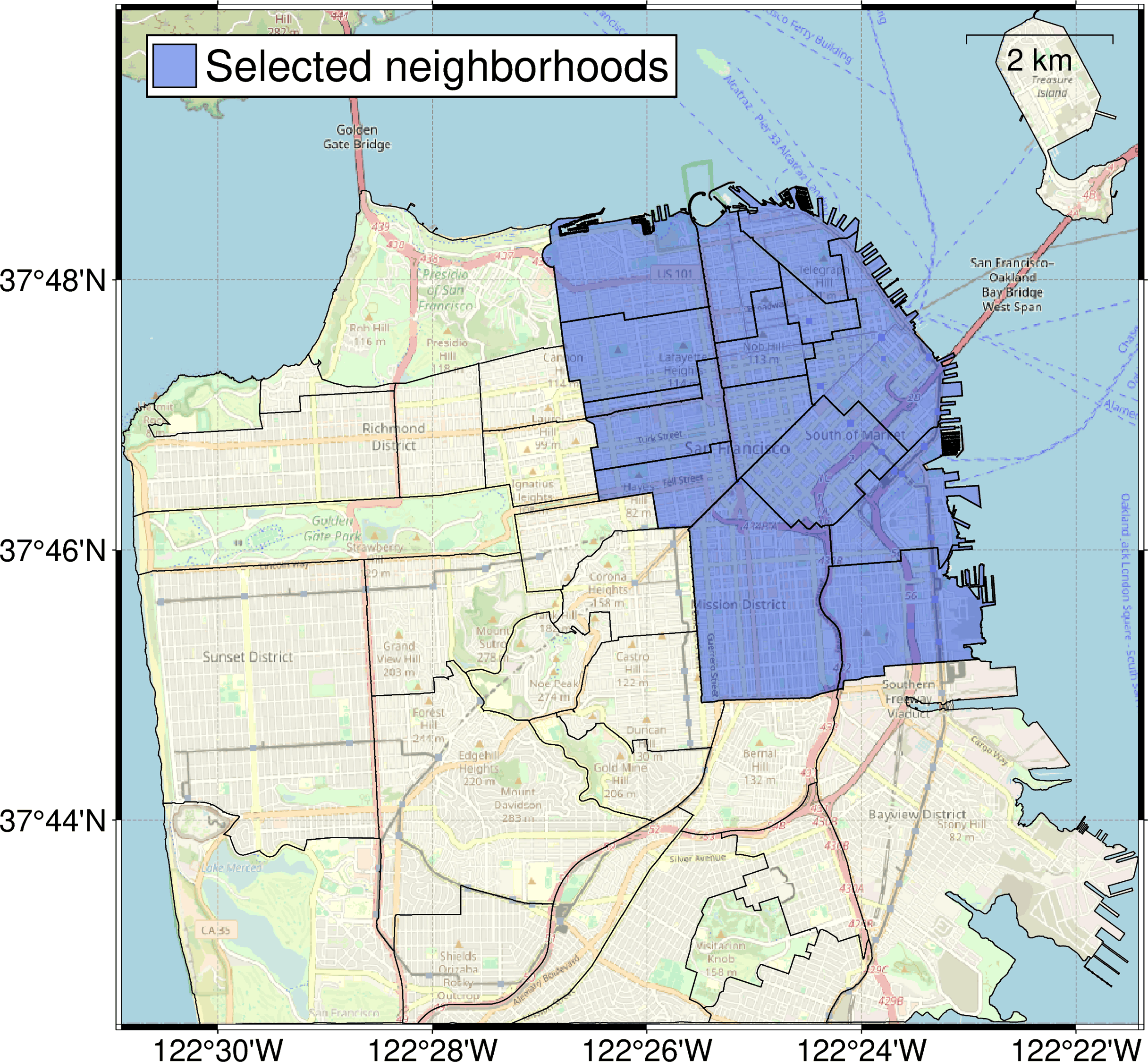}
    \caption{\textbf{Selected neighborhoods in northeastern San Francisco.} 
    The analyzed region includes major commercial districts such as the Financial District and Mission Bay, as well as residential neighborhoods including Pacific Heights and Hayes Valley. The highlighted area represents the subset of neighborhoods selected for the building portfolio analysis.}
    \addcontentsline{sitoc}{subsection}{Supplementary Fig.~\arabic{figure}. Selected neighborhoods in northeastern San Francisco}
    \label{sup_fig:sanfrancisco_selected_neighbors}
\end{figure}

\newpage
\begin{figure}[H]
    \centering
    \includegraphics[width=1.0\linewidth]{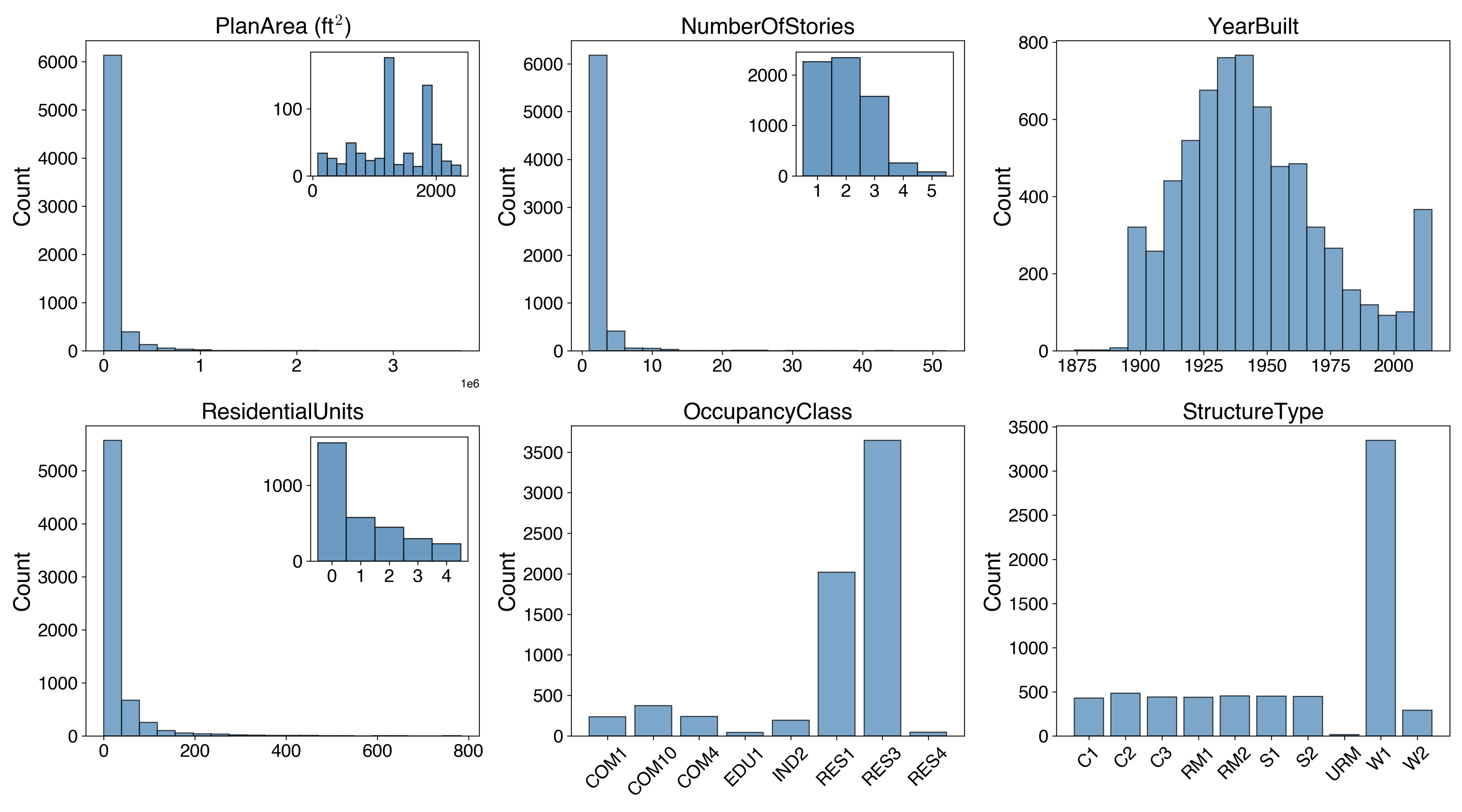}
    \caption{\textbf{Building attribute distributions for the northeastern San Francisco portfolio.} 
    Histograms show the distributions of key building attributes, including plan area, number of stories, construction year, residential units, occupancy class, and structure type. Insets magnify the histograms near small values.}
    \addcontentsline{sitoc}{subsection}{Supplementary Fig.~\arabic{figure}. Building attribute distributions for the northeastern San Francisco portfolio}
    \label{sup_fig:sanfrancisco_attributes_histograms}
\end{figure}

\newpage
\begin{figure}[H]
    \centering
    \includegraphics[width=1.0\linewidth]{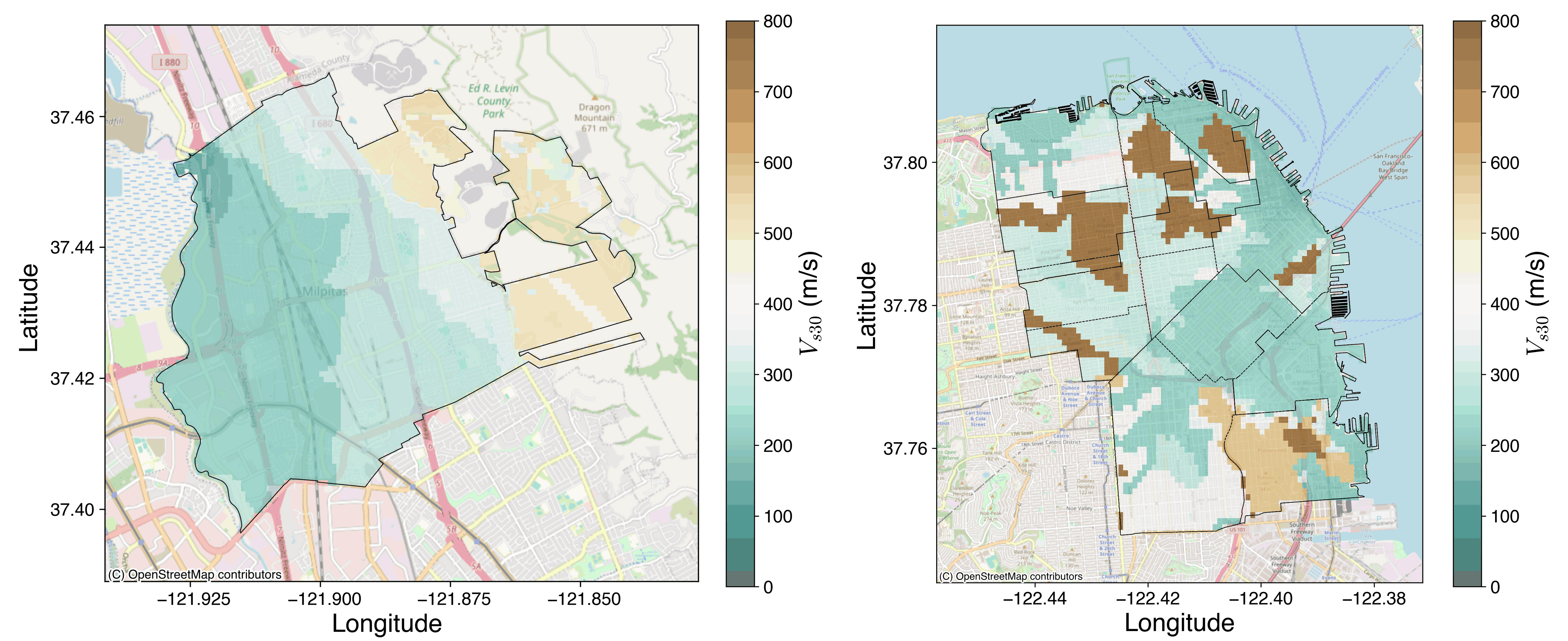}
    \caption{\textbf{Maps of the near-surface shear-wave velocity $V_{s30}$ for Milpitas (left) and northeastern San Francisco (right).}
    The values are retrieved from~\citet{thompson_updated_2018}. The selected San Francisco region exhibits greater heterogeneity in this geomechanical property, which strongly influences seismic response, as well as greater variation in building types.}
    \addcontentsline{sitoc}{subsection}{Supplementary Fig.~\arabic{figure}. Maps of the near-surface shear-wave velocity \texorpdfstring{$V_{s30}$}{Vs30} for Milpitas (left) and northeastern San Francisco (right)}
    \label{sup_fig:vs30_milpitas_sf}
\end{figure}

\newpage
\begin{figure}[H]
    \centering
    \includegraphics[width=0.7\linewidth]{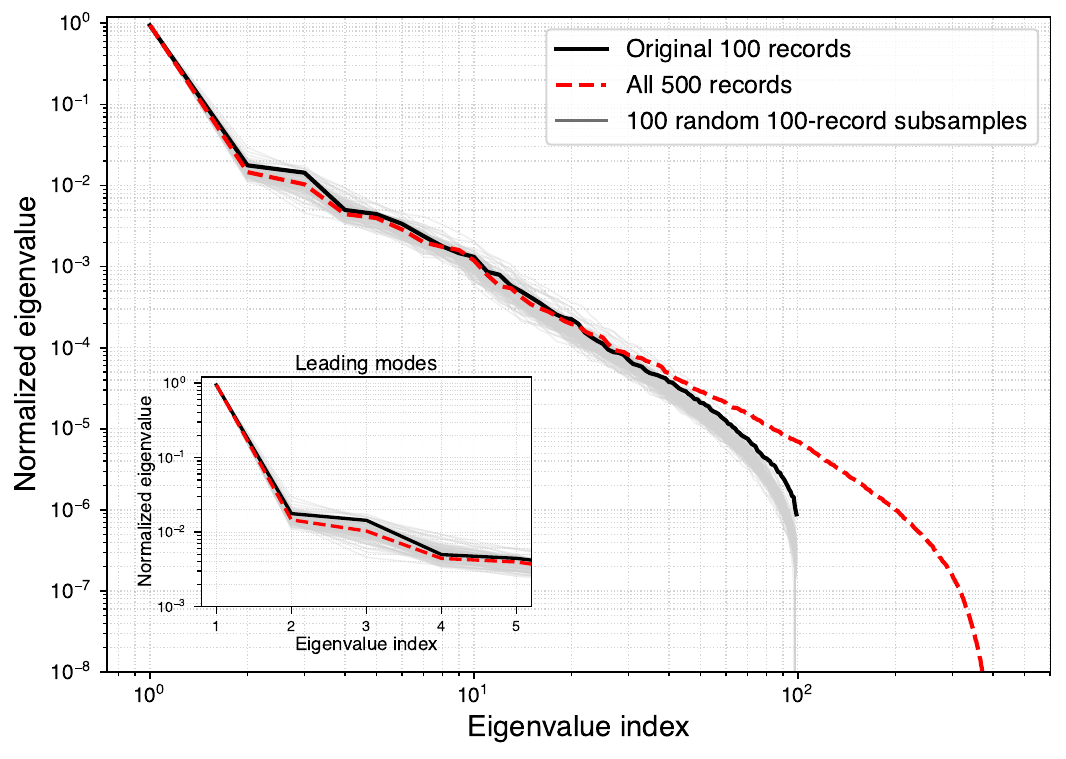}
    \caption{\textbf{Sampling stability of the IDA-derived capacity-dependence spectrum.}
    Normalized non-zero eigenvalue spectra estimated from the original 100 ground-motion records (black), all 500 records (red dashed), and 100 random 100-record subsamples drawn from the full 500-record pool (gray). Eigenvalues are ordered from largest to smallest and normalized by the leading eigenvalue. The inset magnifies the first five modes. In all cases, the spectra are dominated by the first few modes, and the original 100-record spectrum lies within the spread of the subsampled spectra, indicating that the dominant dependence structure is stable.}
    \addcontentsline{sitoc}{subsection}{Supplementary Fig.~\arabic{figure}. Sampling stability of the IDA-derived capacity-dependence spectrum}
    \label{sup_fig:capa_eigen_spectrum}
\end{figure}

\newpage
\begin{figure}[H]
    \centering
    \includegraphics[width=0.6\linewidth]{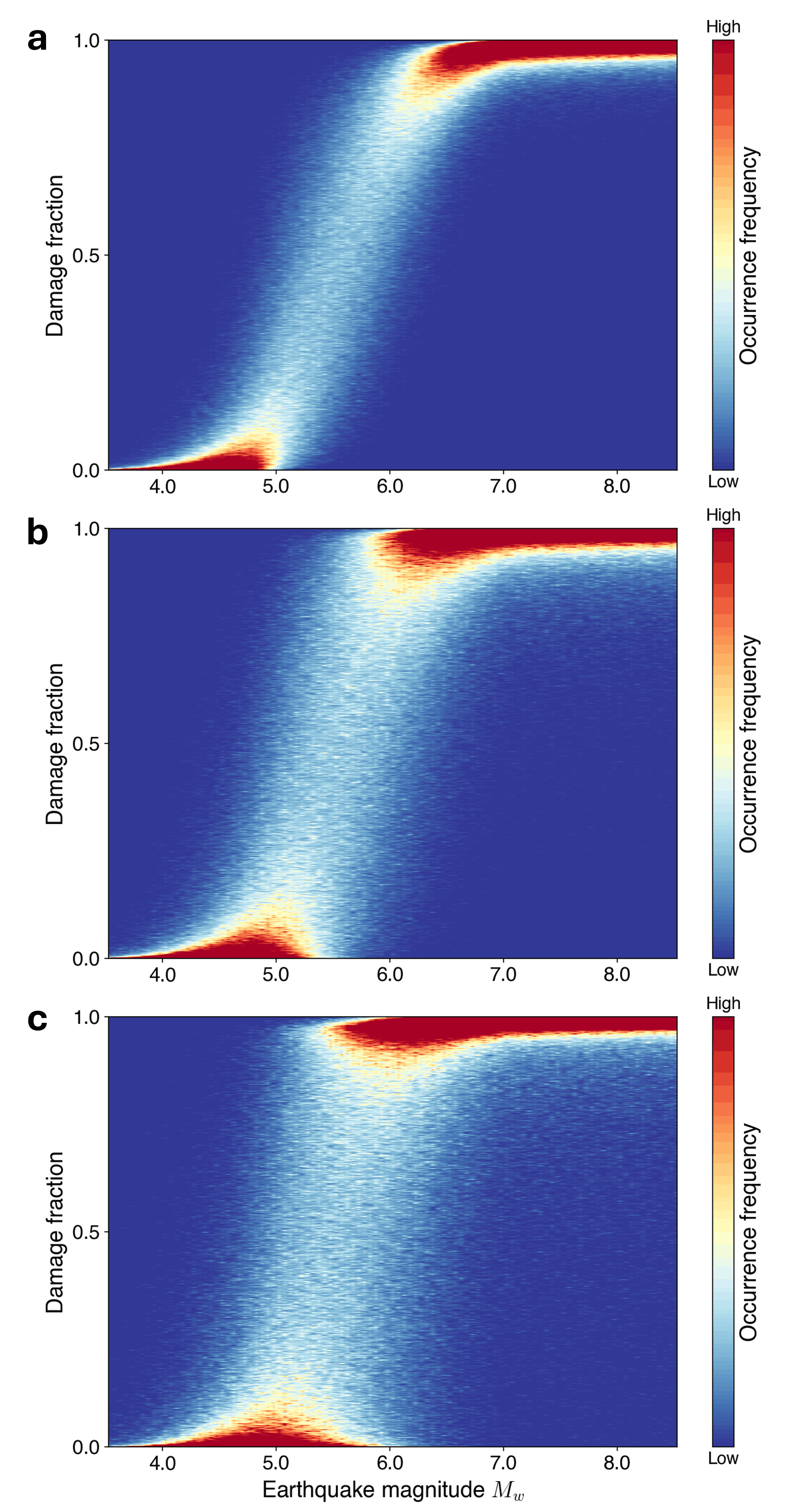}
    \caption{\textbf{Sensitivity of regional damage distributions to the capacity-dependence model.}
    Occurrence-frequency distributions of the regional damage fraction with increasing earthquake magnitude for \textbf{a}, conditionally independent capacities, \textbf{b}, a prescribed equicorrelation model with $\rho=0.5$, and \textbf{c}, the IDA-derived capacity-dependence model. Building-level marginal fragility distributions, the building inventory, the spatially correlated demand model, and the Monte~Carlo settings are held fixed across the three cases. All panels use the same histogram bins and color normalization. As capacity dependence increases from \textbf{a} to \textbf{c}, the damage-fraction distribution broadens and the bimodal, abrupt transition becomes more pronounced. The equicorrelation case is included as a simplified sensitivity case and is not intended as a calibrated representation of the portfolio.}
    \addcontentsline{sitoc}{subsection}{Supplementary Fig.~\arabic{figure}. Sensitivity of regional damage distributions to the capacity-dependence model}
    \label{sup_fig:capa_dependence_ablation}
\end{figure}

\newpage
\begin{figure}[H]
    \centering
    \includegraphics[width=1.0\linewidth]{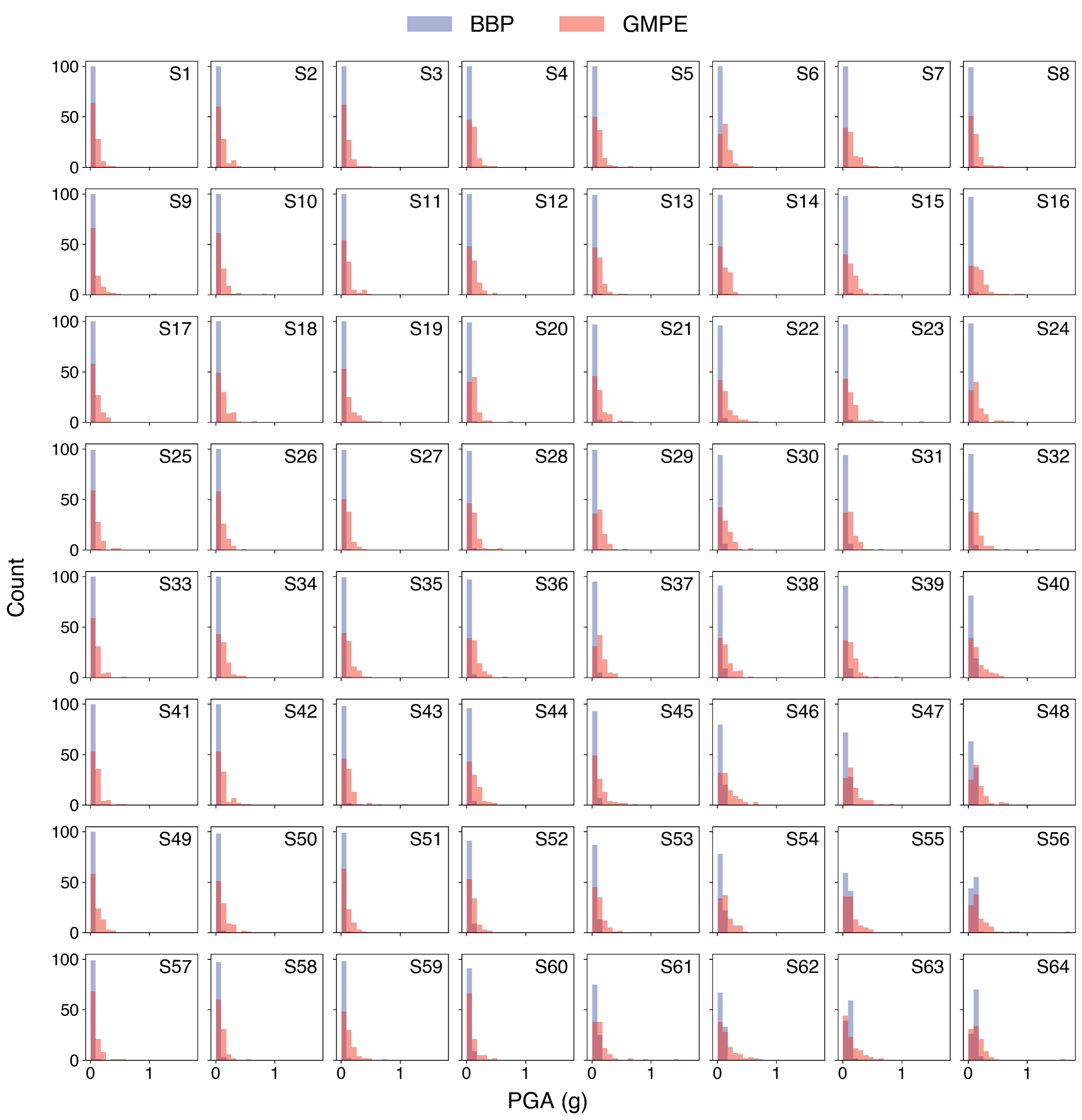}
    \caption{\textbf{Comparison of peak ground acceleration (PGA) distributions between physics-based and empirical models at $M_w=5.0$ across recording stations.}
    Histograms compare PGA values obtained from the physics-based SCEC Broadband Platform (BBP) simulations and the empirical Chiou and Youngs ground-motion prediction equation (GMPE). Both datasets were sampled at the same 64 recording stations (S1--S64) uniformly distributed across Milpitas, with 100 realizations generated for each method.}
    \addcontentsline{sitoc}{subsection}{Supplementary Fig.~\arabic{figure}. Comparison of peak ground acceleration (PGA) distributions between physics-based and empirical models at \texorpdfstring{$M_w=5.0$}{Mw=5.0} across recording stations}
    \label{sup_fig:pga_comparison_per_station_Mw5}
\end{figure}

\newpage
\begin{figure}[H]
    \centering
    \includegraphics[width=1.0\linewidth]{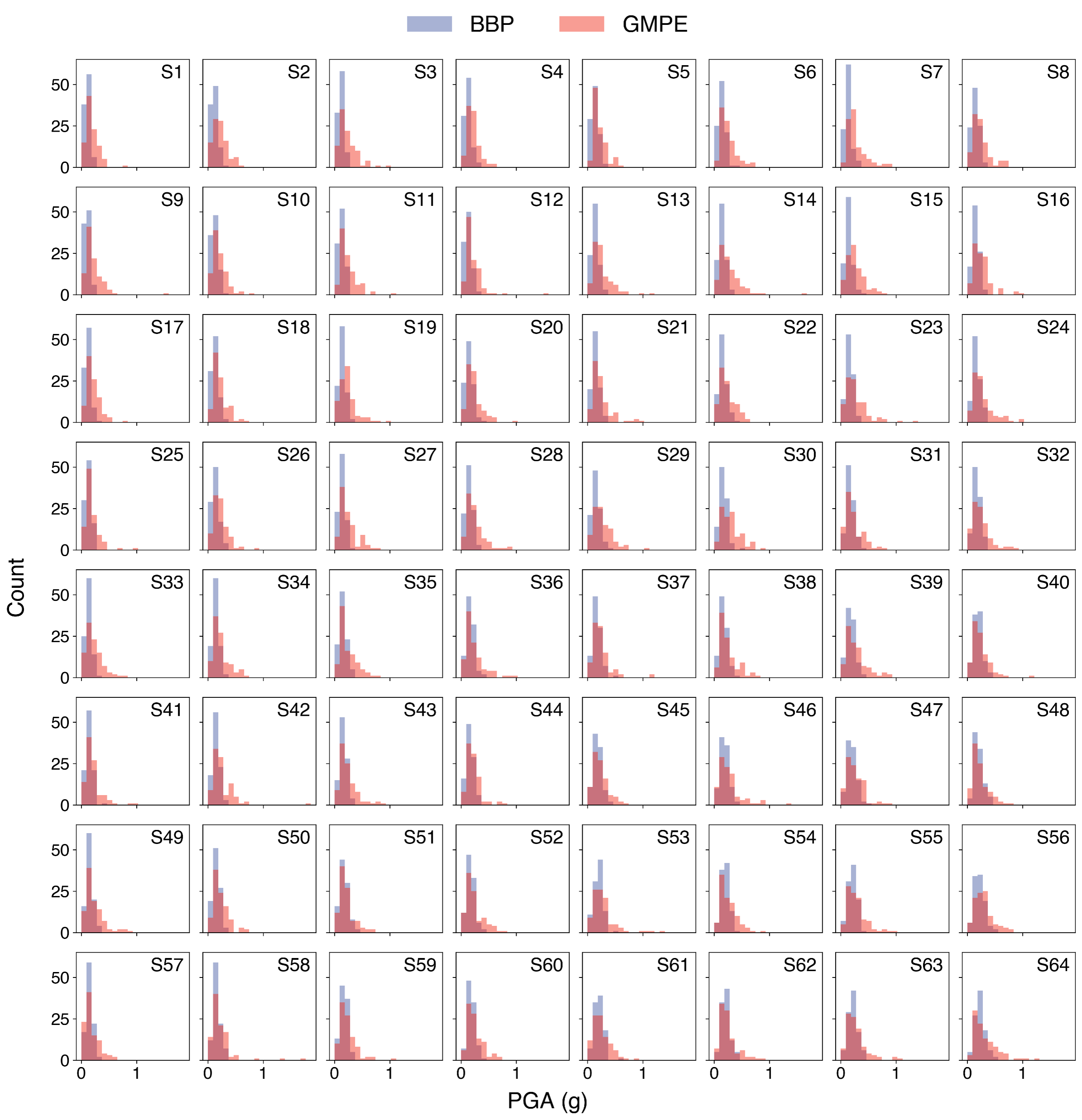}
    \caption{\textbf{Comparison of peak ground acceleration (PGA) distributions between physics-based and empirical models at $M_w=6.0$ across recording stations.}
    Histograms compare PGA values obtained from the physics-based SCEC Broadband Platform (BBP) simulations and the empirical Chiou and Youngs ground-motion prediction equation (GMPE). Both datasets were sampled at the same 64 recording stations (S1--S64) uniformly distributed across Milpitas, with 100 realizations generated for each method.}
    \addcontentsline{sitoc}{subsection}{Supplementary Fig.~\arabic{figure}. Comparison of peak ground acceleration (PGA) distributions between physics-based and empirical models at \texorpdfstring{$M_w=6.0$}{Mw=6.0} across recording stations}
    \label{sup_fig:pga_comparison_per_station_Mw6}
\end{figure}

\newpage
\begin{figure}[H]
    \centering
    \includegraphics[width=1.0\linewidth]{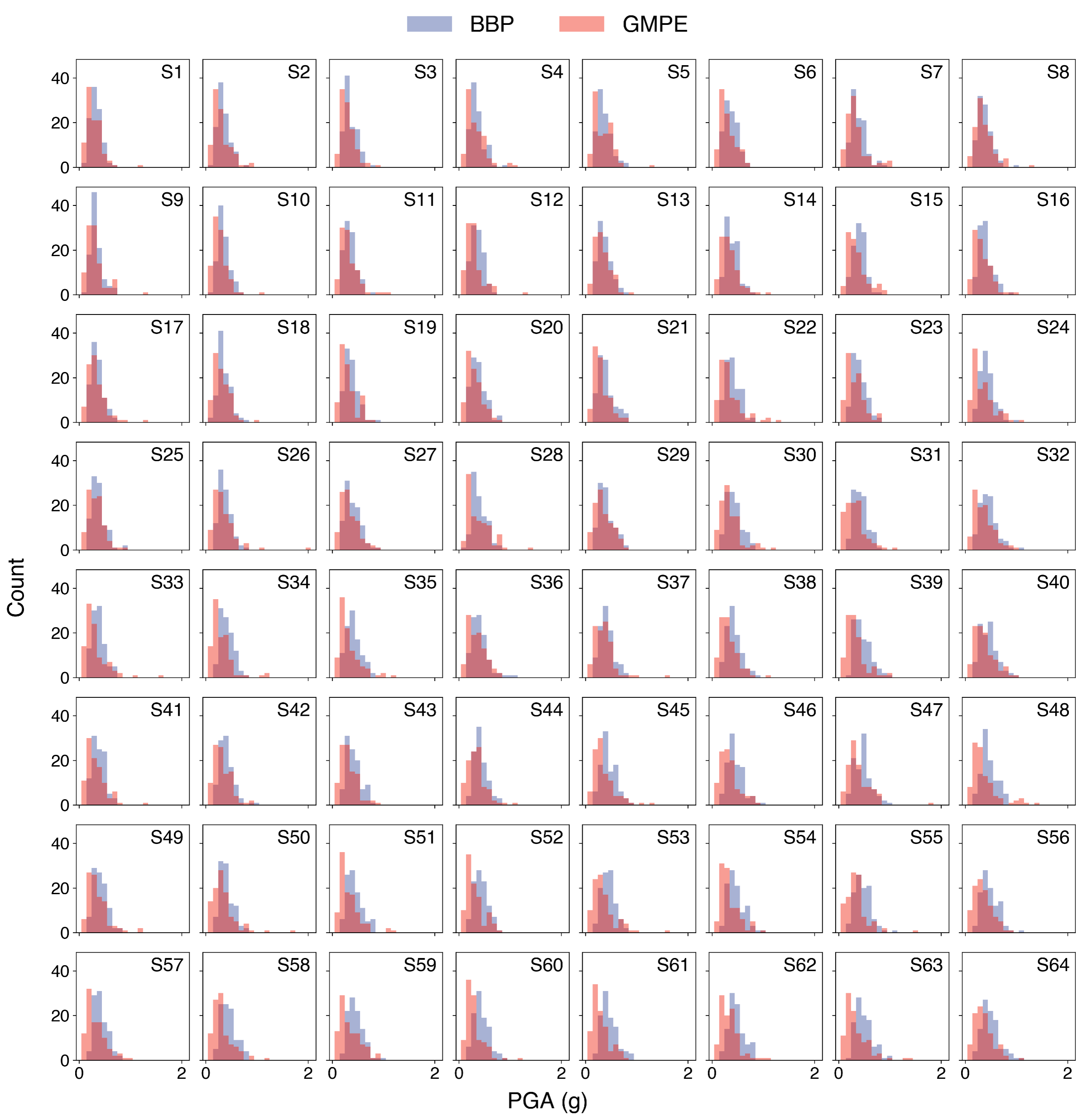}
    \caption{\textbf{Comparison of peak ground acceleration (PGA) distributions between physics-based and empirical models at $M_w=7.0$ across recording stations.}
    Histograms compare PGA values obtained from the physics-based SCEC Broadband Platform (BBP) simulations and the empirical Chiou and Youngs ground-motion prediction equation (GMPE). Both datasets were sampled at the same 64 recording stations (S1--S64) uniformly distributed across Milpitas, with 100 realizations generated for each method.}
    \addcontentsline{sitoc}{subsection}{Supplementary Fig.~\arabic{figure}. Comparison of peak ground acceleration (PGA) distributions between physics-based and empirical models at \texorpdfstring{$M_w=7.0$}{Mw=7.0} across recording stations}
    \label{sup_fig:pga_comparison_per_station_Mw7}
\end{figure}

\newpage
\begin{figure}[H]
    \centering
    \includegraphics[width=1.0\linewidth]{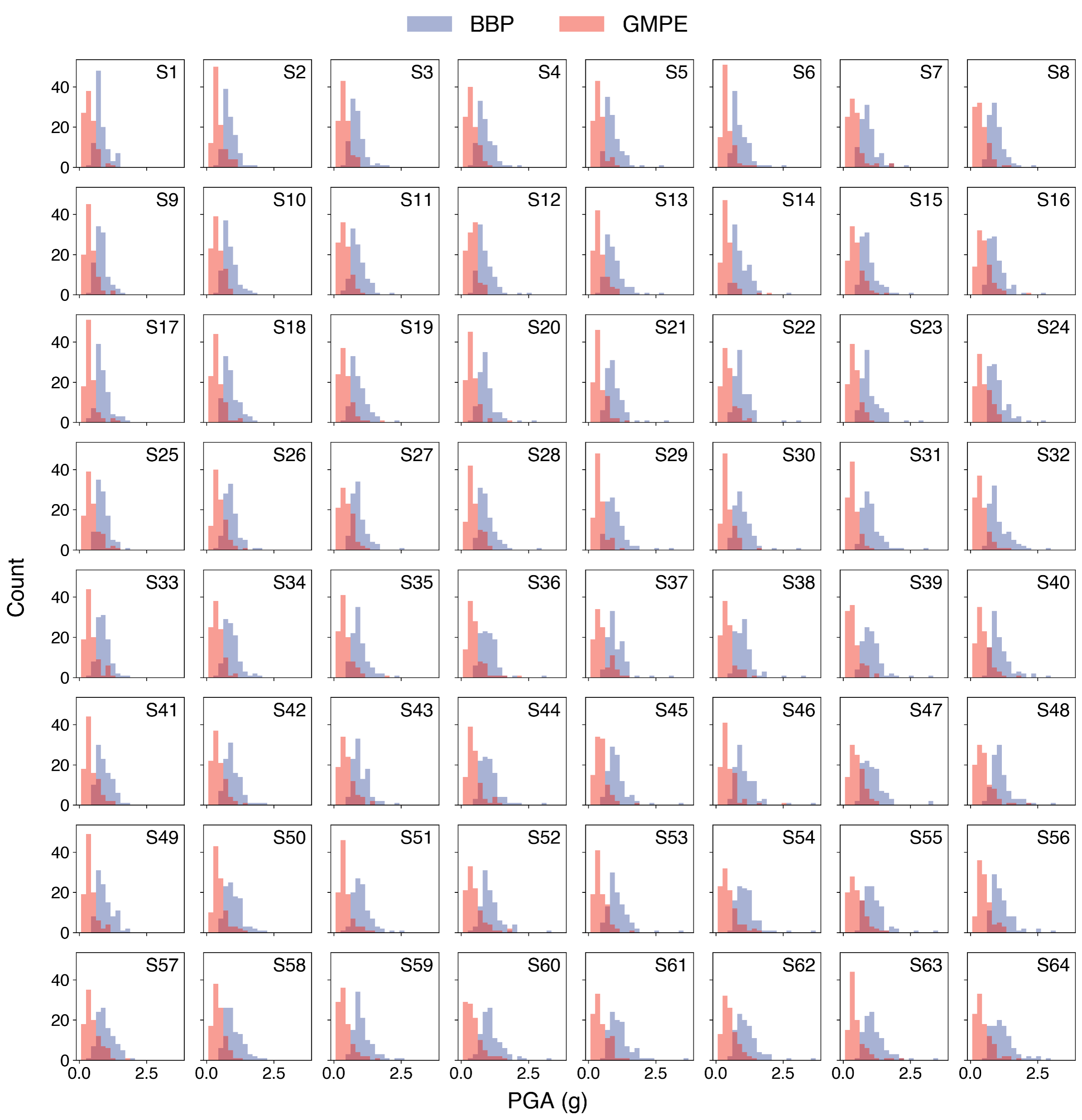}
    \caption{\textbf{Comparison of peak ground acceleration (PGA) distributions between physics-based and empirical models at $M_w=8.0$ across recording stations.}
    Histograms compare PGA values obtained from the physics-based SCEC Broadband Platform (BBP) simulations and the empirical Chiou and Youngs ground-motion prediction equation (GMPE). Both datasets were sampled at the same 64 recording stations (S1--S64) uniformly distributed across Milpitas, with 100 realizations generated for each method.}
    \addcontentsline{sitoc}{subsection}{Supplementary Fig.~\arabic{figure}. Comparison of peak ground acceleration (PGA) distributions between physics-based and empirical models at \texorpdfstring{$M_w=8.0$}{Mw=8.0} across recording stations}
    \label{sup_fig:pga_comparison_per_station_Mw8}
\end{figure}

\newpage
\begin{figure}[H]
    \centering
    \includegraphics[width=1.0\linewidth]{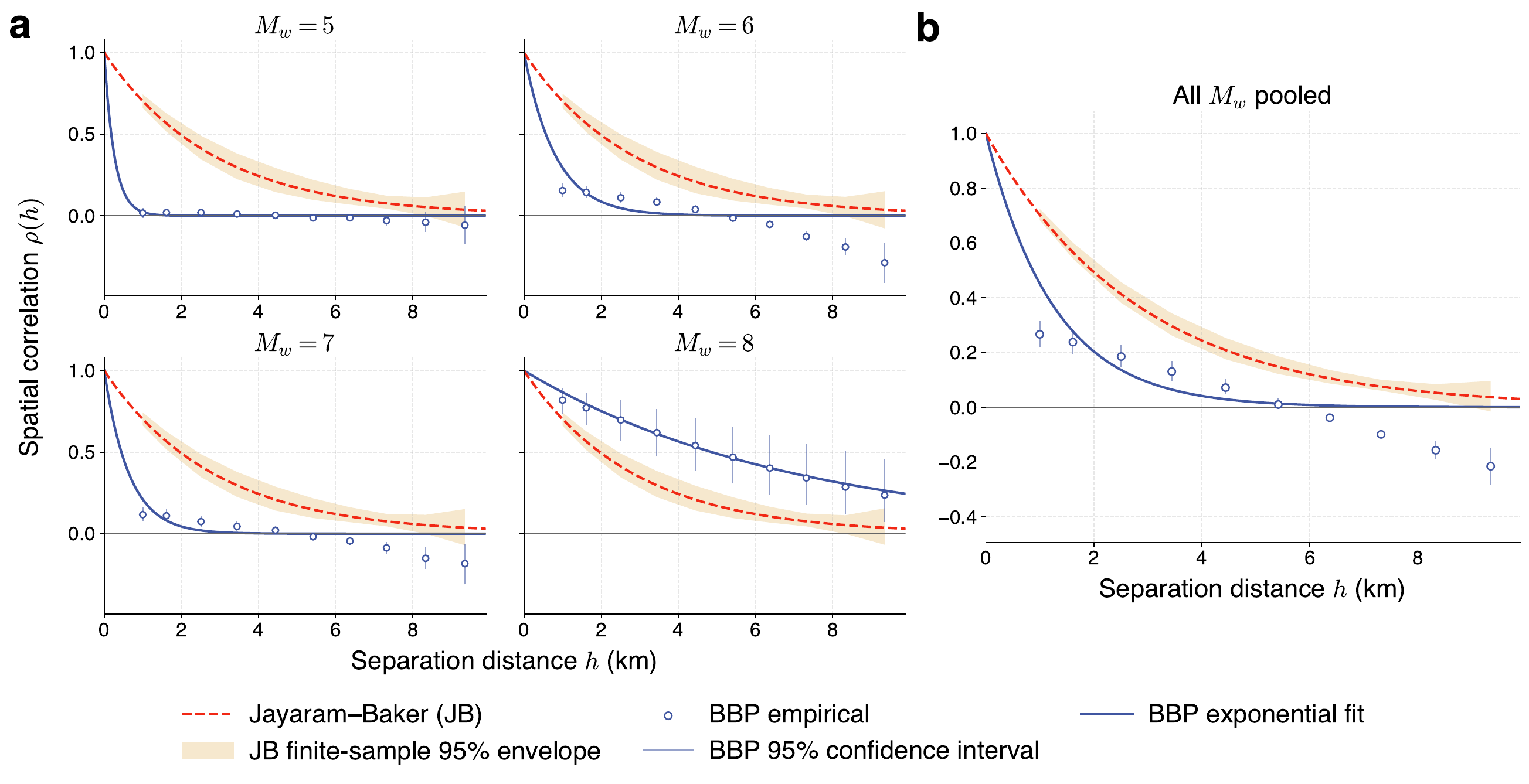}
    \caption{\textbf{Comparison of log-PGA spatial correlations between SCEC-BBP simulations and the Jayaram--Baker model.}
    \textbf{a}, Spatial correlations estimated from SCEC-BBP log-PGA fields at $M_w=5$, 6, 7, and 8 using the same 64 recording stations as in Supplementary Figs.~\ref{sup_fig:pga_comparison_per_station_Mw5}--\ref{sup_fig:pga_comparison_per_station_Mw8}. For each magnitude, the station-specific ensemble mean was removed from log PGA, and correlations were evaluated as a function of station separation distance $h$. Open circles show the distance-binned empirical BBP correlations, vertical bars indicate 95\% bootstrap confidence intervals based on 2,000 resamples of complete BBP spatial fields, and solid blue curves show exponential fits to the BBP estimates. Dashed red curves denote the Jayaram--Baker model used in the GMPE-based simulations, $\rho(h)=\exp(-3h/8.5\,\mathrm{km})$. Shaded regions show the finite-sample 95\% envelopes obtained from correlated Gaussian fields generated with the same station geometry and number of realizations as the corresponding BBP ensembles.
    \textbf{b}, Spatial correlation pooled across all four magnitudes after removing the station-specific mean separately within each $M_w$ and weighting each magnitude by its number of realizations. The BBP correlations generally decay faster than the Jayaram--Baker model for $M_w=5$--7, whereas the $M_w=8$ fields exhibit substantially longer-range dependence.}
    \addcontentsline{sitoc}{subsection}{Supplementary Fig.~\arabic{figure}. Comparison of log-PGA spatial correlations between SCEC-BBP simulations and the Jayaram--Baker model}
    \label{sup_fig:compare_bbp_gmpe_spatial_correlation}
\end{figure}

% =========================
% One bibliography for main+SI (recommended)
% =========================
\clearpage
\phantomsection
\addcontentsline{sitoc}{section}{Supplementary References}
\renewcommand{\refname}{Supplementary References}
\renewcommand{\bibname}{Supplementary References}
\putbib[references_si]
\end{bibunit}

\end{document}